\documentclass{article}
\usepackage{graphicx}  
\usepackage{amsmath}   
\usepackage[compress]{cite}
\usepackage{amssymb}   
\usepackage{bm} 
\usepackage{dcolumn}
\usepackage{color}
\usepackage{mathrsfs}
\usepackage{amsfonts}
\usepackage{enumerate}
\usepackage{varioref}
\RequirePackage[colorlinks,citecolor=blue,urlcolor=magenta,linkcolor=blue]{hyperref}
\def\LL{Lanczos-Lovelock }
\def\gr{general relativity}
\labelformat{section}{Section #1} 
\labelformat{subsection}{Section #1} 
\labelformat{subsubsection}{Section #1}
\labelformat{subsubsubsection}{Section #1}
\labelformat{equation}{Eq.~(#1)} 
\labelformat{figure}{Fig.~#1} 
\labelformat{subfigure}{Fig.~\thefigure#1} 
\labelformat{table}{Tab.~#1} 
\labelformat{appendix}{Appendix #1}
\title{Thermodynamical interpretation  of the geometrical variables associated with null surfaces}
\author{Sumanta Chakraborty
\footnote{sumantac.physics@gmail.com; \hskip 0.05 true cm sumanta@iucaa.in}
\hskip 0.1 true cm and 
T. Padmanabhan
\footnote{paddy@iucaa.in}\\
\\
{\small{IUCAA, Post Bag 4, Ganeshkhind,}}\\
{\small{Pune University Campus, Pune 411 007, India}}
}
\begin{document}
  
\maketitle
\begin{abstract}

The emergent gravity paradigm interprets gravitational field equations as describing the thermodynamic limit of the underlying statistical mechanics of microscopic degrees of freedom of the spacetime. The connection is established by attributing a heat density $Ts$ to the null surfaces  where $T$ is the appropriate Davies-Unruh temperature and $s$ is the entropy density. The field equations can be obtained from a thermodynamic variational principle which extremises the total heat density of all null surfaces. The explicit form of $s$ determines the nature of the theory. We explore the consequences of this paradigm for an arbitrary null surface and highlight the \textit{thermodynamic} significance of various \textit{geometrical} quantities. In particular, we show that: (a) A conserved current, associated with the time development vector in a natural fashion, has direct thermodynamic interpretation in all Lanczos-Lovelock models of gravity. (b) One can generalize the notion of gravitational momentum, 
introduced in arXiv 1506.03814 to all Lanczos-Lovelock models of gravity such that the conservation of the total momentum leads to the relevant field equations. (c) The thermodynamic variational principle which leads to the field equations of gravity can also be expressed in terms of the gravitational momentum in all Lanczos-Lovelock models. (d) Three different projections of gravitational momentum related to an arbitrary null surface in the spacetime lead to three different equations, all of which have  thermodynamic interpretation. The first one reduces to a Navier-Stokes equation for the transverse drift velocity. The second can be written as a thermodynamic identity $TdS = dE + P dV$. The third describes the time evolution of the null surface in terms of suitably defined surface and bulk degrees of freedom. The implications are discussed. 

\end{abstract}
\newpage
\tableofcontents
\newpage 
\section{Introduction}

The dynamical evolution of a fluid or a  gas can be studied without any direct reference to the fact that they are made of microscopic degrees of freedom, viz., atoms and molecules. Such a description was known for centuries before physicists realized that matter is made of discrete entities. But the existence of the microscopic degrees of freedom had always left a clear signature \textit{even at the macroscopic scales} in the form of the heat content of matter. While even the cave men knew the distinction between a hot body and a cold one, the real nature of heat was not well understood until Boltzmann pointed out that the heat content of matter is a direct evidence for the existence of the microscopic degrees of freedom. Boltzmann essentially said: ``If you can heat it, it must have microstructure''. In other words, the microscopic degrees of freedom make their presence felt even at the macroscopic scales (in the form of  heat) and the correct (thermodynamic) description  had taken this into account 
phenomenologically in terms of temperature etc., even before Boltzmann explained to us what it really is. The key new element in thermodynamics which is absent in, say, the Newtonian mechanics of point particles, is the heat content $TS$ of the matter, which is the difference $(F-E)$ between the free energy and the internal energy of the system.  In  terms of densities, for systems with zero chemical potential which we will be interested in, the heat density is $Ts = P+\rho$ where $s$ is the entropy density, $\rho$ is the energy density and $P$ is pressure. 

Over the decades, it has been realized that spacetimes --- through the existence of the null surfaces  which act as horizons to certain class of observers --- also  possess the heat density $Ts$ \cite{Bekenstein:1973ur,Bekenstein:1974ax,Hawking:1974sw,Hawking:1976de,Davies:1974th,Unruh:1976db,Gibbons:1977mu,Iyer:1994ys}. This connection between thermodynamics and spacetime dynamics forms the core of the emergent gravity paradigm \cite{Padmanabhan:2009vy,Padmanabhan:2003gd,Padmanabhan:2014jta,Gravitation,Padmanabhan:2013xyr,Padmanabhan:2007en,
Padmanabhan:2002sha,Padmanabhan:2007xy}. This paradigm  rests on  the results obtained over the last decade or so which suggest that the field equations of gravity, in a large class of theories, have the same status as the equations describing, say, a fluid or an elastic solid. That is, gravity emerges in the thermodynamic limit of the statistical mechanics of the atoms of spacetime \cite{Padmanabhan:2009kr,Padmanabhan:2010xh,Padmanabhan:2013nxa,Chakraborty:2014rga}. The heat 
density of spacetime is an evidence for the existence of the atoms of spacetime just as the heat density of matter was successfully interpreted by Boltzmann as evidence for the atomic structure of matter. There is considerable amount of evidence which suggests that this is indeed a useful and correct point of view to pursue \cite{Parattu:2013gwb,Chakraborty:2014joa}. 

The development of emergent gravity paradigm has helped us to understand several interesting features of classical gravity itself \cite{Padmanabhan:2010xe}, further bolstering our confidence in the veracity of this approach. In particular, the following results are of specific interest to the current work.
\begin{enumerate}

\item The first one has to do with the curious relationship between Einstein's field equations and the structure of null surfaces in the spacetime. Previous works have shown that they manifest in \textit{three} different ways: 
\begin{itemize}

\item In the most general situation, there arises an identification between Navier-Stokes equation and Einstein's equation. 
Einstein's equations, when projected on an \emph{arbitrary} null surface, in \textit{any} spacetime, leads to Navier-Stokes equation of fluid dynamics \cite{Padmanabhan:2010rp,Kolekar:2011gw}. (This  generalizes the previously known results for black hole horizons \cite{Damour:1979,Damour:1982}.)

\item From the   Einstein equations applied to  a null surface, one can get \cite{Padmanabhan:2002sha,Cai:2005ra,Akbar:2006er,Kothawala:2007em,Paranjape:2006ca} a thermodynamic identity of the form $T\delta _{\lambda}S=\delta _{\lambda}E+P\delta _{\lambda}V$  in which the symbols have their usual meanings and the variation can be interpreted as changes due to virtual displacement of the null surface along null geodesics parametrized by the affine parameter $\lambda$ off the surface. Initially proved for a few  configurations with a high level of symmetry \cite{Paranjape:2006ca,Kothawala:2009kc} this result has now been generalized  for arbitrary null surfaces in both \gr\ and \LL theories of gravity \cite{Chakraborty:2015aja,Chakraborty:2015wma}.

\item A comparatively more approximate relationship between the null surfaces and Einstein's equation emerged from the early work \cite{Jacobson:1995ab}, which `derived' Einstein's field equations using the local Rindler horizon as a null surface and  the Clausius relation. This  relies heavily on the structure of Raychaudhuri equation as well as the assumptions: (a) the entropy density is one quarter of the transverse area and, more importantly, (b) the quadratic terms in the Raychaudhuri equation (involving the squares of shear and expansion) can be set to zero. Since neither the Raychaudhuri equation nor the assumption that entropy is proportional to the horizon area hold for theories more general than Einstein's gravity,  this approach could not be generalized in a simple manner to more general class of theories.

\end{itemize}

\item The second result pertains to the derivation of gravitational field equations from a thermodynamic variational principle \cite{Padmanabhan:2007en,Padmanabhan:2007xy}. It turns out that maximizing the sum of gravitational heat density and matter heat density, associated with every  null surface in the spacetime, leads to the appropriate gravitational field equations. This derivation of the field equations is consistent with the emergent gravity paradigm and is much more general than  the other approaches (like, for example, \cite{Jacobson:1995ab}) used in the literature to obtain the same. Not only does the approach has a high level of logical simplicity, it also generalizes smoothly to all \LL models.  

\end{enumerate}
In this paper, we shall revisit these issues and elaborate further on them using another recent development, which was also motivated by the emergent gravity paradigm. This was the introduction of  the notion of the momentum attributed to the spacetime by a class of observers,  proposed in \cite{Padmanabhan:2013nxa,Padmanabhan:2015lla}. This proposal was in the context of Einstein's theory. But, since virtually every result of emergent gravity paradigm could be generalized in a meaningful way  to \LL models, we would expect the notion of gravitational momentum  to possess a similar generalization. We will show that this is indeed the case. 

Further, it seems reasonable to expect that: (a) The appropriate projections of the gravitational momentum must allow us to obtain the above results --- describing Einstein's equations in a thermodynamic language --- in a straightforward manner. (b) The thermodynamic variational principle should have a simple representation in terms of the gravitational momentum.

We will see that these expectations are also borne out. Our analysis will also highlight the thermodynamic significance of several variables which were originally considered purely geometrical.

To achieve these goals, we will use two different foliations of the spacetime. The first one is the standard (1+3) foliation in terms of space-like hypersurfaces determined by constant values of a suitable time function $t(x)$. The second is a coordinate system which is adapted to a  fiducial null surface. This adaptation is closely related to the notion of a local Rindler horizon \cite{Jacobson:1995ab} and will  prove to be quite useful in illustrating the relationship between field equations and the null surface thermodynamics. We will see that the  Noether current, gravitational momentum and related constructs, associated with  the time evolution vector field in the two foliations  allow us to obtain the results mentioned above.

The paper is organized as follows: In \ref{Paper06_Noether} we introduce  the Noether current, gravitational momentum and a closely related entity which we call the reduced gravitational momentum in \gr\ and \LL gravity. The next section deals with the two coordinate systems, one associated with the (1+3) foliation of the spacetime and the other being the  Gaussian null coordinates adapted to an arbitrary null surface. In \ref{Paper06_NoetherThermo} we discuss the Noether current and the associated thermodynamic results for both the (1+3) foliation and the Gaussian null coordinates. The thermodynamic interpretation of the reduced gravitational momentum is given in \ref{Paper06_Reduced} and the thermodynamic variational principle for null surfaces is presented in \ref{Paper06_ThermoVar}. The final section, i.e., \ref{Paper06_Project} discusses the three different projections of the gravitational momentum vis-a-vis a null surface and their thermodynamic interpretation. We end the paper with a 
short discussion on our results. 

We will set $c,\hbar$ and $16\pi G$ to be unity for most part of our discussion. (Sometimes, when we switch to $G=1$ units, it  will be mentioned specifically.) Thus Einstein's field equations in this notation takes the form $2G_{ab}=T_{ab}$. The Latin letters, $a,b,c,\ldots$ runs over all the spacetime indices, the Greek letters, $\mu,\nu,\alpha, \ldots$ runs over all the spatial indices and the upper case Latin letters, $A,B,C,\ldots$ runs over the co-dimension two surface.
\section{Noether Current and the Gravitational Momentum}\label{Paper06_Noether}

Given an arbitrary vector field $v^a$ in the spacetime, one can define three other geometrical objects which depend on it and have direct thermodynamic significance under certain circumstances.  The first is a conserved current, $J^a[v]$  which can be introduced without invoking any symmetry or invariance  principle. (This happens to be the usual Noether current; but the usual way of deriving it using diffeomorphism invariance of Hilbert action is misleading since it suggests, $J^a[v]$ has something to do with action and gravitational dynamics. It has nothing to do with either and its conservation is a trivial algebraic identity.) The second is the gravitational momentum associated with a vector field $P^a[v]$ which was introduced in Ref. \cite{Padmanabhan:2015lla}. The third is a closely related vector (that appears in the definitions of both $J^{a}$ and $P^{a}$) which we will call the reduced gravitational momentum $\mathcal{P}^{a}$. Each of these vector fields, associated with a given vector field $v^a$, 
can be defined for both Einstein's gravity as well as for \LL models. We will now introduce these vector fields.
\subsection{Noether current}

Given an arbitrary vector field $v^{a}$, we can immediately construct a conserved current  $J^{a}=\nabla _{b}J^{ab}$
from the antisymmetric 2-tensor $ J^{ab}=\nabla ^{a}v^{b}-\nabla ^{b}v^{a}$. (The normalization of this current is arbitrary; we have chosen it so as to make the later results transparent and simple. To match the conventional description, the left hand side should be multiplied by $16\pi G$ which we have set to unity.) This construction does not require any mention of diffeomorphism invariance or action principles and is completely devoid of any dynamical content at this stage. Elementary algebra now leads to the expression:
\begin{equation}
\sqrt{-g}\, J^{a}(v)=\sqrt{-g}[\nabla _{b}\left(\nabla ^{a}v^{b}-\nabla ^{b}v^{a}\right)]
=2\sqrt{-g}\, R^{a}_{b}v^{b}+f^{bc}\pounds _{v}N^{a}_{bc}
\label{Paper06_Sec03_Eq02}
\end{equation}
where we have defined 
\begin{align}\label{Paper06_Sec_01_Eq01}
f^{ab}=\sqrt{-g}g^{ab};\qquad N^{c}_{ab}=-\Gamma ^{c}_{ab}+\frac{1}{2}\left(\delta ^{c}_{a}\Gamma ^{d}_{db}+\delta ^{c}_{b}\Gamma ^{d}_{ad}\right)
\end{align}
These variables contain the same amount of information as the metric and the connection but has more direct thermodynamic interpretation; see ref. \cite{Parattu:2013gwb}. These expressions are generally covariant because the Lie derivative of the connection $\pounds_v\Gamma^c_{ab}$, given by
\begin{align}\label{Paper06_Sec04_Eq04}
\pounds _{v}\Gamma ^{a}_{bc}=\nabla _{b}\nabla _{c}v^{a}+R^{a}_{~cmb}v^{m}
\end{align}
is generally covariant, making $\pounds_vN^c_{ab}$ a generally covariant object. 

There is a natural generalization of this current which can be introduced as follows. We begin by noting that the $J^{ab}$ and $J^a$ can be expressed in an equivalent form as 
\begin{align}
J_{ab}=2P_{ab}^{cd}\nabla _{c}v_{d};\qquad
J_{a}=2P_{ab}^{cd}\nabla ^{b}\nabla _{c}v_{d};\quad
P_{ab}^{cd}\equiv(1/2)(\delta _{a}^{c}\delta _{b}^{d}-\delta_{a}^{d}\delta_{c}^{b})
\label{Paper06_SecLL_09}
\end{align}
where $P^{ab}_{cd}$ is a tensor (which we will call the entropy tensor for reasons which will become clear later on) called determinant tensor. This tensor gives us the Ricci scalar from the curvature tensor:
\begin{align}\label{Paper06_NewF01}
R=\frac{1}{2}(\delta ^{a}_{c}\delta ^{b}_{d}-\delta ^{a}_{d}\delta ^{c}_{b})R^{cd}_{ab}=P^{ab}_{cd}R^{cd}_{ab}
\end{align}
which shows if we take $R=F(R^{ab}_{cd}, g_{ik})$ to be a function of the (2,2) Riemann tensor $R^{ab}_{cd}$ and the metric tensor $g_{ik}$, then it is actually independent of the metric tensor. That is, treating $R^{ab}_{cd}$ and $g_{ab}$ to be algebraically independent we can also define $P^{ab}_{cd}$ tensor through the following relations:
\begin{equation}\label{Paper06_SecLL_02}
P^{ab}_{cd}=\left(\frac{\partial R}{\partial R_{ab}^{cd}} \right)_{g_{ik}}; \quad \left(\frac{\partial R}{\partial g_{ab}}\right)_{R^{ab}_{cd}}=0;\qquad 
\nabla _{a}P^{ab}_{cd}=0.
\end{equation}
Clearly, $P^{ab}_{cd}$ has the symmetries of the curvature tensor and is divergence-free in all the indices. 

An obvious generalization of this tensor can be obtained by replacing $R$ by some other arbitrary function $F(R^{ab}_{cd},\delta ^{p}_{q})$ constructed out of $R^{ab}_{cd}$ and Kronecker delta function, and then define $P^{ab}_{cd}$ in an analogous manner. That is, we define $P^{ab}_{cd}$ by the relations 
\begin{equation}\label{Paper06_SecLL_02TP}
P^{ab}_{cd}=\left(\frac{\partial F}{\partial R_{ab}^{cd}} \right)_{g_{ik}}; \quad \left(\frac{\partial F}{\partial g_{ab}}\right)_{R^{ab}_{cd}}=0;\qquad 
\nabla _{a}P^{ab}_{cd}=0.
\end{equation}
This requires finding the most general scalar function $F(R^{ab}_{cd},\delta ^{p}_{q})$ for which the condition $\nabla _{a}P^{ab}_{cd}=0$ in \ref{Paper06_SecLL_02TP} is identically satisfied. This problem can be completely solved \cite{Lovelock:1971yv,Lanczos:1932zz,Lanczos:1938sf}. It turns out that the most general function which satisfies this criterion can be expressed as the sum:
\begin{equation}\label{Paper06_SecLL_06}
F=\sum _{m}c_{m}F_{m}
\end{equation}
where $c_m$s are constants and $F_{m}$ is given by
\begin{align}\label{Paper06_NewF02}
F_{m}=\frac{1}{2^{m}}\delta ^{aba_{2}b_{2}\ldots a_{m}b_{m}}_{cdc_{2}d_{2}\ldots c_{m}d_{m}}R^{cd}_{ab}
R^{c_{2}d_{2}}_{a_{2}b_{2}}\ldots R^{c_{m}d_{m}}_{a_{m}b_{m}}
\end{align}
where $\delta ^{aba_{2}b_{2}\ldots a_{m}b_{m}}_{cdc_{2}d_{2}\ldots c_{m}d_{m}}$ is the completely antisymmetric $m$-dimensional determinant tensor. The scalar $F$ is constructed out of $R^{ab}_{cd}$ and the Kronecker delta function $\delta ^{c}_{d}$, without any metric $g_{ab}$ present in it. Then $P^{ab}_{cd}$ for the $m$-th term in the sum in \ref{Paper06_SecLL_06} can be obtained directly using \ref{Paper06_SecLL_02TP} and \ref{Paper06_NewF02}, which leads to,
\begin{equation}\label{Paper06_SecLL_07}
P^{ab}_{cd}=\frac{\partial F_{m}}{\partial R^{cd}_{ab}}
=\frac{m}{2^{m}}\delta ^{aba_{2}b_{2}\ldots a_{m}b_{m}}_{cdc_{2}d_{2}\ldots c_{m}d_{m}}
R^{c_{2}d_{2}}_{a_{2}b_{2}}\ldots R^{c_{m}d_{m}}_{a_{m}b_{m}}\equiv mQ^{ab}_{cd}
\end{equation}
The $m$th order term $F_{m}$ can be expressed as $F_{m}=Q^{ab}_{cd}R^{cd}_{ab}$; for $m=1$, $P^{ab}_{cd}$ and $Q^{ab}_{cd}$ coincides. 

Using the definition of $P^{abcd}$ from \ref{Paper06_SecLL_07}, in \ref{Paper06_SecLL_09} leads to a natural generalization of $J^{ab}$ and $J^a$ which, as we shall see later, will be closely related to the \LL models of gravity. In particular the Noether current in \ref{Paper06_SecLL_09} now becomes
\begin{align}\label{Paper06_nLL01}
J^{a}(v)=2P^{ab}_{cd}\nabla _{b}\nabla^{c}v^{d}=2\mathcal{R}^{a}_{b}v^{b}+2P_{p}^{~qra}\pounds _{v}\Gamma ^{p}_{qr}
\end{align}
where $\mathcal{R}^{a}_{b}$ is defined as: $\mathcal{R}^{a}_{b}\equiv P^{ai}_{jk}R_{bi}^{jk}$. 

The relationship between $J^{ab}$ and $J^a$ has an obvious electromagnetic analogy which is useful in some calculations to get an intuitive grasp of the results. Given any vector field $v^{a}$, we can always construct the Noether potential $J^{ab}(v)$ in \gr, which is exactly analogous to the construction of the electromagnetic field tensor $F^{ab}$ starting from the vector potential $A^{j}=v^{j}$ (see \ref{Paper06_SecLL_09}). The Noether current $J^a = \nabla_b J^{ab}$ bears the same relation to $J^{ab}$ as the electromagnetic current does to $F^{ab}$. The dual of the tensor $J^{ab}$ is $\tilde{J}^{ab}=\epsilon ^{abcd}J_{cd}$ and the corresponding (magnetic) current 
\begin{align}\label{Paper06_SecLL_11}
 \tilde{J}^{a}= \nabla _{b}\tilde{J}^{ab}=2\epsilon ^{abcd}\nabla _{b}\nabla _{c}v_{d}
=\epsilon ^{abcd}\left(\nabla _{b}\nabla _{c}v_{d}-\nabla _{c}\nabla _{b}v_{d}\right)
=-\epsilon ^{abcd}R_{ibcd}v^{i}
\end{align}
should vanish identically, because this is just electromagnetism in disguise. This is indeed true since we have the identity,
\begin{align}\label{Paper06_SecLL_12}
\epsilon ^{abcd}R_{ibcd}&=\frac{1}{3}\epsilon ^{abcd}R_{ibcd}+\frac{1}{3}\epsilon ^{adbc}R_{idbc}+\frac{1}{3}\epsilon ^{acdb}R_{icdb}
\nonumber
\\
&=\frac{1}{3}\epsilon ^{abcd}\left(R_{ibcd}+R_{idbc}+R_{icdb}\right)=0
\end{align}
Further, as in the case of electromagnetism, we can also define an ``electric'' field and a ``magnetic'' field as measured by an observer with four-velocity $u_{a}$ as 
\begin{align}
E^{a}(v\vert u)&= u_{b}J^{ab}(v)=u_{b}\left(\nabla ^{a}v^{b}-\nabla ^{b}v^{a}\right)
\label{Paper06_SecLL_13}
\\
B^{a}(v\vert u)&=\frac{1}{2}\epsilon ^{abcd}u_{b}J_{cd}(v)=\epsilon ^{abcd}u_{b}\nabla _{c}v_{d}
\label{Paper06_SecLL_14}
\end{align}
where $E^{a}(v\vert u)$ stands for ``electric'' part of $J^{ab}[v]$ as measured by an observer with four-velocity $u_{a}$. Just as in electromagnetism, we have $u_{a}E^{a}(v\vert u)=0=u_{a}B^{a}(v\vert u)$ so that $E^a, B^a$ are purely spatial.
The generalization of the electric field to \LL gravity is straightforward with $P^{abcd}$ of \gr\ being replaced by that of \LL model. This leads to the  expression for electric field given by:
\begin{align}
E^{a}(v\vert u)&= u_{b}J^{ab}(v)=2P^{abcd}u_{b}\nabla _{c}v_{d}
\label{Paper06_SecLL_New01}
\end{align}
The situation with magnetic field is somewhat more complicated in $d>4$ dimensions. The $J^{ab}$ has $d(d-1)/2$ components of which the electric vector $E_a\equiv u_bJ^{ab}$, (with the constraint $u^aE_a=0$) contains information about $d-1$ components. The magnetic field contains the information about the remaining $(d-1)(d-2)/2$ independent components which is
greater than $(d-1)$ for $d>4$.  Since in \LL gravity we are working in $d$ dimensions $(d>4)$ there is no vector field that can account for all the independent components of magnetic field.  In d-dimensions the magnetic field is given by completely antisymmetric $(d-3)$ rank tensor $B^{n_{1}\ldots n_{d-3}}(v\vert u)=\epsilon ^{n_{1}\ldots n_{d-3}abc}u_{a}J_{bc}(v)$. Thus total number of independent components is $^{d-1}C_{d-3}=(d-1)(d-2)/2$ as to be expected. Because of this rather complicated structure we shall not  discuss the magnetic field and shall concentrate on electric field, which, fortunately, turns out to be adequate.
\subsection{Gravitational momentum}

The second structure we want to associate with an arbitrary vector field $v^{a}$ is the gravitational four-momentum density $P^{a}(v)$, defined --- in the context of \gr\ --- as 
\begin{equation}\label{Paper06_Sec03_Eq03}
 P^{a}(v)=-Rv^{a}-g^{ij}\pounds _{v}N^{a}_{ij}
\end{equation}
From \ref{Paper06_Sec03_Eq02} we can substitute for the Lie variation term in \ref{Paper06_Sec03_Eq03}, which relates the gravitational momentum and the Noether current as
\begin{equation}\label{Paper06_Sec03_Eq04}
J^{a}(v)=-P^{a}(v)+ 2G^{a}_{b}v^{b}
\end{equation}
The physical meaning of the gravitational momentum can be understood from the following result. (This  was motivated and discussed in detail in \cite{Padmanabhan:2013nxa,Padmanabhan:2015lla}. We will not repeat the motivation and logic behind this definition here). Consider the special case in which $v^a$ is the  velocity of an arbitrary observer, who will attribute to the matter, with energy momentum tensor $T_{ab}$, the momentum density $\mathcal{M}^{a}=-T^{a}_{b}v^{b}$. We would expect the total momentum associated with matter plus gravitation to be conserved \cite{Padmanabhan:2015lla} in nature, for all observers. This condition requires: 
\begin{align}\label{Paper06_New_01}
0=\nabla _{a}\left(P^{a}+\mathcal{M}^{a}\right)&=\nabla _{a}\left(-J^{a}+2G^{a}_{b}v^{b}-T^{a}_{b}v^{b}\right)
\nonumber
\\
&=\nabla _{a}\left(2G^{a}_{b}v^{b}-T^{a}_{b}v^{b}\right)=\left(2G^{a}_{b}-T^{a}_{b}\right)\nabla _{a}v^{b}\equiv \mathcal{S}^{ab}\nabla _{a}v_{b}
\end{align}
where $\mathcal{S}^{ab}\equiv (2G^{ab}-T^{ab})$ is a symmetric tensor and in the last line we have used Bianchi identity and the fact that $J^{a}$ and $T^{a}_{b}$ are conserved. The above relation should hold for any normalised time like vector field $v^{a}$, which requires   $\mathcal{S}^{ab}=0$, i.e., $G_{ab}=8\pi T_{ab}$, which are  the field equation for gravity. (This result should be obvious from the fact that $\nabla _{a}v_{b}$ can be chosen to be arbitrary at any event even for normalised timelike vector fields. A more formal proof, suggested by S. Date, goes as follows: Choose first $v^a$ to be  a normalized \textit{geodesic} velocity field with $v^a v_a =-1$ and $v^a\nabla_a v^b =0$. Then the most general $\mathcal{S}^{ab}$ which satisfies $\mathcal{S}^{ab}\nabla_a v^b=0$ has the form  $\mathcal{S}^{ab}=\alpha (X^{a}v^{b}+X^{b}v^{a})+\beta v^{a}v^{b}$ with two arbitrary functions $\alpha$ and $\beta$ and an arbitrary vector $X^{a}$ which can be chosen without loss of generality to be purely spatial,
 i.e, $v_aX^a=0$. Choose next the velocity field to be $u_{a}=-N\nabla _{a}t$. Using the form of $\mathcal{S}^{ab}\nabla_au_b =0$ leads to $\alpha =\beta =0$. This immediately gives $\mathcal{S}^{ab}=0$.)

We will now generalise the notion of the gravitational momentum in exact analogy with the way we generalised the Noether current, by the relation:
\begin{align}\label{Paper06_SecLL_24}
 P^{a}(v)=-\mathcal{R}v^{a}-2P_{p}^{~qra}\pounds _{v}\Gamma ^{p}_{qr}
\end{align}
where 
\begin{equation}\label{Paper06_SecLL_nNew02}
m\mathcal{R}=P^{ab}_{cd}R_{ab}^{cd}=\delta ^{a}_{b}\mathcal{R}^{b}_{a}                            
\end{equation} 
Simple algebra gives the equivalent form:
\begin{align}\label{Paper06_SecLL_nNew01}
 P^{a}(v)=-J^{a}(v)+2E^{a}_{b}v^{b}
\end{align}
where $J^{a}(v)$ is the Noether current defined in \ref{Paper06_SecLL_09} and
\begin{equation}
E^{a}_{b}\equiv P^{ai}_{jk}R_{bi}^{jk}-\frac{1}{2}\delta^{a}_{b}\mathcal{R};\qquad
m\mathcal{R}\equiv P^{ab}_{cd}R_{ab}^{cd}
\end{equation} 
It is possible to prove that $E_{ab}$ is symmetric  \cite{Padmanabhan:2011ex} and that $\nabla_aE^a_b =0$. To demonstrate the latter, note that the covariant derivative of $E_{ab}$ involves two parts,
\begin{align}\label{Paper06_New_13}
\nabla _{a}\left(P^{aijk}R_{bijk}\right)&=P^{aijk}\nabla _{a}R_{jkbi}=P^{aklm}\left(-\nabla _{k}R_{lmab}-\nabla _{b}R_{lmka}\right)
\nonumber
\\
&=-\nabla _{k}\left(P^{kalm}R_{balm}\right)+P^{pq}_{rs}\nabla _{b}R_{pq}^{rs}
\\
\partial _{b}\mathcal{R}&=P^{pq}_{rs}\nabla _{b}R^{rs}_{pq}
\end{align}
which can be obtained by   working in a local inertial frame. These two can be combined to yield $\nabla _{a}E^{a}_{b}=- \nabla_{a}E^{a}_{b}$, thereby leading to $\nabla_{a}E^{a}_{b} =0$.

With this definition of $P^a$, the conservation of total momentum of matter plus gravity leads to,
\begin{align}\label{Paper06_SecLL_25}
0=\nabla _{a}\left(P^{a}+\mathcal{M}^{a}\right)&=\nabla _{a}\left(-J^{a}+2E^{a}_{b}v^{b}-T^{a}_{b}v^{b}\right)
\nonumber
\\
&=\left(2E^{a}_{b}-T^{a}_{b}\right)\nabla _{a}v^{b}
\end{align}
Invoking the same argument as in the case of \gr\ and requiring the above relation to hold for all timelike vector fields $v^{a}$ lead to 
\begin{equation}
E^{a}_{b}\equiv P^{ai}_{jk}R_{bi}^{jk}-\frac{1}{2}\delta^{a}_{b}\mathcal{R}
=\frac{1}{2}T^{a}_{b},                  
\end{equation} 
which are indeed the field equations in \LL gravity. Thus imposing the condition that total momentum, of matter plus gravity is conserved, with the gravitational momentum given by \ref{Paper06_SecLL_24}, leads to the gravitational field equations in all \LL theories of gravity. This also allows us to associate the Noether current in \ref{Paper06_nLL01} with the \LL models.
\subsection{Reduced gravitational momentum}

We notice that the combinations 
\begin{equation}
\mathcal{P}^{a}=-g^{ij}\pounds _{v}N^{a}_{ij};\qquad \mathcal{P}^{a}=-2P_{p}^{~qra}\pounds _{v}\Gamma ^{p}_{qr}
\end{equation}
appear quite naturally in both Noether current and in the gravitational momentum in Einstein's theory and in \LL models. We shall call this combination \textit{reduced} gravitational momentum $\mathcal{P}^{a}$. (We will see later that $\mathcal{P}^{a}$ is closely related to the rate of  production of heat per unit area on null surfaces.)  

The algebraic reason for the occurrence of this combination is as follows. It turns out that, in the thermodynamic interpretation of gravity, the combination $\bar{T}_{ab}=T_{ab}-(1/2)Tg_{ab}$ occurs more naturally than the energy momentum tensor  $T_{ab}$ with the field equations often arising \cite{Padmanabhan:2013nxa,Chakraborty:2014rga} in the form $2R_{ab}=\bar{T}_{ab}$ rather than as $2G_{ab}={T}_{ab}$ in Einstein's gravity. The total momentum of gravity plus matter can be expressed, \textit{on-shell}, in the form:
\begin{align}\label{Paper06_Sec_03_Eq06}
 \left(P^{a}+\mathcal{M}^{a}\right)&=-g^{ij}\pounds _{v}N^{a}_{ij}-Rv^{a}- T^{a}_{b}v^{b}
\nonumber
\\
&=-g^{ij}\pounds _{v}N^{a}_{ij}+\frac{1}{2} T\delta ^{a}_{b}v^{b}- T^{a}_{b}v^{b}
\nonumber
\\
&=-g^{ij}\pounds _{v}N^{a}_{ij}- \bar{T}^{a}_{b}v^{b}\equiv  (\mathcal{P}^{a}+\bar{\mathcal{M}}^{a})
\end{align}
where $\bar{\mathcal{M}^{a}}=-\bar{T}^{a}_{b}v^{b}$ is the matter momentum associated with $\bar{T}^{a}_{b}$.
This shows that the vector $\mathcal{P}^{a}=-g^{ij}\pounds _{v}N^{a}_{ij}$ bears the same relation to $\bar{T}^{a}_{b}$ as $P^a$ does with ${T}^{a}_{b}$. Just as $\bar{T}^{a}_{b}$ appears more naturally in the emergent gravity paradigm, the $\mathcal{P}^{a}$ will also appear repeatedly in our discussions. 

The notion of reduced gravitational momentum can also be generalized to the \LL models as well. For pure $m$- th order \LL gravity we have the relation $2[m-(D/2)]L= T$. So we can write,
\begin{align}\label{Paper06_SecLL_26}
 \left(P^{a}+\mathcal{M}^{a}\right)&=-2P_{p}^{~qra}\pounds _{v}\Gamma ^{p}_{qr}-Lv^{a}- T^{a}_{b}v^{b}
\nonumber
\\
&=-2P_{p}^{~qra}\pounds _{v}\Gamma ^{p}_{qr}-\frac{1}{2[m-(D/2)]}T\delta ^{a}_{b}v^{b}- T^{a}_{b}v^{b}
\nonumber
\\
&=-2P_{p}^{~qra}\pounds _{v}\Gamma ^{p}_{qr}- \bar{T}^{a}_{b}v^{b}\equiv  \mathcal{P}^{a}+ \bar{\mathcal{M}}^{a}
\end{align}
where the reduced gravitational momentum is  naturally defined as
\begin{equation}\label{Paper06_nNew10}
\mathcal{P}^{a}=-2P_{p}^{~qra}\pounds _{v}\Gamma ^{p}_{qr}
\end{equation}
In the case of \gr\ the gravitational momentum $\mathcal{P}^{a}$ defined in \ref{Paper06_nNew10} goes over to the $-g^{ij}\pounds _{v}N^{a}_{ij}$ term, as it should. 
\section{Coordinate systems and the associated vector fields}\label{sec:coordsys}

The Noether current and gravitational momentum require a vector field $v^a$ for their definition and we expect them to have simple physical interpretations when we use naturally defined vector fields in the spacetime, associated with the foliations we use to describe the geometry. Our next task is to introduce these foliations and the associated vector fields. We will be using two different foliations and  the corresponding coordinate systems in our work. The first one is based on the standard (1+3) foliation while the second one is adapted to a particular null surface. Both of these will turn out to be useful in understanding the thermodynamic interpretation of the spacetime dynamics.
\subsection{(1+3) Foliation and the Associated Vector Fields}\label{Paper06_Foliation}

This is completely straight forward and we will follow the conventions introduced in \cite{Padmanabhan:2013nxa}. Our primary interest is in the case of a coordinate system adapted to a null surface (which we will discuss in the next \ref{Paper06_GNC}) and the purpose of this section is just to recall the key formulas needed later for comparison.

In the given  spacetime, we  introduce an \textit{arbitrary} $1+(d-1)$ foliation based on a time function $t(x^a)$, with a unit normal $u_a(x^i)\propto \nabla_a t$. This  splits the metric $g_{ab}$ into  the lapse ($N$), shift ($N_\alpha$) and the $(d-1)$-metric $h_{ab} = g_{ab} + u_au_b$. The unit normal to these  hypersurfaces is $u_{a}=-N\nabla _{a}t$  which reduces to $-N\delta ^{0}_{a}$ in the natural coordinate system with $t$  as  the time  coordinate. Observers with four velocity $u_a$  will be called fundamental observers. (These observers follow the world lines  $x^\alpha =$ constant, i.e., they have the same spatial coordinates, if we choose a gauge with  $N_{\alpha}=0$. We will, however, keep our discussion  general and shall keep $N_{\alpha}$ non-zero unless explicitly mentioned.)  This foliation also introduces another natural vector field:
\begin{equation}\label{Paper06_New_03}
\xi _{a}=Nu_{a}\to-(N^{2},\mathbf{0}); \qquad \xi ^{a}=Nu^{a}\to(1,-N^\alpha)
\end{equation}
where the components are in the preferred coordinate system. This vector corresponds to the standard timelike Killing vector if  the spacetime is static. Further, in \emph{any} spacetime, it has the time component which is unity, i.e.,  $\xi ^{0}=1$. We will call $\xi ^{a}$ the \textit{time evolution vector field}.

The fundamental observers will have   (in general, nonzero) acceleration vector $a_i\equiv u^j\nabla_ju_i=h^j_i(\nabla_jN/N)$ which is purely spatial (i.e., $u^ia_i=0$) and has the magnitude $a\equiv\sqrt{a_ia^i}$. The conditions $t(x)=$ constant, $N(x)=$ constant, taken together, define the codimension-2 surface $\mathcal{S}$ with an induced metric $q_{ab}$ the area element $\sqrt{q}d^{d-2}x$ and the binormal  $\epsilon_{ab}\equiv r_{[a}u_{b]}$  where  $r^\alpha=\epsilon(a^\alpha/a)$ is essentially the unit vector along the acceleration. (The factor $\epsilon=\pm 1$ ensures that the normal $r^\alpha$ is pointing outwards irrespective of the direction of acceleration. We will usually assume $\epsilon=1$.) Note that $a_{i}$ and $\nabla _{i}N$ projected on the $t=\textrm{constant}$ surface coincides. 

We will next consider the construction of an appropriate coordinate system associated with an arbitrary null surface which we will use extensively in this paper. 
\subsection{Gaussian Null Coordinates and the associated vector fields}\label{Paper06_GNC}

The construction of a coordinate system associated with an arbitrary null surface has already been discussed in detail in \cite{Moncrief:1983,Morales,Parattu:2015gga}, which we will briefly review. This coordinate system will have the following properties: (a) All the redundant gauge degrees of freedom are eliminated, leaving only 6 free functions in the metric tensor. (b) The null surface we are interested in is chosen to be a surface determined by $r=0$ where $r$ is one of the spatial coordinates. The other $r=$ constant (but non-zero) valued surfaces will represent timelike surfaces and the null surface can be obtained as a limit $r\to 0$. (e.g., this is what will happen in Schwarzschild spacetime if we choose $(r-2M)$ as one of the coordinates with the event horizon being the chosen null surface).

This coordinate system is known as Gaussian null coordinates (henceforth referred to as GNC), constructed in analogy with the Gaussian normal coordinates. To handle the fact that the normals are null, we need to introduce another auxiliary null vector $k^{a}$ and then construct the coordinates by moving away from the null surface along the appropriate null geodesics. After such a construction the line element adapted to an arbitrary null surface (identified with $r=0$) takes the following form in GNC:
\begin{equation}\label{Paper06_Sec02_Eq01}
ds^{2}=-2r\alpha du^{2}+2dudr-2r\beta _{A}dudx^{A}+q_{AB}dx^{A}dx^{B}
\end{equation}
This line element contains six independent functions $\alpha$, $\beta _{A}$ and $q_{AB}$ as advocated, all dependent on the coordinates $\left(u,r,x^{A}\right)$. We shall restrict to $\alpha>0$, which is adequate for our purposes. The metric on the two-surface (i.e. $u=\textrm{constant}$ and $r=\textrm{constant}$) is represented by $q_{AB}$. The surface $r=0$ is the fiducial null surface but surfaces with $r=$ non-zero constant are not null.

We will now introduce the time development vector $\xi^a$ appropriate for this coordinate system as the one with the components $\xi ^{a}=\delta ^{a}_{0}$ in the GNC; that is:
\begin{align}\label{Paper06_Sec02_Eq05}
\xi ^{a}=\left(1,0,0,0\right);\qquad
\xi _{a}=\left(-2r\alpha ,1,-r\beta _{A}\right)
\end{align} 
It can be easily shown that $\xi^a$ will be identical to the timelike Killing vector corresponding to the  Rindler time coordinate if we rewrite  the standard Rindler metric in the GNC form. Therefore, we can think of $\xi^a$ as a natural generalization of the time development vector corresponding to the Rindler-like observers in the GNC; of course, it will not be a Killing vector in general. Since $\xi ^{2}=-2r\alpha$,  we see that, in the $r\to 0$ limit, $\xi ^{a}$ becomes  null. Given $\xi ^{a}$, we can construct the four-velocity $u_{a}$ for a comoving  observer by dividing $\xi ^{a}$ by its norm $\sqrt{2r\alpha}$ obtaining: 
\begin{align}\label{Paper06_Sec02_Eq06}
u^{i}=\left(\frac{1}{\sqrt{2r\alpha}},0,0,0\right);\qquad u_{i}=\left(-\sqrt{2r\alpha},\frac{1}{\sqrt{2r\alpha}},\frac{-r\beta _{A}}{\sqrt{2r\alpha}}\right)
\end{align}
The form of $u^i$ shows that the comoving observers can also be thought of as  observers with 
$(r,x^A) = $ constant. This proves to be convenient for probing the properties of the null surface.

This four-velocity, has  the four-acceleration  $a^{i}=u^{j}\nabla _{j}u^{i}$. The magnitude of the acceleration $\sqrt{a_{i}a^{i}}$, multiplied by the redshift factor $\sqrt{2r\alpha}$, has a finite result in the null limit, (i.e., $r\to 0$ limit):
\begin{align}\label{Paper06_Sec02_Eq07}
Na |_{r\to 0} = \left(\sqrt{2r\alpha}\right)a\vert _{r\to 0}=\alpha -\frac{\partial _{u}\alpha}{2\alpha}
\end{align}
When the acceleration $\alpha$ varies slowly in time (i.e., $\partial_u \alpha/\alpha^2 \ll 1$) the second term is negligible and $Na\to \alpha$. The redshifted Unruh-Davies \cite{Davies:1974th,Unruh:1976db} temperature associated with the $r=0$ surface, as measured by $(r,x^{A})=\textrm{constant}$ observer is given \cite{Padmanabhan:2013nxa} by \ref{Paper06_Sec02_Eq07}. We will call this temperature as the ``acceleration temperature''.

We will next introduce the relevant null vectors associated with the GNC. Given the four-velocity $u_{a}$ and four-acceleration $a_{i}$ we can construct two null vectors $\bar{\ell}^{i}$ and $\bar{k}^{i}$ as:
\begin{align}
\bar{\ell}^{a}=\frac{\sqrt{2r\alpha}}{2}\left(u^{a}+r^{a}\right);
\qquad
\bar{k}^{a}=\frac{1}{\sqrt{2r\alpha}}\left(u^{a}-r^{a}\right)
\label{Paper06_Sec02_Eq08a}
\end{align}
where $r_{i}$ is the unit vector in the direction of the acceleration, i.e., $r_{i}=a_{i}/a$. These two vectors $\bar{\ell}^{i}$ and $\bar{k}^{i}$ satisfy: $\bar{\ell}^{2}=0$, $\bar{k}^{2}=0$ and $\bar{\ell}^{a}\bar{k}_{a}=-1$ and we have the following components on the null surface:
\begin{align}
\bar{\ell}_{i}\Big \vert _{r\to 0}&=\left(0,1,0,0\right)
\label{Paper06_Sec02_Eq09a}
\\
\bar{k}_{i}\Big \vert _{r\to 0}&=\left(-1,\frac{q_{AB}a^{A}a^{B}}{4r\alpha a^{2}},-\frac{a_{A}}{\sqrt{2r\alpha}~a}\right)
\label{Paper06_Sec02_Eq09b}
\end{align}
Since we are essentially interested only in the $r\to 0$ limit, it is more convenient to work with a simpler vector field
$\ell _{i}\equiv\nabla _{i}r$ \textit{everywhere}, which reduces to this $\bar{\ell}_{i}$ on the null surface and defines the  natural null normal to the $r=0$ surface as a limiting case. Similarly, we can introduce another vector $k_a$ in place of $\bar k_a$ to simplify the computations. Using the non-uniqueness in the definition of $\bar{k}_{a}$, we can change it to another vector $k_{a}$ such that,
\begin{align}
\bar{k}_{a}=k_{a}+A\ell _{a}+B_{A}e^{A}_{a}
\label{Paper06_Sec02_Eq10}
\end{align}
where $e^{A}_{a}$ are basis vectors on the null surface  and $\ell _{a}=\bar{\ell}_{a}$. From the property $\ell _{a}e^{a}_{A}=0$ and $\ell ^{2}=0$ we get $\ell _{a}k^{a}=-1$, since $\ell _{a}\bar{k}^{a}=-1$. The condition $k^{2}=0$, leads to a condition between A and $B_{A}$ as: $2A=q_{CD}B^{C}B^{D}$. Choosing $A=(q_{AB}a^{A}a^{B}/4r\alpha a^{2})$ and $B_{A}=-(a_{A}/\sqrt{2r\alpha}~a)$ leads to the simple form $k_{a}=(-1,0,0,0)$. Thus the vector $\ell_a=\nabla _{a}r$ and the auxiliary $k_a=-\nabla_au$  have the following components in the GNC:
\begin{subequations}
\begin{align}
\ell _{a}&=\left(0,1,0,0\right),\qquad \ell ^{a}=\left(1,2r\alpha +r^{2}\beta ^{2},r\beta ^{A}\right)
\label{Paper06_Sec02_Eq11a}
\\
k_{a}&=\left(-1,0,0,0\right),\qquad k^{a}=\left(0,-1,0,0\right)
\label{Paper06_Sec02_Eq11b}
\end{align}
\end{subequations}
Along with these two vectors we also have the vector $\xi ^{a}$, which is the time development  vector 
introduced earlier. Thus, through the GNC, we have introduced three  vectors $\ell _{a}$, $k_{a}$ and $\xi _{a}$.  The binormal associated with the null surface can be obtained in terms of $\ell _{a}$ and $k_{a}$ as $\epsilon _{ab}=\ell _{a}k_{b}-\ell _{b}k_{a}$. 

There are a few more geometrical quantities associated with the null vectors which we will introduce for ready reference later on. The first one corresponds to the induced metric $q_{ab}$ on the null surface, defined as,
\begin{align}\label{Paper06_nNew01}
q_{ab}\equiv g_{ab}+\ell _{a}k_{b}+\ell _{b}k_{a};\qquad q^{a}_{b}\equiv \delta ^{a}_{b}+\ell ^{a}k_{b}+\ell _{b}k^{a}
\end{align}
Note that both $\ell _{a}q^{a}_{b}$ and $k_{a}q^{a}_{b}$ identically vanishes on the null surface, thanks to the relation $\ell _{a}k^{a}=-1$; we can think of $q^{a}_{b}$ as a projector on to the $r=0$ surface, which is two-dimensional. Using this projector, we can define extrinsic curvature on a null surface:
\begin{align}\label{Paper06_nNew02}
\Theta _{ab}\equiv q^{m}_{a}q^{n}_{b}\nabla _{m}\ell _{n}=\frac{1}{2}q^{m}_{a}q^{n}_{b}\pounds _{\ell}q_{mn}
\end{align}
If $\lambda$ is the parameter along the null generator $\ell _{a}$ on the null surface, we the only nonzero components of $\Theta _{ab}$ are (see \ref{Paper06_AppNew02} of \ref{Paper06_App03_General}),
\begin{align}\label{Paper06_nNew03}
\Theta _{AB}=\frac{1}{2}\dfrac{d}{d\lambda}q_{AB}
\end{align}
The trace of $\Theta _{ab}$, is $\Theta =q_{ab}\Theta ^{ab}$ and it is useful to define the trace free \textit{shear tensor} $\sigma _{ab}$ as,
\begin{align}\label{Paper06_nNew04}
\sigma _{ab}\equiv\Theta _{ab}-\frac{1}{2}q_{ab}\Theta
\end{align}
Then, as described in Ref. \cite{Kolekar:2011gw} we can introduce shear viscosity coefficient, $\eta =(1/16\pi)$ and the bulk viscosity coefficient $\zeta=-(1/16\pi)$ as well as the dissipation term  by:
\begin{align}\label{Paper06_nNew05}
\mathcal{D}\equiv 8\pi \left(2\eta \sigma _{ab}\sigma ^{ab}+\zeta \Theta ^{2}\right)=\Theta _{ab}\Theta ^{ab}-\Theta ^{2}
\end{align}
The importance of $\sigma_{ab}$ and $\mathcal{D}$ --- which will occur repeatedly in our discussion --- arises from the following fact: It turns out that Einstein's equations, when projected on to any  null surface in any spacetime, takes the form of a Navier-Stokes equations \cite{Padmanabhan:2010rp} with $\sigma_{ab}$ acting as a viscous tensor and $\eta,\zeta$ acting as bulk and shear viscosity coefficients, In that case, the apparent viscous dissipation is given by $\mathcal{D}$. (The conceptual issues related to this `dissipation without dissipation' since there can be no real, irreversible, drain of energy are clarified in \cite{Padmanabhan:2010rp}). In the GNC, we have on the null surface,
\begin{align}\label{Paper06_NewEq01}
\mathcal{D}=\frac{1}{4}q^{ac}q^{bd}\partial _{u}q_{ab}\partial _{u}q_{cd}-\left(\partial _{u}\ln \sqrt{q}\right)^{2}
\end{align}
which vanishes when $\partial _{u}q_{ab}=0$ on the null surface. We will have occasion to comment on these results later on.
\section{Noether current for the time evolution vector field and spacetime thermodynamics}\label{Paper06_NoetherThermo}
 
In \ref{sec:coordsys} we introduced two natural foliations of the spacetime, one based on the (1+3) split and the other adapted to a fiducial null surface. These coordinate charts  also come with certain natural vector fields. In the case of the (1+3) split, the time evolution is related to the vector field $\xi^a$ we introduced in \ref{Paper06_New_03}. In the case of the foliation based on a null surface, we again have a natural time evolution field $\xi^a$ given by \ref{Paper06_Sec02_Eq05} as well as the null vector $\ell_a $ which is tangent to the null congruence defining the null surface. It would be interesting to study the Noether current and the gravitational momenta corresponding to these vector fields which turn out to have direct thermodynamic significance. In this section we shall consider the Noether currents; we will take up the properties of gravitational momenta in \ref{Paper06_Project}.
\subsection{Noether current adapted to (1+3) foliation}
 
We will begin with a brief description of several properties of the Noether current associated with the vector field $\xi ^{a}$, developing further the ideas introduced in Ref. \cite{Padmanabhan:2013nxa}. 
  
Let us start with the situation corresponding to \gr. The form of the Noether potential $J^{ab}$ and current $J^a$ is most easily analysed in terms of the electromagnetic analogy introduced earlier. Using the  foliation vectors $u_{a}=-N\nabla _{a}t$ and $\xi ^{a}=Nu^{a}$ we readily obtain the following expressions for ``electric'' and ``magnetic'' fields of $J^{ab}[\xi]$ as,
\begin{equation}
E^{a}(\xi \vert u)=u_{b}\left[\nabla ^{a}\left(Nu^{b}\right)-\nabla ^{b}\left(Nu^{a}\right)\right]=-2Na^{a};\qquad B^{a}(\xi \vert u)=0
\label{Paper06_SecLL_16}
\end{equation}
 For $J^{ab}[\xi]$ the  magnetic field vanishes and the electric field is proportional to the acceleration. 
So $J^{ab}$  can be expressed in terms of $E^a$ and $u^b$ as $16\pi J^{ab}(v)=u^{a}E^{b}(v\vert u)-u^{b}E^{a}(v\vert u)$ whose divergence gives (with $16\pi$ inserted; we are now using units with $G=1$):
\begin{align}\label{Paper06_SecLL_17}
16\pi J^{a}(v)=16\pi \nabla _{b}J^{ab}(v)&=u^{a}\nabla _{b}E^{b}(v\vert u)+E^{b}(v\vert u)\nabla _{b}u^{a}-E^{a}(v\vert u)\nabla _{b}u^{b}-u^{b}\nabla _{b}E^{a}(v\vert u)
\end{align}
Projecting along $u_{a}$  leads to the Noether charge density on a $t=\textrm{constant}$ surface given by:
\begin{align}\label{Paper06_SecLL_18}
16\pi u_{a}J^{a}(\xi)=-\nabla _{b}E^{b}(\xi \vert u)-u^{a}u^{b}\nabla _{a}E_{b}(\xi \vert u)=-D_{b}E^{b}(\xi \vert u)=D_{\alpha}\left(2Na^{\alpha}\right)
\end{align}
where $D_i$ is the covariant derivative operator on the spatial slices. This  matches with earlier results \cite{Padmanabhan:2013nxa} and is analogous to $\nabla \mathbf{\cdot E} = \rho$ in electromagnetism. Incidentally, there is a similar result for $J^{a}(u)$; one can show that $16\pi u_{a}J^{a}(u)=D_{\alpha}(Na^{\alpha})$.

The thermodynamic interpretation of the result $16\pi u_{a}J^{a}(\xi)= D_{\alpha}\left(2Na^{\alpha}\right)$ can be extracted by integrating the Noether charge density $ u_{a}J^{a}(\xi)$ over a 3-dimensional region $\mathcal{V}$ of a  $t=\textrm{constant}$ surface with integration measure $d^{3}x\sqrt{h}$ to obtain the total Noether charge in that volume. We get:
\begin{align}\label{Paper06_SecLL_22TP}
\int _{\mathcal{V}}d^{3}x\sqrt{h}u_{a}J^{a}(\xi)=\int _{\partial \mathcal{V}}q^{\alpha}r_{\alpha}d^2x
\end{align}
where $r_{\alpha}$ is the vector normal to $\partial \mathcal{V}$ and $q^{\alpha}$ is the \emph{heat flux} vector defined as,
\begin{align}
q^{\alpha}&=\left(\frac{Na}{2\pi}\right)\left(\frac{\sqrt{q}}{4}\right)\hat a^\alpha = (Ts) \hat a^\alpha
\label{Paper06_SecLL_23b}
\end{align}
where $a^\alpha$ is the unit vector  in the direction of the acceleration. The magnitude of the heat flux vector $q^{\alpha}$ gives the heat density,  $Ts$, where $T=Na/2\pi$ is the local Davies-Unruh temperature \cite{Davies:1974th,Unruh:1976db} and $s=\sqrt{q}/4$ is the entropy density per unit coordinate area. Thus, the total Noether charge contained in an arbitrary \emph{bulk volume} $\mathcal{V}$ within $t=\textrm{constant}$ surface is equal to total heat flux contained within the \emph{boundary} $\partial \mathcal{V}$. When the boundary $\partial \mathcal{V}$ is a $N=\textrm{constant}$ surface then $r^{\alpha}=\hat{a}^{\alpha}$ and  the total heat flux $q^{\alpha}r_{\alpha}$ equals the heat content of the boundary, as derived previously \cite{Padmanabhan:2013nxa}. Also note that
\begin{equation}
\sqrt{q}J^{ab}(\xi)=q^{a}u^{b}-q^{b}u^{a}=2(Ts)\epsilon_{ab}
\label{Paper06_New_17a}
\end{equation}
where the bi-normal is defined as $\epsilon _{ab}=(1/2)(\hat{a}_{a}u_{b}-\hat{a}_{b}u_{a})$. That is, the Noether potential is proportional to the  bi-normal of the equipotential surfaces.

The above analysis uses the fact that $u_aJ^a(\xi)$ is a 3-divergence, so that the spatial volume integral of  $u_aJ^a(\xi)$ can be converted to a surface integral. In this case, it is natural to interpret $u_aJ^a(\xi)$ as a spatial density, viz., charge per unit volume of space. It turns out that similar results can be obtained even for the component of $J^a(\xi)$ in the direction of the normal to the equipotential surface along the following lines.
It can be easily shown that
\begin{equation}\label{Paper06_Newacc}
\hat{a}_{p}J^{p}(\xi)=-\left(g^{ij}-\hat{a}^{i}\hat{a}^{j}\right)\nabla _{i}\left(2Nau_{j}\right)=-g^{ij}_{\perp}\nabla _{i}\left(2Nau_{j}\right)
\end{equation}
where the  tensor $g^{ij}_{\perp}$ acts as a projection tensor transverse to the unit vector $\hat a^i$. However in order to define a surface covariant derivative  we need $\hat{a}^{i}$ to foliate the spacetime, which in turn implies $u^{i}\nabla _{i}N=0$. In this case \ref{Paper06_Newacc} can be written as $\hat{a}_{p}J^{p}(\xi)=-\mathcal{D}_{i}(2Nau^{i})$, where $\mathcal{D}_{i}$ is the covariant derivative operator corresponding to the induced metric $g^{ij}_{\perp}$ on the $N=\textrm{constant}$ surfaces with normal $\hat{a}_{i}$. (When $u^{i}\nabla _{i}N=0$, we have $a_{j}=\nabla _{j}\ln N$.) To obtain an integral version of this result, let us transform from the original $(t,x^{\alpha})$ coordinates to a new coordinate system $(t,N,x^{A})$ using $N$ itself as a ``radial'' coordinate. In this coordinate we have $a_{i}\propto\delta _{i}^{N}$ and thus $u^{N}=0$, thanks to the relation $u^{i}a_{i}=0$. Thus $\mathcal{D}_{i}(2Nau^{i})$ will transform into $\mathcal{D}_{\bar{\alpha}}(2Nau^{\bar{\alpha}})$, 
where $\bar{\alpha}$ stands for the set of coordinates ($t,x^{A}$) on the $N$ = constant surface.  Integrating both sides $\hat{a}_{p}J^{p}(\xi)=-\mathcal{D}_{\bar{\alpha}}(2Nau^{\bar{\alpha}})$, over the $N=\textrm{constant}$ surface will now lead to the result (with restoration of $1/16\pi$ factor):
\begin{align}\label{Paper06_AccNew}
\int d^{2}xdt\sqrt{-g_{\perp}}\,\hat{a}_{p}J^{p}(\xi)=-\int d^{2}x\sqrt{q}N\left(\frac{Na}{8\pi}\right)u^{t}
=\int d^{2}x\left(\frac{Na}{2\pi}\right)\left(\frac{\sqrt{q}}{4}\right)\Bigg|^{t_1}_{t_2} 
\end{align}
where we have used the standard result $\sqrt{-g_{\perp}}=N\sqrt{q}$, with $q$ being the determinant of the two-dimensional hypersurface. The right hand side can be thought of as the difference in the heat content $Q(t_2)-Q(t_1)$ between the two surfaces $t=t_2$ and $t=t_1$ where:
\begin{equation}
Q(t)\equiv \int d^{2}x\left(\frac{Na}{2\pi}\right)\left(\frac{\sqrt{q}}{4}\right)
=\int d^{2}x (Ts)
\end{equation} 
This looks very similar to the result we obtained in the case of the integral over $u_i J^i$ earlier (see \ref{Paper06_SecLL_22TP} with $a^\alpha=r^\alpha$ on the $N=$ constant surface), but there is a difference in the interpretation of the left hand side. While $u_i J^i$ can be thought of as the charge density per unit \textit{spatial} volume, the quantity $\hat{a}_{p}J^{p}(\xi)$ represents the \textit{flux} of Noether current through a time-like surface; therefore, $\hat{a}_{p}J^{p}(\xi)$ should be thought of as a current per unit area per unit time. We will see later that the flux of Noether current through null surfaces leads to a very similar result.

The generalization of these results to \LL gravity is straightforward if we rewrite the expression in \ref{Paper06_SecLL_23b} for the heat flux $q^a$ in the form:
\begin{align}
q^{\alpha}&=\left(\frac{Na}{2\pi}\right)\left(\frac{\sqrt{q}}{4}\right)\left(2P^{\alpha b}_{c \beta}u_{b}u^{c}\hat{a}^{\beta}\right)
\label{Paper06_SecLL_23b1}
\end{align}
(When $P_{ab}^{cd}\equiv(1/2)(\delta _{a}^{c}\delta _{b}^{d}-\delta_{a}^{d}\delta_{c}^{b})$ the $q^{\alpha}$ becomes parallel to $a^{\alpha}$, thanks to $u_{a}a^{a}=0$ and reduces to the previous expression.) To get the results for the \LL gravity, we only need to replace $P^{ab}_{cd}$ for \gr\ with the $P^{ab}_{cd}$ for \LL gravity given in \ref{Paper06_SecLL_07}.
  
In this case, the antisymmetric Noether potential is constructed by using the entropy tensor $P^{ab}_{cd}$ introduced through \ref{Paper06_SecLL_02} as, $16\pi J^{ab}(v)=2P^{ab}_{cd}\nabla^{c}v^{d}$. Then the ``electric'' component of $J^{ab}(\xi)$ turns out to be,
\begin{align}\label{Paper06_SecLL_20}
E^{a}(\xi \vert u)&=2P^{abcd}u_{b}\nabla _{c}\xi _{d}=2P^{abcd}u_{b}\left[N\nabla _{c}u_{d}+u_{d}\nabla _{c}N\right]
\nonumber
\\
&=2P^{abcd}u_{b}u_{d}\left(Na_{c}-u_{c}u^{b}\nabla _{b}N\right)-2NP^{abcd}u_{b}\left(K_{cd}+u_{c}a_{d}\right)
\nonumber
\\
&=4NP^{abcd}u_{b}a_{c}u_{d}\equiv -2N\chi ^{a}
\end{align}
where 
\begin{equation}
\chi ^{a}\equiv -2P^{ab}_{cd}u_{b}a^{c}u^{d}                                   
\end{equation}  
Note that  $u_{a}\chi ^{a}=0$, thanks to the antisymmetry of $P^{ab}_{cd}$ in the first two indices making $\chi^a$ another spatial vector. (However the ``magnetic'' component does not vanish in this case and is algebraically complicated.) The contraction of $J^{a}(\xi)$ with $u_{a}$ will again lead to the simple relation $u_{a}J^{a}(\xi)=-D_{a}E^{a}(\xi \vert u)$, thanks to the Frobenius identity for $u_{a}$ and complete antisymmetric tensor $\epsilon ^{abcd}$. (Of course, this is again to be expected from the electromagnetic analogy related to $\nabla\mathbf{\cdot E} = \rho$.) Thus we  obtain, by using \ref{Paper06_SecLL_20}, the result:
\begin{align}\label{Paper06_SecLL_21}
16\pi u_{a}J^{a}(\xi)&=D_{\alpha}\left(2N\chi ^{\alpha}\right)
\end{align}
Integrating over $(d-1)$-dimensional $t=\textrm{constant}$ hypersurface with $d^{d-1}x\sqrt{h}$ as the volume measure we get, 
\begin{align}\label{Paper06_New_N05}
\int _{\mathcal{V}}d^{d-1}x\sqrt{h}u_{a}J^{a}(\xi)
=\int _{\partial \mathcal{V}}\left(\frac{Na}{2\pi}\right)\left(\frac{1}{2} \sqrt{q} P^{\alpha bc\beta}u_{b}u_{c}\hat{a}_{\beta}\right)r_{\alpha}d^{d-2}x
=\int _{\partial \mathcal{V}}q^{\alpha}r_{\alpha}d^{d-2}x
\end{align}
where the heat flux vector $q^{\alpha}$ is now defined as
\begin{align}
q^{\alpha}=\left(\frac{Na}{2\pi}\right)\left(\frac{1}{2} \sqrt{q} P^{\alpha b}_{c\beta}u_{b}u^{c}\hat{a}^{\beta}\right)
=\left(\frac{Na}{2\pi}\right)\left(\frac{\sqrt{q}}{4}\right)\chi ^{\alpha}
\label{Paper06_SecLL_23a}
\end{align}
Thus, even in \LL gravity we can define a suitable  heat flux vector which  is spatial (though is no longer in the direction of acceleration.) 
If we consider a region bounded by an equipotential surface, then $r_\alpha = \hat a_\alpha$ and the integrand of \ref{Paper06_New_N05} will have the combination $P^{\alpha bc \beta} \hat a_\alpha u_b \hat a_\beta u_c$ which can be re-expressed in terms of the bi-normal of the surface. This allows us to express the integrand
in the form $Ts = (Na/2\pi)s $ and read-off the entropy density as
\begin{eqnarray}\label{Paper06_SecLL_10}
s=-\frac{1}{8} \sqrt{q} P^{abcd}\epsilon _{ab}\epsilon _{cd}
\end{eqnarray}
This is in fact an alternate way of defining $P^{abcd}$ which justifies calling it the \textit{entropy tensor}. 

Using the form of the Noether current it is possible relate the evolution of spacetime to the difference between suitably defined surface and bulk degrees of freedom. (This was discussed in detail in ref.\cite{Padmanabhan:2013nxa} which we will recall here because we will later obtain a similar result for the null surface.) To do this, we start with the relation:
\begin{align}\label{Paper06_New_05}
D_{\alpha}\left(2Na^{\alpha}\right)=8\pi \left(2N\bar{T}^{ab}u_{a}u_{b}\right)+u_{a}g^{bc}\pounds _{\xi}N^{a}_{bc}
\end{align}
where we have used the field equation $R_{ab}=8\pi \bar{T}_{ab}$, in the expression for $u_{a}J^{a}(\xi)$ and \ref{Paper06_SecLL_18}. The Lie variation term \cite{Padmanabhan:2013nxa} is given by (see section A.3 of \cite{Padmanabhan:2013nxa}):
\begin{align}\label{Paper06_Sec03_Eqnew}
\sqrt{h}u_{a}g^{ij}\pounds _{\xi}N^{a}_{ij}=-h_{ab}\pounds _{\xi}p^{ab}
\end{align}
which depends only on the extrinsic curvature through $p^{ab}=\sqrt{h}(Kh^{ab}-K^{ab})$. For a section of a spacelike surface $\mathcal{V}$ with boundary $\partial\mathcal{ V}$, we define the average boundary temperature as:
\begin{equation}
T_{\rm avg}=\frac{1}{A_{\rm sur}}\int_{\partial\mathcal{ V}} d^2x\sqrt{q} \frac{Na}{2\pi}
=\frac{1}{A_{\rm sur}}\int_{\partial\mathcal{ V}} d^2x\sqrt{q}T_{\rm loc} 
\end{equation} 
where $T_{\rm loc}=Na/2\pi$ is the local (Tolman-redshifted) Davies-Unruh temperature \cite{Davies:1974th,Unruh:1976db} and $A_{\rm sur}$ is the area of $\partial\mathcal{ V}$. We also define the surface and bulk degrees of freedom by:
\begin{align}\label{Papper06_NewFin01}
N_{\rm sur}=\int_{\partial\mathcal{ V}} d^{2}x\sqrt{q}=\frac{A_{\rm sur}}{L_P^2};
\quad  N_{\rm bulk}=\frac{1}{(1/2)T_{\rm avg}}\int d^{3}x\sqrt{h}2N\bar{T}_{ab}u^{a}u^{b}=\frac{|E_{\rm Komar}|}{(1/2)T_{\rm avg}}
\end{align}
The $N_{\rm sur}$ counts the number of degrees of freedom on the surface area in Planck units; $N_{\rm bulk}$ is the number of degrees of freedom in a bulk volume \textit{if} an amount of (Komar) energy $E_{\rm Komar}$, is in equipartition at the temperature  $T_{\rm avg}$. (We do \textit{not} assume that the equipartition is reached, of course.)  Using \ref{Paper06_Sec03_Eqnew} in \ref{Paper06_New_05} and integrating it over a bulk volume bounded by $N=\textrm{constant}$ surface can be written in the form\cite{Padmanabhan:2013nxa}:
\begin{align}\label{Paper06_NewFin02}
-\frac{1}{8\pi}\int d^{3}x\sqrt{h}~h_{ab}\pounds _{\xi}p^{ab}=\frac{1}{2}T_{\rm avg}\left(N_{\rm sur}-N_{\rm bulk}\right)
\end{align}
which  shows that the difference between the surface and the bulk degrees of freedom defined in \ref{Paper06_NewFin02} is what drives the time evolution of the spacetime through the Lie variation of the momentum $p^{ab}$. It will turn out that similar result holds for null surfaces as well.

We conclude this section with a discussion of the Newtonian limit of \gr\ treated using the Noether current which has some amusing features. The Newtonian limit is obtained by setting $N^{2}=1+2\phi$, $g_{0\alpha}=0$ and $g_{\alpha \beta}=\delta _{\alpha \beta}$, where $\phi$ is the Newtonian potential \cite{Gravitation}. Then the acceleration of the fundamental observers turn out to be $a_{\alpha}=\partial _{\alpha}\phi$. Since the spatial section of the spacetime is flat, the extrinsic curvature identically vanishes and so does the Lie variation term. Also $2\bar{T}_{ab}u^{a}u^{b}=\rho _{\rm Komar}=\rho$, which immediately leads to (with $G$ inserted, see \ref{Paper06_New_05}):
\begin{align}\label{Paper06_New_06}
\nabla ^{2}\phi =4\pi G\rho
\end{align}
the correct Newtonian limit. The same can also be obtained using the four velocity $u_{a}$. The Noether charge associated with $u_{a}$ turns out to have the following expression \cite{Padmanabhan:2013nxa}
\begin{align}\label{Paper06_New_07}
D_{\alpha}a^{\alpha}=16\pi u_{a}J^{a}(u)=16\pi \bar{T}_{ab}u^{a}u^{b}+u_{a}g^{bc}\pounds _{u}N^{a}_{bc}
\end{align}
In the Newtonian limit the following results hold $2\bar{T}_{ab}u^{a}u^{b}=\rho$ and $u_{a}g^{bc}\pounds _{u}N^{a}_{bc}=-D_{\alpha}a^{\alpha}$ (which follows from the Newtonian limit of the result $u_{a}g^{bc}\pounds _{\xi}N^{a}_{bc}=ND_{\alpha}a^{\alpha}+2a^{\alpha}D_{\alpha}N-Nu_{a}g^{ij}\pounds _{u}N^{a}_{ij}$ and the fact that in spacetime with flat spatial section the term $u_{a}g^{bc}\pounds _{\xi}N^{a}_{bc}$ identically vanishes). This immediately leads to \ref{Paper06_New_06}. 

We also see that the Noether charge is positive as long as $\rho >0$ in the Newtonian limit. In fact, the Noether charge contained inside any equipotential surface is always a positive definite quantity as long as $r^\alpha$ and $a^\alpha$ point in the same direction (which happens when $\bar{T}_{ab}u^{a}u^{b}>0$). To prove this we can integrate the Noether charge over a small region on a $t=\textrm{constant}$ hypersurface to obtain,
\begin{align}
\int _{t=\rm{constant}}d^{3}x\sqrt{h}u_{a}J^{a}(u)=\frac{1}{8\pi}\int _{N,t=\textrm{constant}}d^{2}x\sqrt{q}2Na^{\alpha}r_{\alpha}
=\int _{N,t=\rm{constant}}d^{2}x\left(\frac{Na}{2\pi}\right)\left(\frac{\sqrt{q}}{4}\right)
\end{align}
Since $\rho$ is positive definite in this case the fundamental observers are accelerating outwards and thus $r_{\alpha}a^{\alpha}=a$. The temperature as measured by these fundamental observers is a positive definite quantity and so is the entropy density and hence the positivity of Noether charge follows. 
\subsection{Noether current adapted to GNC}

We shall next consider the corresponding thermodynamic interpretation of the  Noether current when we use the time development vector field adapted to the null surface in the GNC. Let us begin with the above result, i.e., Noether charge contained in a bulk region equals the heat content of the boundary, which was derived for a subregion of a spacelike surface. It turns out that a similar result holds for a null surface as well. Given the fact that  the  Noether current corresponding to the time development vector  led to a nice thermodynamic interpretation, we will consider the object $\ell _{a}J^{a}(\xi)$. (In GNC $\xi ^{a}$ has the components in \ref{Paper06_Sec02_Eq05} and  $\ell _{a}$ is given by \ref{Paper06_Sec02_Eq11a} but of course our results are covariant.). It then turns out that (see \ref{Paper06_App05_Eq15b} of \ref{Paper06_App05_Derv}),
\begin{align}\label{Paper06_Sec06_Eq10}
16\pi \ell _{a}J^{a}(\xi)=J^{r}(\xi)=\frac{1}{\sqrt{q}}\dfrac{d}{d\lambda}\left(2\alpha \sqrt{q}\right)
\end{align}
where $\lambda$ is the parameter along the null generator $\ell ^{a}$, which in GNC is simply $u$. This equation can be integrated over the null surface with integration measure $\sqrt{q}d^{2}xd\lambda$ and leads to, 
\begin{align}\label{Paper06_Sec06_Eq11}
\int d^{2}xd\lambda~ \sqrt{q}\ell _{a}J^{a}(\xi)=\int d^{2}x \left(\frac{\alpha}{2\pi}\right)\left(\frac{\sqrt{q}}{4}\right)\Big \vert _{\lambda =\lambda _{1}}^{\lambda =\lambda _{2}}=Q(\lambda_2)-Q(\lambda_1)
\end{align}
where
\begin{equation}\label{Paper06_HeatNew}
Q(\lambda)=\int d^{2}x \left(\frac{\alpha}{2\pi}\right)\left(\frac{\sqrt{q}}{4}\right)
=\int d^{2}x Ts
\end{equation} 
is the heat content of the null surface at a given $\lambda$.

This again shows that total Noether charge density for the vector field $\xi ^{a}$ integrated over the null surface  equals the difference of the heat content $Q$ of the two dimensional boundaries located at $\lambda =\lambda _{2}$ and $\lambda =\lambda _{1}$. Previously the connection between bulk Noether charge to surface heat density was derived in the context of spacelike surfaces. The result  in \ref{Paper06_Sec06_Eq11}  generalizes the previous connection --- between bulk Noether charge and surface heat density in the context of spacelike surfaces ---  by showing that the total Noether charge on a null surface is also expressible as  the heat content of the boundary. 

The following aspect of this result is worth highlighting. We obtained earlier two results (see \ref{Paper06_SecLL_22TP} and \ref{Paper06_AccNew}) of similar nature. The first one (in \ref{Paper06_SecLL_22TP}) was for a spatial region $\mathcal{V}$ contained in a space-like hypersurface.  In that case, the normal to the surface was $u_a$ and the natural integration measure, for integrating the normal component of a vector field is $u_a\sqrt{h}\, d^3x$. We computed the integral $J^a u_a\sqrt{h}\, d^3x$ and found that it is given by a boundary term; we could have also computed the integral in a region contained within two boundary surfaces like, for example, in the shell-like region between two spherical surfaces of radii $R_1$ and $R_2$. We would have then found that the Noether charge in the bulk region is the difference between the heat contents of the two surfaces. 

In the second case, (in \ref{Paper06_AccNew}) we considered the \textit{flux} of Noether current through a timelike surface with normal $\hat a^i$. In this case we calculated the integral of $\hat{a}_{p}J^{p}(\xi)$ with the measure $d^{2}xdt\sqrt{-g_{\perp}}$ on a timelike surface and got a similar result. We also mentioned that in this case, the integrand $\hat{a}_{p}J^{p}(\xi)$ is to be thought of as heat flux per unit time.

In the case of a null surface, our result is similar to the second one, given by \ref{Paper06_AccNew}. Now the corresponding integration measure for integrating a vector field over a null surface is given by $\ell_a \sqrt{q}\, d^2x\, d\lambda$ where $\ell_a = dx_a/d\lambda$ is the tangent vector to the null congruence defining the null surface. Therefore, in this case, we calculate the integral over $J^a \ell_a \sqrt{q}\, d^2x\, d\lambda$.  This leads to the difference $\Delta~Q$ of the heat content at the two boundaries corresponding to $\lambda = \lambda_1$ and $\lambda = \lambda_2$.  In the integrand in this case, $J^a \ell_a \sqrt{q}\, d^2x\, d\lambda$ one of the coordinates $\lambda$ is similar to a time coordinate rather than a spatial coordinate. So we cannot think of $J^a \ell_a$ as charge per unit \textit{volume}; instead it represents charge per unit area (flux) per unit time and more appropriately, it is the \emph{rate of production of heat per unit area} of the null surface. 

It is possible to proceed further and relate the change in the heat content with the matter energy flux through the null surface. To do this, we use \ref{Paper06_Sec03_Eq02} and field equation $G_{ab}=8\pi T_{ab}$ to obtain (on the null surface):
\begin{align}\label{Paper06_Sec06_Eq12}
16\pi T_{ab}\ell ^{a}\ell ^{b}=16\pi \ell _{a}J^{a}(\xi)-\ell _{a}g^{bc}\pounds _{\xi}N^{a}_{bc}
\end{align}
The Lie derivative term can be computed directly to give (see \ref{Paper06_New_10} of \ref{Paper06_App05_Derv}):
\begin{align}\label{Paper06_New_N03_1}
\sqrt{q}\ell _{a}g^{ij}\pounds _{\xi}N^{a}_{ij}=-q_{ab}\pounds _{\xi}\Pi ^{ab}+\dfrac{d^{2}\sqrt{q}}{d\lambda ^{2}}
\end{align}
where $\Pi ^{ab}=\sqrt{q}[\Theta ^{ab}-q^{ab}(\Theta+\kappa)]$ is the momentum conjugate to $q_{ab}$ and $\Theta ^{ab}=q^{a}_{m}q^{b}_{n}\nabla ^{m}\ell ^{n}$ where $\Theta$ is the trace of $\Theta ^{ab}$.
Integrating this result over the null surface between $\lambda=\lambda_1$ and $\lambda=\lambda_1$, and assuming for simplicity that the boundary terms at $\lambda=\lambda_1,\lambda_1$ do not contribute (which assumes $d\mathcal{A}/d\lambda=0$ at the end points where $\mathcal{A}$ is the area of the null surface), we get:
\begin{align}\label{Paper06_Sec06_Eq13}
-\frac{1}{16\pi}\int d^{2}xd\lambda~q_{ab}\pounds _{\xi}\Pi ^{ab}
=\left[Q(\lambda_2)-Q(\lambda_1)\right]
-\int d\lambda d^{2}x\sqrt{q}T_{ab}\ell ^{a}\ell ^{b}
\end{align}
This expression  shows  that the evolution of the spacetime, which is encoded by the evolution of the momentum $\Pi ^{ab}$ is driven by the difference between (i) heat content $Q$ at the boundary and (ii) the matter heat flux flowing into the null surface. We can rewrite this in a nicer manner as follows. We define the surface degrees of freedom as (in units with $G=1$):
\begin{equation}
 N_{\rm sur}\equiv \frac{A}{L_P^2}=A=\int d^{2}x \sqrt{q}
\end{equation} 
and the average temperature as:
\begin{equation}
T_{\rm avg}=\frac{1}{A}\int d^{2}x~\sqrt{q}\left(\frac{\alpha}{2\pi}\right)                                                                           \end{equation} 
We also introduce the effective bulk degrees of freedom by:
\begin{align}
N_{\rm bulk}=\frac{1}{(1/2)T_{\rm avg}}\int d\lambda d^{2}x\sqrt{q}2\bar{T}_{ab}\ell ^{a}\ell ^{b}
\end{align}
If the matter heat flux, given by the integral in the right hand side, thermalizes at the average temperature of the null surface then, $N_{\rm bulk}$ will represent the the effective equipartition degrees of freedom. We now rewrite \ref{Paper06_Sec06_Eq13} as,
\begin{align}\label{Paper06_New_Int}
-\frac{1}{8\pi}\int d^{2}xd\lambda~q_{ab}\pounds _{\xi}\Pi ^{ab}
=\frac{1}{2}\int _{\lambda _{2}} d^{2}x \left(\frac{\alpha}{2\pi}\right)\sqrt{q}
-\frac{1}{2}\int _{\lambda _{1}} d^{2}x \left(\frac{\alpha}{2\pi}\right)\sqrt{q}
-2\int d\lambda d^{2}x\sqrt{q}T_{ab}\ell ^{a}\ell ^{b}
\end{align}
which, on using our definitions, becomes:
\begin{align}\label{Paper06_New_Int02}
-\frac{1}{8\pi}\int d^{2}xd\lambda~(q_{ab}\pounds _{\xi}\Pi ^{ab})=\frac{1}{2}T_{\rm avg}\left[\left(N_{\rm sur}\right)^{\lambda _2}_{\lambda _{1}}-N_{\rm bulk}\right]
\end{align}
This has the interpretation that  and hence the evolution of spacetime on a null surface, encoded in the Lie variation of the momentum $\Pi ^{ab}$ along the time development vector, can be thought of as due to the difference between surface degrees of freedom and the bulk degrees of freedom. This an exact analogy to the corresponding result in $(1+3)$ foliation presented in \ref{Paper06_NewFin02}, which was originally obtained in \cite{Padmanabhan:2013nxa}.

For the sake of completeness, we clarify the notion of $T_{ab}\ell ^{a}\ell ^{b}$ being the heat flux  through the null surface. This concept has been introduced in \cite{Jacobson:1995ab} and arises as follows: Let there be  a matter field,  in the spacetime, with energy momentum tensor $T_{ab}$. Around any given spacetime event $\mathcal{P}$, we can construct local inertial and hence local Rindler frames. We then  have an approximate Killing vector field $\xi ^{a}$, generating boosts, which  coincides with the null normal $\ell ^{a}$ at the null surface (see \ref{Paper06_GNC}).  The heat flow vector is defined as the boost energy current obtained by projecting $T_{ab}$ along $\xi ^{b}$ yielding $T_{ab}\xi ^{b}$. Thus the energy (heat) flux 
through the null surface will be:
\begin{align}\label{Paper06_New_11}
 Q=\int \left(T_{ab}\xi ^{b}\right)d\Sigma ^{a}=\int T_{ab}\xi ^{b}\ell ^{a}\sqrt{q}d^{2}x d\lambda
\end{align}
where $\sqrt{q}d^{2}xd\lambda$ is the integration measure on a null surface generated by null vectors $\ell ^{a}$, parametrized by $\lambda$. Hence, in the null limit, $T_{ab}\ell ^{a}\ell ^{b}$ (when $\xi ^{a} \to  \ell ^{a}$ on the null surface) represents the heat flux through the null surface. 

The same argument can also be presented along the following lines. On a null surface we can decompose $T_{ab}\xi ^{b}$ in canonical null basis as, $T_{ab}\xi ^{b}=A\ell _{a}+Bk_{a}+C_{A}e^{A}_{a}$. Then the heat flux \emph{through} the surface is given by the component $B$ along $k^{a}$, which is off the null surface. This component $B$ is obtained by contracting with $\ell _{a}$ (since $\ell ^{2}=0$ and $\ell _{a}k^{a}=-1$). This  leads to the heat flux through the null surface to be $T_{ab}\ell ^{a}\ell ^{b}$. 
\section{Reduced Gravitational momentum associated with the time development vector}\label{Paper06_Reduced}

The expression for the Noether current contains a Lie derivative term which was defined earlier as the reduced gravitational momentum. In this section, we shall describe some key properties of this reduced gravitational momentum vector in different contexts, emphasising the results in GNC. 

In the case of (1+3) foliation using spacelike surfaces with normal $u_{a}=-N\nabla _{a}t$ (N is the lapse function) and induced metric $h_{ab}=g_{ab}+u_{a}u_{b}$ the reduced gravitational momentum can be related \cite{Padmanabhan:2013nxa} to the Lie variation of $p^{ab}=\sqrt{h}(Kh^{ab}-K^{ab})$ by:
\begin{align}\label{Paper06_Sec03_Eq07}
-\sqrt{h}u_a\mathcal{P}^{a}(\xi)=\sqrt{h}u_{a}g^{ij}\pounds _{\xi}N^{a}_{ij}=-h_{ab}\pounds _{\xi}p^{ab}
\end{align}
In the case of null surfaces, one can obtain a  similar relation. For a general, non-affine parametrization, i.e., when the null generator $\ell ^{a}$ satisfy the relation $\ell ^{b}\nabla _{b}\ell ^{a}=\kappa \ell ^{a}$, we find that 
the Lie variation term is given by (see \ref{Paper06_New_10} of \ref{Paper06_App05_Derv}):
\begin{align}\label{Paper06_New_N03}
-\sqrt{q}\ell_a\mathcal{P}^{a}(\xi)=\sqrt{q}\ell _{a}g^{ij}\pounds _{\xi}N^{a}_{ij}=-q_{ab}\pounds _{\xi}\Pi ^{ab}+\dfrac{d^{2}\sqrt{q}}{d\lambda ^{2}}
\end{align}
where $\Pi ^{ab}=\sqrt{q}[\Theta ^{ab}-q^{ab}(\Theta +\kappa)]$ is the momentum conjugate to $q_{ab}$. Thus as in the case of the spacelike surface, for null surfaces as well, the Lie variation of $N^{c}_{ab}$ is directly related to the Lie variation of the momentum conjugate to the induced metric on the null surface. But in the case of null surfaces there is an extra term which  contributes only at the boundaries $\lambda=\lambda_1,\lambda_2$.

It can be seen from straightforward algebra (see \ref{Paper06_App03_Eq37} of \ref{Paper06_App03_General}) that $q_{ab}\pounds _{\xi}\Pi ^{ab}$ is directly related to the object $\mathcal{D}=\Theta _{ab}\Theta ^{ab}-\Theta ^{2}$ defined in \ref{Paper06_nNew05} which, as we mentioned earlier, can be interpreted as  dissipation \cite{Kolekar:2011gw}.
This leads to an alternative expression on the null surface for the Lie variation term in the adapted coordinate system as (see \ref{Paper06_App05_Eq57} and \ref{Paper06_App05_NewLie} in \ref{Paper06_App05_Derv}),
\begin{align}\label{Paper06_Sec06_Eq16}
\ell _{a}g^{ij}\pounds _{\xi}N^{a}_{ij}=2\mathcal{D}+\frac{2}{\sqrt{q}}\partial _{\lambda}^{2}\left(\sqrt{q}\right)+2\partial _{\lambda}\alpha
\end{align}
Integrating this expression over the $r=0$ null surface with integration measure $d^{2}xdu~\sqrt{q}$, neglecting total divergence and dividing by $16\pi$ leads to,
\begin{align}\label{Paper06_LieNew}
\frac{1}{16\pi}\int d^{2}xd\lambda~\sqrt{q}\ell _{a}g^{ij}\pounds _{\xi}N^{a}_{ij}=\frac{1}{8\pi}\int d^{2}xd\lambda \sqrt{q}\mathcal{D}+\int d^{2}x \left(\frac{\sqrt{q}}{4}\right)d\left(\frac{\alpha}{2\pi}\right)
\end{align}
which explicitly shows that the Lie variation term integrated over the null surface, leads to the dissipation and the $sdT$ term (interpreted as $dT=(dT/d\lambda)d\lambda$). Hence the reduced gravitational momentum on the null surface, can be given a natural thermodynamic interpretation. 

What is important regarding the above result is that for $\partial _{\lambda}q_{ab}=0$ (i.e., the induced metric on the null surface is independent of the parameter along the null generator) the dissipation term vanishes and thus \ref{Paper06_LieNew} can be written as:
\begin{align}\label{Paper06_LieNew02}
\frac{1}{16\pi}\int d^{2}xd\lambda~\sqrt{q}\ell _{a}g^{ij}\pounds _{\xi}N^{a}_{ij}=Q(\lambda_2)-Q(\lambda_1)
\end{align}
Hence the Lie variation term in this particular situation is equal to difference in the heat content allowing us to relate $\mathcal{P}^{a}$ to the rate of heating of the null surface. 
\section{The thermodynamic variational principle for the field equations}\label{Paper06_ThermoVar}

It has been shown earlier \cite{Padmanabhan:2007en,Padmanabhan:2007xy} that the gravitational field equations can be obtained by extremising the total heat density of all the null surfaces in the spacetime. The purpose of this section and the next is to show that this variational principle takes a simple form in terms of the gravitational momentum for both \gr\ and the \LL models.
\subsection{General relativity}

In \ref{Paper06_Noether} we have defined the gravitational momentum $P^{a}(v)$ and the matter momentum $\mathcal{M}_{a}(v)$ associated with the vector field $v^{a}$. Using these two we can construct a thermodynamic variational principle leading to the field equations for gravity. We start by defining a suitable expression for the heat density associated with a null surface as:
\begin{align}\label{Paper06_New_21}
\mathcal{Q}=\int  d^{2}xd\lambda \sqrt{q}\left[\left\lbrace -\ell _{a}P^{a}(\ell)\right\rbrace+\left\lbrace -\ell _{a}\mathcal{M}^{a}(\ell)\right\rbrace \right]
\end{align}
where $\ell _{a}$ is the null vector, such that on the null surface $\ell _{a}\ell ^{a}=0$. Then from \ref{Paper06_Sec03_Eq04} we observe that the Ricci scalar does not contribute since $\ell ^{a}\ell _{a}=0$ on the null surface. From the definition of matter momentum as $\mathcal{M}^{a}(\ell)=-T^{a}_{b}\ell^{b}$, we find that:
\begin{align}\label{Paper06_New_22}
\mathcal{Q}=\int d^{2}xd\lambda \sqrt{q}\left(T_{ab}\ell ^{a}\ell ^{b}+\frac{1}{16\pi}\ell _{a}g^{bc}\pounds _{\ell}N^{a}_{bc}\right)
\end{align}
To understand the interpretation of this quantity as the heat content of the null surface, note that the matter term  involves an integral over $d^2x d\lambda \sqrt{q}\, \ell_a(T^a_b\ell^a)$ on the null surface. As we argued earlier, $T^{ab}\ell_a\ell_b$ can be related to be heat flux due to the matter and hence this integral represents the heat content of the null surface contributed by matter. (In the case of an ideal fluid in GNC, $T^{ab}\ell_a\ell_b$ will be equal to $\rho +p$, which is same as the heat density $Ts$ when we use the Gibbs relation.) Similarly, the expression obtained by replacing $\mathcal{M}^a$ by the gravitational momentum $P^a$ can be thought of as the heat density due to gravity. The explicit form involving reduced gravitational momentum supports this interpretation. Since $\ell _{a}=\nabla _{a}\phi$, we have $J^{a}(\ell)=0$. Hence we have $\ell _{a}g^{bc}\pounds _{\ell}N^{a}_{bc}=-2R_{ab}\ell ^{a}\ell ^{b}$, which allows us to write:
\begin{align}\label{Paper06_New_23}
\mathcal{Q}=\int d^{2}xd\lambda \sqrt{q}\left(T_{ab}\ell ^{a}\ell ^{b}-\frac{1}{8\pi}R_{ab}\ell ^{a}\ell ^{b}\right)
\end{align}
Varying the integrand with respect to all the null vectors fields $\ell ^{a}$, after adding a Lagrange multiplier term $\lambda(x)\ell^a\ell_a $ to enforce the condition $\ell^2=0$ will lead to the field equation $G_{ab} = 8\pi T_{ab} + \Lambda g_{ab}$ where the cosmological constant $\Lambda$ appears as an integration constant. (The details of this derivation has been given in several previous works \cite{Padmanabhan:2007en,Padmanabhan:2007xy,Padmanabhan:2013nxa} and hence is not repeated here.) So, starting from the projection of gravitational momentum and matter momentum along the null normal $\ell ^{a}$, we can derive the field equations for gravity. 

The same result arises even if we have started using the contractions $-\ell _{a}P^{a}(\xi)$ and $-\ell _{a}\mathcal{M}^{a}(\xi)$. As we have shown in \ref{Paper06_App03_Eq21} in \ref{Paper06_App03_General} the Lie variation term and $R_{ab}\ell ^{a}\ell ^{b}$ differs only by a total divergence. Hence after neglecting the surface contributions, the action would be identical to \ref{Paper06_New_23} and thus on variation of $\ell _{a}$ it would lead to the field equations for gravity.
\subsection{\LL gravity}

The power of this analysis becomes apparent when we realise that the same expression as presented in \ref{Paper06_New_21} leads to the \LL field equations when we use the corresponding momentum $P^{a}(\ell)$ given by \ref{Paper06_SecLL_24}. Using this, we can construct a variational principle associated with a null surface, leading to the field equations for \LL theories of gravity. This is achieved by using exactly the same expression for the heat content:
\begin{align}\label{Paper06_New_LL_01}
\mathcal{Q}=\int  d^{D-2}xd\lambda \sqrt{q}\left[\left\lbrace -\ell _{a}P^{a}(\ell)\right\rbrace+\left\lbrace -\ell _{a}\mathcal{M}^{a}(\ell)\right\rbrace \right]
\end{align}
where $\ell _{a}$ is the null vector, such that on the null surface $\ell _{a}\ell ^{a}=0$. As in the case of \gr, the scalar $\mathcal{R}$ present in gravitational momentum does not contribute since $\ell ^{a}\ell _{a}=0$ on the null surface. From the definition of matter momentum as $\mathcal{M}^{a}(\ell)=-T^{a}_{b}\ell^{b}$, we  arrive at the explicit form:
\begin{align}\label{Paper06_New_LL_02}
\mathcal{Q}=\int d^{D-2}xd\lambda \sqrt{q}\left(T_{ab}\ell ^{a}\ell ^{b}+\frac{1}{16\pi}\ell _{a}P_{m}^{~nqa}\pounds _{\ell}\Gamma ^{m}_{nq}\right)
\end{align}
Since $\ell _{a}=\nabla _{a}\phi$, we have $J^{a}(\ell)=0$. Hence $\ell _{a}P_{m}^{~nqa}\pounds _{\ell}\Gamma ^{m}_{nq}=-2\mathcal{R}_{ab}\ell ^{a}\ell ^{b}$, which gives the variational principle for the null surface to be:
\begin{align}\label{Paper06_New_LL_03}
\mathcal{Q}=\int d^{D-2}xd\lambda \sqrt{q}\left(T_{ab}\ell ^{a}\ell ^{b}-\frac{1}{8\pi}\mathcal{R}_{ab}\ell ^{a}\ell ^{b}\right)
\end{align}
Again varying the integrand with respect to all the null vectors fields $\ell ^{a}$, with a Lagrange multiplier $\lambda \ell_a\ell^a$ to impose the constraint $\ell^2=0$, we  obtain the field equation to be $E_{ab}=8\pi T_{ab} + \Lambda g_{ab}$ as before.  Hence starting from the gravitational momentum and matter momentum along the normal, null vector field $\ell ^{a}$, we can derive the field equations for gravity. 

Thus, one can write down a thermodynamic variational principle directly in terms of the gravitational momentum. The logical simplicity of this result and the fact that it holds for \LL models without any modification is note worthy. Because of this feature, this procedure seems to be the natural way of obtaining the field equations in the emergent gravity paradigm.
\section{Projections of Gravitational Momentum on the Null Surface}\label{Paper06_Project}

We shall now take up a further application of the concept of gravitational momentum. Given the thermodynamic significance of the null surfaces, we would expect the flow of gravitational momentum vis-a-vis a given null surface to be of some importance. To explore this, we have to first choose a suitable vector field using which the gravitational momentum can be defined. The most natural choice --- as before --- is the time evolution field $\xi^a$. 
Further, in the canonical null basis $(\ell ^{a},k^{a},e^{a}_{A})$ the gravitational momentum can be decomposed as: $P^{a}=A\ell ^{a}+Bk^{a}+C^{A}e^{a}_{A}$. These components $A,B$ and $C^{A}$ are related to the three projections of $P^a$ by  $A=-P^{a}(\xi)k_{a}$, $B=-P^{a}(\xi)\ell _{a}$ and $C^{A}=P^{a}(\xi)e^{A}_{a}$. So we need to consider the following three components $q^{a}_{b}P^{b}(\xi), k_{a}P^{a}(\xi)$ and $\ell_{a}P^{a}(\xi)$ to get the complete picture.  We will see that each of them lead to interesting thermodynamic interpretation.  In view of the rather involved calculations, we will first provide  a summary of the thermodynamic interpretations of these projections:
\begin{itemize}

\item The component $q^{b}_{a}P^{a}(\xi)$ leads to the Navier-Stokes equation for fluid dynamics, using which we can obtain  yet another justification for the dissipation term introduced in \ref{Paper06_nNew05}. This is described in \ref{Paper06_NS} 

\item The component $k_{a}P^{a}(\xi)$, when evaluated on an arbitrary null surface leads to a result which can be stated in the form: $TdS=dE+PdV$, i.e., as a thermodynamic identity. This helps us to identify a notion of energy associated with an arbitrary null surface. We obtain this result in \ref{Paper06_FirstLaw}.

\item Finally, the component $\ell _{a}P^{a}(\xi)$ yields the evolution of null surface, which involves both $ds/d\lambda$ and $dT/d\lambda$, where $s$ is the entropy density and $T$ is the temperature associated with an arbitrary null surface and $\lambda$ is the parameter along the null generator $\ell _{a}$. This is studied  in \ref{Paper06_Evo}.
\end{itemize}

We shall now show how all these three results  tie up with the notion of gravitational momentum and arise naturally from the three different projections of the gravitational momentum.

\subsection{Navier-Stokes Equation}\label{Paper06_NS}

The equivalence of field equations for gravity, when projected on a null surface, and the Navier-Stokes equation is an important result in the emergent paradigm of gravity. This result, which generalizes the previous one \cite{Damour:1982} in the context of black holes,   shows that when Einstein's equations are projected on any null surface and viewed in local inertial frame, they become \emph{identical} to the Navier-Stokes equation of fluid dynamics \cite{Padmanabhan:2010rp}. Here we will project the gravitational momentum $P^{a}$ on an arbitrary null surface parametrized by GNC and  show that the Navier-Stokes equation is obtained. For this purpose we will consistently use the vector field $\xi ^{a}$. Even though we will present the results for the adapted coordinates to a null surface (i.e., in GNC), the same result continues to hold in any other parametrization as well. 

In order to project the gravitational momenta $16\pi P^{c}(\xi)=g^{ab}\pounds _{\xi}N^{c}_{ab}+R\xi ^{c}$ on the null surface we need to determine the induced metric $q^{a}_{b}$, which \cite{Parattu:2015gga} turns out to be $\textrm{diag}(0,0,1,1)$. Hence it is the angular part which is going to contribute. The vectors $\ell ^{a}$ and $k^{a}$ are already given in \ref{Paper06_Sec02_Eq11a} and \ref{Paper06_Sec02_Eq11b} respectively. In the Navier-Stokes equation two other vectors play a crucial role. These vectors and their components in the adapted GNC system has the following expressions:
\begin{align}\label{Paper06_Sec04_Eq01}
\omega _{b}=\ell ^{m}\nabla _{m}k_{b}=\left(\alpha ,0,\frac{1}{2}\beta _{A}\right);\qquad 
\omega ^{a}=\left(0,\alpha ,\frac{1}{2}\beta ^{A}\right)
\end{align}
and
\begin{align}\label{Paper06_Sec04_Eq02}
\Omega _{a}=\omega _{a}+\alpha k_{a}=\left(0,0,\frac{1}{2}\beta _{A}\right);\qquad \Omega ^{a}=\left(0,0,\frac{1}{2}\beta ^{A}\right)
\end{align}
From \cite{Padmanabhan:2010rp} we know that $\beta _{A}$ can be interpreted as the transverse velocity of observers on the null surface. (In particular it can be interpreted as velocity drift of local Rindler observers parallel to the Rindler horizon.) With this physical motivation, let us now start calculating the projection of $P^{a}$ on the null surface. Using the coordinates adapted to a given null surface, we arrive at:
\begin{align}\label{Paper06_Sec04_Eq03}
-16\pi q^{a}_{b}P^{b}(\xi)=q^{a}_{b}g^{pq}\pounds _{\xi}N^{b}_{pq}
=2\pounds _{\xi}N^{A}_{ur}+q^{BC}\pounds _{\xi}N^{A}_{BC}
\end{align}
where  we have used the identity $\ell _{a}q^{a}_{b}=0$. We next need to find the Lie variation of $N^{c}_{ab}$. For that we use the  expression in \ref{Paper06_Sec04_Eq04} for the Lie variation of Christoffel symbol with respect to an arbitrary vector field $v^{a}$ and the result: $N^{a}_{bc}=Q^{ad}_{be}\Gamma ^{e}_{cd}+Q^{ad}_{ce}\Gamma ^{e}_{bd}$. Then the Lie variation of $N^{a}_{bc}$ becomes
\begin{align}\label{Paper06_Sec04_Eq05}
\pounds _{v}N^{a}_{bc}&=Q^{ad}_{be}\pounds _{v}\Gamma ^{e}_{cd}+Q^{ad}_{ce}\pounds _{v}\Gamma ^{e}_{bd}
\nonumber
\\
&=\frac{1}{2}\left(\delta ^{a}_{b}\nabla _{c}\nabla _{d}v^{d}+\delta ^{a}_{c}\nabla _{b}\nabla _{d}v^{d}\right)-\frac{1}{2}\left(\nabla _{b}\nabla _{c}v^{a}+\nabla _{c}\nabla _{b}v^{a}\right)
-\frac{1}{2}\left(R^{a}_{~bmc}+R^{a}_{~cmb}\right)v^{m}
\end{align}
We obtain the expressions for Lie variation of $N^{A}_{ur}$ and $N^{A}_{BC}$ on the null surface (located at $r=0$) to be (see \ref{Paper06_App05_Eq18a} of \ref{Paper06_App05_Derv})
\begin{align}
\pounds _{\xi}N^{A}_{ur}&=\frac{1}{2}\partial _{u}\beta ^{A}=\frac{1}{2}q^{AB}\partial _{u}\beta _{B}+\frac{1}{2}\beta _{B}\partial _{u}q^{AB}
\label{Paper06_Sec04_Eq06a}
\\
\pounds _{\xi}N^{A}_{BC}&=\frac{1}{2}\delta ^{A}_{B}\partial _{C}\Theta +\frac{1}{2}\delta ^{A}_{C}\partial _{B}\Theta -\partial _{u}\hat{\Gamma}^{A}_{BC}
\label{Paper06_Sec04_Eq06b}
\end{align}
Substituting these results in \ref{Paper06_Sec04_Eq03} we arrive at,
\begin{align}\label{Paper06_Sec04_Eq07}
-16\pi q^{p}_{b}P^{b}=q^{p}_{a}\left(g^{bc}\pounds _{\xi}N^{a}_{bc}\right)&=q^{PB}\partial _{u}\beta _{B}+\beta _{B}\partial _{u}q^{PB}+q^{PC}\partial _{C}\Theta -q^{BC}\partial _{u}\hat{\Gamma}^{P}_{BC}
\end{align}
In order to get the Navier-Stokes equation in its familiar form we need to lower the free index in \ref{Paper06_Sec04_Eq07} and multiply both sides by $(-1/2)$. Using Noether current for $\xi _{a}$ we have from \ref{Paper06_Sec03_Eq02}
\begin{align}\label{Paper06_New_N06}
q_{ab}(g^{pq}\pounds _{\xi}N^{b}_{pq})=q_{ab}J^{b}(\xi)-2R_{mn}\ell ^{m}q^{n}_{a}
\end{align}
On using Einstein's equations $R_{ab}=8\pi (T_{ab}-(1/2)Tg_{ab})$ and the result $\ell _{a}q^{a}_{b}=0$, \ref{Paper06_Sec04_Eq07} leads to the following result:
\begin{align}\label{Paper06_New_14}
8\pi T_{mn}\ell ^{m}q^{n}_{a}&=\frac{1}{2}\beta _{A}\partial _{u}\ln \sqrt{q}+\frac{1}{2}\partial _{u}\beta _{A}-\partial _{A}\alpha -\frac{1}{2}\partial _{A}\partial _{u}\ln \sqrt{q}+\frac{1}{2}q_{AB}q^{PQ}\partial _{u}\hat{\Gamma}^{B}_{PQ}
\nonumber
\\
&=\frac{1}{2}\beta _{A}\partial _{u}\ln \sqrt{q}+\frac{1}{2}\partial _{u}\beta _{A}-\partial _{A}\alpha -\frac{1}{2}\partial _{A}\partial _{u}\ln \sqrt{q}
\nonumber
\\
&-\frac{1}{2}q_{AB}\hat{\Gamma} ^{B}_{PQ}\partial _{u}q^{PQ}+\frac{1}{2}\partial _{D}\left(q^{CD}\partial _{u}q_{AC}\right)
+\frac{1}{2}\partial _{u}q^{CF}\partial _{C}q_{AF}-\frac{1}{2}q_{AB}\partial _{u}\left(q^{BD}\partial _{D}\ln \sqrt{q}\right)
\nonumber
\\
&=\frac{1}{2}\beta _{A}\partial _{u}\ln \sqrt{q}+\frac{1}{2}\partial _{u}\beta _{A}-\partial _{A}\alpha -\partial _{A}\partial _{u}\ln \sqrt{q}-\frac{1}{2}q_{AB}\partial _{u}q^{BD}\left(\partial _{D}\ln \sqrt{q}\right)
\nonumber
\\
&+\frac{1}{2}\partial _{D}\left(q^{CD}\partial _{u}q_{AC}\right)+\frac{1}{2}\partial _{u}q^{CF}\partial _{C}q_{AF}
-\frac{1}{2}\partial _{C}q_{AF}\partial _{u}q^{CF}-\frac{1}{2}q^{CD}\hat{\Gamma}^{E}_{AC}\partial _{u}q_{ED}
\end{align}
where in the second line we have used the relation \ref{Paper06_App05_Eq10b} and in the third line \ref{Paper06_App05_Eq10c} as presented in \ref{Paper06_App05_Derv}. From \ref{Paper06_App05_Eq09} of \ref{Paper06_App05_Derv} we obtain:
\begin{align}
\frac{1}{2}\partial _{u}\beta _{A}&+\partial _{B}\left(\frac{1}{2}q^{BC}\partial _{u}q_{AC}\right)+\frac{1}{2}q^{CD}\partial _{u}q_{AD}\partial _{C}\ln \sqrt{q}-\frac{1}{2}q^{BD}\partial _{u}q_{CD}\hat{\Gamma}^{C}_{AB} 
\nonumber
\\
&+\partial _{u}\ln \sqrt{q}\frac{1}{2}\beta _{A}-\partial _{A}\partial _{u}\ln \sqrt{q}-\partial _{A}\alpha 
\nonumber
\\
&=q^{n}_{a}\pounds _{\ell}\Omega _{n}+D_{m}\sigma ^{m}_{a}+\Theta \Omega _{a}-D_{a}\left(\frac{\Theta}{2}+\alpha \right)
\label{Paper06_Sec04_Eq08b}
\end{align}
After some trivial manipulations in \ref{Paper06_New_14} and using \ref{Paper06_Sec04_Eq08b} the following final expression is obtained:
\begin{align}\label{Paper06_Sec04_Eq11}
8\pi T_{mn}\ell ^{m}q^{n}_{a}=q^{n}_{a}\pounds _{\ell}\Omega _{n}+D_{m}\sigma ^{m}_{a}+\Theta \Omega _{a}-D_{a}\left(\frac{\Theta}{2}+\alpha \right)
\end{align}
which can be interpreted as the Navier-Stokes equation for fluid dynamics. 

The correspondence between \ref{Paper06_Sec04_Eq08b} and \ref{Paper06_Sec04_Eq11} with Navier-Stokes equation for fluid dynamics is based on the following identifications of various geometric quantities on the null surface: (i) The momentum density is given by $-\Omega _{a}/8\pi$. In the coordinates adapted to the null surface, $\Omega_a$ has only transverse components which are given by $(1/2)\beta_A$. This reinforces our interpretation of $\beta_A$ as the transverse fluid velocity. Moreover, we can identify  (ii) the pressure $\kappa/8\pi$, (iii)the shear tensor defined as $\sigma _{mn}$ (see \ref{Paper06_nNew04}), (iv) the shear viscosity coefficient $\eta = (1/16\pi)$, (v) the bulk viscosity coefficient $\zeta = -1/16\pi$ and finally (vi) an external force given by $F_{a}=T_{ma}\ell ^{m}$. Thus \ref{Paper06_Sec04_Eq08b} has the form of a Navier-Stokes equation for a fluid with 
the convective derivative replaced by the Lie derivative. Since this equation and its interpretation have been extensively discussed in the references cited earlier, we will not repeat them and will confine ourselves to highlighting the dissipation term.

In order to interpret the dissipation term we  start from the heat density $q=-4P^{ab}_{cd}\nabla _{a}\ell ^{c}\nabla _{b}\ell ^{d}$ (i.e., heat content per unit null surface volume $\sqrt{q}d^2x d\lambda$) used in the variational principle, where $P^{ab}_{cd}=(1/2)(\delta ^{a}_{c}\delta ^{b}_{d}-\delta ^{a}_{d}\delta ^{b}_{c})$ and $\ell ^{a}$ is the null generator of the null surface. To connect up the heat density with the dissipation term \cite{Kolekar:2011gw} we consider a virtual displacement of the null surface in which the volume changes by $\delta A\delta \lambda$, where $\delta A$ is a small area element on the two-surface (i.e., $\sqrt{q}d^{2}x$). Then $q\delta Ad\lambda$ represents the change in heat content of the null surface, which is obtained by multiplying the heat density $q$ with the infinitesimal 3-volume element on the null surface. Expanding the expression for the heat density and introducing temperature through local Rindler horizon \cite{Kolekar:2011gw} we obtain,
\begin{align}
\dfrac{q\delta Ad\lambda}{8\pi}=-\delta Ad\lambda \left(2\eta \sigma _{ab}\sigma ^{ab}+\zeta \Theta ^{2}\right)+\frac{1}{2}\frac{\kappa}{2\pi}d\delta A
\label{Paper06_Sec04_Eq12}
\end{align}
where the first term represents the  (virtual) dissipation of a viscous fluid during the evolution of the small area element $\delta A$ of the null surface along the null generator and has the following expression,
\begin{align}\label{Paper06_Sec04_Eq13}
dE=\delta Ad\lambda \left(2\eta \sigma _{ab}\sigma ^{ab}+\zeta \Theta ^{2}\right)
=\frac{1}{8\pi}\delta Ad\lambda ~\mathcal{D} 
\end{align}
So the combination $\mathcal{D}=\left(\Theta _{ab}\Theta ^{ab}-\Theta ^{2}\right)$ represents the  dissipation as we have mentioned earlier. The second term in \ref{Paper06_Sec04_Eq12} can be interpreted as $(1/2)kT\ dN$, where $T=(\kappa /2\pi)$ and $dN$ is the degrees of freedom on the null surface corresponding to the change in area $\delta A$. For affinely parametrized null generator $\kappa =0$, which would lead to,
\begin{align}
\dfrac{q\delta Ad\lambda}{8\pi}
=-\delta Ad\lambda \left(2\eta \sigma _{ab}\sigma ^{ab}+\zeta \Theta ^{2}\right)
=\frac{1}{8\pi}\delta Ad\lambda ~\mathcal{D} 
\end{align}
Thus for affinely parametrized null generators the change in the heat content is solely due to the dissipation term $\mathcal{D}$.

Having derived Navier-Stokes equation from the projection of gravitational momentum $P^{a}(\xi)$  on the null surface, we will now take up the task of projecting it along $k_{a}$ and $\ell _{a}$ respectively and retrieve the thermodynamic information encoded in them. 
\subsection{A Thermodynamic Identity for the null surface}\label{Paper06_FirstLaw}

It has been shown that in a wide class of gravity theories, the gravitational field equations near a horizon imply a thermodynamic identity $T\delta_{\bar{\lambda}}S=\delta_{\bar{\lambda}}E+P\delta_{\bar{\lambda}}V$ where the variations are interpreted as being due to virtual displacement of the null surface along the affine parameter $\bar{\lambda}$ of $k^{a}$. This result was first obtained for \gr\ and \LL theories of gravity when (i) the spacetime admits some symmetry, e.g. staticity or spherical symmetry and (ii) has a horizon. This may suggest that this connection --- between the field equations and a thermodynamic identity --- is a specific phenomenon that occurs only in  solutions containing horizons. But this illusion is broken in \cite{Chakraborty:2015aja} for \gr\ and in \cite{Chakraborty:2015wma} for \LL theories of gravity. There it was shown that 
gravitational field equations near any generic null surface in both \gr\ and \LL theories of gravity lead to a relation: $T\delta_{\bar{\lambda}}S=\delta_{\bar{\lambda}}E+P\delta_{\bar{\lambda}}V$. 

Here we will show that this thermodynamic identity is also contained in the gravitational momentum for $\xi ^{a}$ and can be retrieved from its component along $\ell ^{a}$ which is contained in the projection $P^{a}(\xi)k _{a}$ (which picks out the component along $\ell^a$  because $k^2=0$ and $\ell^ak_a=-1$). Since all the details are similar to the one in \cite{Chakraborty:2015aja} we will be quite brief and just indicate the manner in which the result can be obtained.
(For detailed derivation see Appendix C of \cite{Chakraborty:2015aja}). 
From \ref{Paper06_Sec03_Eq04} we obtain,
\begin{align}\label{Paper06_Sec05_Eq01}
-k_{a}P^{a}(\xi)= k_{a}J^{a}(\xi)-2G^{a}_{b}\xi ^{b}k_{a}
\end{align}
In the second term we can use the field equations, i.e., $2G_{ab}=T_{ab}$ leading to  the combination $T_{ab}\xi ^{a}k^{b}$, which is  the work function (effective pressure) for the matter (see e.g.,\cite{Hayward:1997jp,Kothawala:2010bf}) which, when integrated over the two-surface, will yield the average force $\bar{F}$ due to matter flux on the surface. 
The  first term, viz.,  the projection of the Noether current  $k_{a}J^{a}(\xi)$ has been evaluated explicitly in the earlier work \cite{Chakraborty:2015aja}. Using this result from ref.\cite{Chakraborty:2015aja} for the projection of the  Noether current we arrive at:
\begin{align}\label{Paper06_nNew06}
\bar{F}\delta \bar{\lambda}=T\delta _{\bar{\lambda}}S-\delta _{\bar{\lambda}}E~,
\end{align}
In this expression, $\bar{F}$ stands for the average force on the null surface, $T$ corresponds to the null surface temperature obtained using local Rindler observers, $S=(A/4)$ is the entropy  and $E$ represents the energy given by:
\begin{equation}\label{Paper06_AppA_Eq.14a}
E=\frac{1}{2}\int d\bar{\lambda} \left(\frac{\chi}{2}\right)-\frac{1}{8\pi}\int d^{2}x\partial _{u}\sqrt{q}-\frac{1}{16\pi}\int d \bar{\lambda} \int d^{2}x \sqrt{q}\left\lbrace \frac{1}{2}\beta_A\beta^A \right\rbrace~.
\end{equation}
where $\chi$ stands for the Euler characteristics of the two-surface. (For a detailed discussion see \cite{Chakraborty:2015aja}). A simpler expression covering most of the interesting cases is obtained by setting (a) $\beta _{A}\vert _{r=0}=0$  and (ii) $\partial _{A}\alpha =0$ on the null surface \cite{Chakraborty:2015aja}. On imposing these conditions we arrive at the following simpler expression for energy as, 
\begin{equation}\label{Paper06_nNew09}
E=\frac{1}{2}\int d\bar{\lambda} \left(\frac{\chi}{2}\right)-\frac{1}{8\pi}\int d^{2}x\partial _{u}\sqrt{q}
=\frac{1}{2}\int d\bar{\lambda} \left(\frac{\chi}{2}\right)-\frac{1}{8\pi}\partial _{u}A
\end{equation}
Hence the projection of the gravitational momentum along $\ell _{a}$ is equivalent to thermodynamic identity. 
\subsection{Evolution of the null surface}\label{Paper06_Evo}

Finally, we consider the component of the gravitational momentum $P^a[\xi]$ along $k^a$. If we   expand any vector in the  basis $(\ell ^{a},k^{a},e^{a}_{A})$ as $v^{a}=A\ell ^{a}+Bk^{a}+C^{A}e^{a}_{A}$, the component along $k^{a}$, is obtained by the contraction $\ell _{a}v^{a}$, since $\ell ^{2}=0$ and $\ell _{a}k^{a}=-1$ on the null surface. For a given  null surface we will use our adapted coordinates, i.e., GNC and will show that this component is intimately connected with spacetime evolution. 

In the adapted coordinate system the scalar $-\ell _{a}P^{a}(\xi)$ has the following expression,
\begin{align}\label{Paper06_Sec06_Eq01}
-16\pi \ell _{a}P^{a}(\xi)=\ell _{a}g^{bc}\pounds _{\xi}N^{a}_{bc}
=2\pounds _{\xi}N^{r}_{ur}+q^{AB}\pounds _{\xi}N^{r}_{AB}
\end{align}
where the first line follows from the fact that on the null surface $\ell ^{2}=0$. Both the Lie variation terms have been calculated in \ref{Paper06_App05_Derv} explicitly. From the expressions obtained there the projection of gravitational momentum  turns out to be (see \ref{Paper06_App05_Eq57} and \ref{Paper06_App05_NewLie} of \ref{Paper06_App05_Derv}),
\begin{align}\label{Paper06_Sec06_Eq02}
-16\pi \ell _{a}P^{a}(\xi)=\ell _{a}g^{bc}\pounds _{\xi}N^{a}_{bc}=2\partial _{\lambda}\alpha+2\mathcal{D} + \frac{2}{\sqrt{q}} \partial_\lambda ^{2} \sqrt{q}
\end{align}
Here $\mathcal{D}$ represents the dissipation term obtained through the Navier-Stokes equation and has the definition $\mathcal{D}=(\Theta _{ab}\Theta ^{ab}-\Theta ^{2})$. 

To get a physical interpretation we  integrate this expression over the null surface with volume measure $\sqrt{q}d^{2}xd\lambda$ (for the null vector $\ell ^{a}$ the parameter $\lambda$ is just $u$) and divide by proper factors of $\pi$, and ignore the surface term (which arises from the third term) to obtain:
\begin{align}\label{Paper06_Sec06_Eq06}
\frac{1}{16\pi}\int d\lambda d^{2}x\sqrt{q}\ell _{a}g^{ij}\pounds _{\xi}N^{a}_{ij}
&=\frac{1}{8\pi}\int d\lambda d^{2}x\sqrt{q}\mathcal{D}+\int d^{2}x\ s dT
\end{align}
where $s=\sqrt{q}/4$ and $dT\equiv (dT/d\lambda)d\lambda$. This leads to the result that $\ell_a P^a(\xi)$ and $\ell _{a}g^{bc}\pounds _{\xi}N^{a}_{bc}$ can be interpreted as the heating rate of the null surface per unit area. Integrated over the affine parameter $d\lambda$ and the proper area $\sqrt{q}\, d^2x$, it leads to the heating due to the dissipation (given by the first term in the right hand side) and the integral of  $sdT$ which is the second term. 

Incidentally, a similar interpretation can be given for the matter energy flux which crosses the null surface as well. Using a corresponding expression for $R_{ab}\ell^a\xi^b$ and using the field equations, we can easily show that (see \ref{Paper06_App05_Eq58} of \ref{Paper06_App05_Derv}) 
\begin{align}\label{Paper06_Sec06_Eq04}
T_{ab}\ell ^{a}\ell ^{b}=\frac{1}{\sqrt{q}}\left(\frac{\alpha}{2\pi}\right)\frac{\partial}{\partial u}\left(\frac{\sqrt{q}}{4}\right)
-\frac{1}{8\pi}\mathcal{D} - \frac{1}{8\pi} \frac{1}{\sqrt{q}}\frac{\partial^2 \sqrt{q}}{\partial u^2}
\end{align}
Integrating both sides over the null surface and ignoring boundary contributions at the ends of integration of affine parameter, we get
\begin{align}\label{Paper06_Sec06_Eq07}
\int d^{2}xd\lambda ~\sqrt{q}T_{ab}\ell ^{a}\ell ^{b} + \frac{1}{8\pi}\int d^{2}xd\lambda \sqrt{q}\mathcal{D}
=\int d^{2}xTds 
\end{align}
This tells us that the heating due to matter flux plus the heat generated by the dissipation is equal to the integral of $T\partial_u s$ over the null surface. This reconfirms the earlier interpretation  of the projection of the momentum contributing to the heating of the null surface. 

These results can be used to re-express the heat content of the null surface which was used in the thermodynamic variational principle. Two equivalent forms of the variational  principle, which differ by a total divergence can be given 
based on $R_{ab}\ell ^{a}\ell ^{b}$ and the Lie variation term. These two variational principles (neglecting surface contributions) have the following expressions 
\begin{align}\label{Paper06_Sec06_Eq08}
Q_{1}\equiv\int d\lambda d^{2}x\sqrt{q}\left(-\frac{1}{8\pi}R_{ab}\ell ^{a}\ell ^{b}
+T_{ab}\ell ^{a}\ell ^{b}\right)
=\int d\lambda d^{2}x\sqrt{q}\Big[\frac{1}{8\pi}\mathcal{D}
+T_{ab}\ell ^{a}\ell ^{b}\Big]-\int d^{2}x~Tds
\end{align}
and,
\begin{align}\label{Paper06_Sec06_Eq09}
Q_{2}\equiv \int d\lambda d^{2}x \sqrt{q}\left[\frac{1}{16\pi}\ell _{a}g^{ij}\pounds _{\xi}N^{a}_{ij}
+T_{ab}\ell ^{a}\ell ^{b}\right]
=\int d\lambda d^{2}x \sqrt{q}\Big[\frac{1}{8\pi}\mathcal{D}
+T_{ab}\ell ^{a}\ell ^{b}\Big]+\int d^{2}x~sdT
\end{align}
Note that both these variational principles have the dissipation term $\mathcal{D}$ and matter energy flux through the null surface $T_{ab}\ell ^{a}\ell ^{b}$ in common. However $Q_{1}$ is connected to $TdS$ while $Q_{2}$ is connected to $SdT$. Thus both the variational principles have thermodynamic interpretation.
\section{Conclusions}

Since we have described the physical consequences of the results in the various sections themselves, we shall limit ourselves to summarising the key conclusions in this section. 
\begin{itemize}

\item There is considerable amount of evidence to suggest that gravitational field equations have the same status as, say, the equations of fluid mechanics. They describe the macroscopic, thermodynamic, limit of an underlying statistical mechanics of the microscopic degrees of freedom of the spacetime. The macro and micro descriptions are connected  through the heat density $Ts$ of the spacetime. Here, the temperature $T$ arises  from the interpretation of the null surfaces as local Rindler horizons. The entropy density is a phenomenological input, the form of which determines the theory. For a very wide class of theories, it can be defined in terms of a function $F( R^{cd}_{ab}, \delta^i_j)$ built from (2,2) curvature tensor $ R^{cd}_{ab}$ and the Kronecker deltas, as: 
\begin{equation}
s = -\frac{1}{8}\sqrt{q}P^{ab}_{cd} \epsilon_{ab}\epsilon^{cd}; \quad 
P^{ab}_{cd} = \frac{\partial F}{\partial R^{cd}_{ab}}; \quad 
\nabla_a P^{ab}_{cd} = 0 
\end{equation}

\item 
Given a Vector field $v^{a}$, one can construct three currents: (a) the Noether current $J^{a}(v)$, (b) the gravitational momentum $P^{a}(v)$ and (c) the reduced gravitational momentum $\mathcal{P}^{a}(v)$. Interestingly enough, one can attribute  \textit{thermodynamic} meaning to these  quantities which  are usually considered to be \textit{geometrical}. For example, the conserved current $J^{a}$, associated with the time-development vector $\xi ^{a}$ of the spacetime, leads to a conserved charge (i.e., integral of $u_{a}J^{a}(\xi)$ defined either on a spacelike surface or on  a null surface) that can be related to the boundary heat density $Ts$, where $T$ is the Unruh-Davies temperature and $s$ stands for entropy density.

\item One can also define the notion of gravitational momentum $P^a$ for all the \LL models of gravity such that $\nabla_a (P^a + M^a) = 0$ (where $M^a$ is the momentum density of matter) for all observers,  leads to the field equation of the \LL model. This generalizes a  previous result for \gr.

\item The field equations can also be derived from a thermodynamic variational principle, which essentially extremises the total heat density of all the null surfaces in the spacetime. This variational principle can be expressed directly in terms of the total gravitational momentum, thereby providing it with a simple physical interpretation. 

\item One can associate with any null surface the two null vector fields $\ell_a, k_a$ with $\ell_a k^a = -1$ and $\ell_a$ being the tangent vector to the congruence defining the null surface, as well as the 2-metric $q_{ab} = g_{ab} + \ell_a k_b + \ell_b k_a$. These structures define three natural projections of the gravitational momentum ($P^a \ell_a, P^a k_a, P^a q_{ab}$), all of which have  thermodynamic significance. The first one leads to the description of time evolution of the null surface in terms of suitably defined bulk and surface degrees of freedom; the second leads to  a thermodynamic identity which can be written in the form $TdS = dE + PdV$; the third leads to a Navier-Stokes equation for the transverse degrees of freedom on the null surface which can be interpreted as a drift velocity.

\end{itemize}
These results again demonstrate that the emergent gravity paradigm enriches our understanding of the spacetime dynamics and the structure of null surfaces, by allowing a rich variety of thermodynamic backdrops for the geometrical variables.
\section*{Acknowledgements}

Work of S.C. is funded by a SPM fellowship from CSIR, Government of India. Research of T.P. is partially supported by J.C. Bose research grant of DST, Government of India. Authors thank Shyam Date, Dawood Kothawala, Kinjalk Lochan, Krishnamohan Parattu and Suprit Singh for helpful discussions. 
\appendix
\labelformat{section}{Appendix #1} 
\section*{Appendices}
\section{General Analysis Regarding Null Surfaces}\label{Paper06_App03_General}

We will start with a null vector $\ell _{a}=A\nabla _{a}B$, which satisfies the condition $\ell ^{2}=0$ only over a single surface, which is the null surface under our consideration. Then we obtain
\begin{equation}\label{Paper06_App03_Eq01}
\ell ^{a}\nabla _{a}\ell ^{b}=\kappa \ell ^{b}
\end{equation}
where we have the following expression for $\kappa$
\begin{equation}\label{Paper06_App03_Eq02}
\kappa =\ell ^{a}\partial _{a}\ln A+\tilde{\kappa};\qquad 
\tilde{\kappa}=-\frac{1}{2}k^{a}\partial _{a}\ell ^{2}
\end{equation}
The last relation defining $\tilde{\kappa}$ can also be written as $\nabla _{a}\ell ^{2}=2\tilde{\kappa}\ell _{a}$. The derivation of the result goes as follows, let us expand $\nabla _{b}\ell ^{2}$ in canonical null basis, i.e., $\nabla _{a}\ell ^{2}=C\ell _{a}+Dk_{a}+E_{A}e^{A}_{a}$. Then both $E_{A}=e^{a}_{A}\nabla _{a}\ell ^{2}$ and $D=-\ell ^{a}\nabla _{a}\ell ^{2}$ vanishes, since variation of $\ell ^{2}$ along the null surface vanishes. This shows that the only non-zero component of $\nabla _{a}\ell ^{2}$ is along $\ell _{a}$. Then it turns out that \cite{Parattu:2015gga}
\begin{equation}\label{Paper06_App03_Eq03}
\nabla _{i}\ell ^{i}=\Theta +\kappa +\tilde{\kappa}
\end{equation}
where, $\Theta=q^{ma}q_{mb}\nabla _{a}\ell ^{b}$. Note that the term $\tilde{\kappa}$ enters the picture as $\ell ^{2}= 0$ only on the null surface. With this setup let us now find out $R_{ab}\ell ^{a}\ell ^{b}$ in detail, which leads to,
\begin{align}\label{Paper06_App03_Eq04}
R_{ab}\ell ^{a}\ell ^{b}&=\ell ^{j}\left(\nabla _{i}\nabla _{j}\ell ^{i}-\nabla _{j}\nabla _{i}\ell ^{i}\right)
\nonumber
\\
&=\nabla _{i}\left(\ell ^{j}\nabla _{j}\ell ^{i}\right)-\nabla _{j}\left(\ell ^{j}\nabla _{i}\ell ^{i}\right)
-\nabla _{i}\ell ^{j}\nabla _{j}\ell ^{i}+\left(\nabla _{i}\ell ^{i}\right)^{2}
\nonumber
\\
&=\nabla _{i}\left(\ell ^{j}\nabla _{j} \ell ^{i}-\ell ^{i}\nabla _{j}\ell ^{j}\right)
-\left(\nabla _{i}\ell ^{j}\nabla _{j}\ell ^{i}-\left(\nabla _{i}\ell ^{i}\right)^{2}\right)
\end{align}
However in general, $\ell ^{j}\nabla _{j}\ell ^{i}=\kappa \ell ^{i}$ is not true, it only holds on the null surface (when $\ell ^{2}=0$ everywhere this relation is also true everywhere). Since we were doing the calculation for the most general case, $\ell ^{2}\ne 0$ in the above expression we cannot substitute $\ell ^{j}\nabla _{j}\ell ^{i}=\kappa \ell ^{i}$, since it appears inside the derivative. Thus for the special case when $\ell ^{2}=0$ everywhere, we will arrive at the following result
\begin{align}\label{Paper06_App03_Eq05}
R_{ab}\ell ^{a}\ell ^{b}=-\nabla _{i}\left(\Theta \ell ^{i}\right)
-\left(\nabla _{i}\ell ^{j}\nabla _{j}\ell ^{i}-\left(\nabla _{i}\ell ^{i}\right)^{2}\right)
\end{align} 
In order to simplify things quiet a bit we will compute the last term $\left(\nabla _{i}\ell ^{j}\nabla _{j}\ell ^{i}-\left(\nabla _{i}\ell ^{i}\right)^{2}\right)$ which we designate by $\mathcal{S}$. Then we start by calculating the following object on the null surface
\begin{align}\label{Paper06_App03_Eq06}
\Theta ^{ab}&=q^{a}_{m}q^{b}_{n}\nabla ^{m}\ell ^{n}
\nonumber
\\
&=\left(\delta ^{a}_{m}+\ell ^{a}k_{m}+k^{a}\ell _{m}\right)
\left(\delta ^{b}_{n}+\ell ^{b}k_{n}+k^{b}\ell _{n}\right)\nabla ^{m}\ell ^{n}
\nonumber
\\
&=\nabla ^{a}\ell ^{b}+\ell ^{a}k_{m}\nabla ^{m}\ell ^{b}+\kappa k^{a}\ell ^{b}+\ell ^{b}k_{n}\nabla ^{a}\ell ^{n}+\ell ^{a}\ell ^{b}k_{m}k_{n}\nabla ^{m}\ell ^{n}
\nonumber
\\
&-\kappa k^{a}\ell ^{b}+\tilde{\kappa}\ell ^{a}k^{b}-\tilde{\kappa}\ell ^{a}k^{b}+\kappa k^{a}k^{b}\ell ^{2}
\nonumber
\\
&=\nabla ^{a}\ell ^{b}+\ell ^{a}k_{m}\nabla ^{m}\ell ^{b}+\ell ^{b}k_{n}\nabla ^{a}\ell ^{n}+\ell ^{a}\ell ^{b}k_{m}k_{n}\nabla ^{m}\ell ^{n}
\end{align}
In arriving at the third line we have used the following results: $\ell ^{a}\nabla _{a}\ell ^{i}=\kappa \ell ^{i}$ and $\ell _{a}\nabla _{b}\ell ^{a}=\tilde{\kappa}\ell _{b}$. Then we can reverse the above equation leading to
\begin{align}\label{Paper06_App03_Eq07}
\nabla _{a}\ell _{b}=\Theta _{ab}-\ell _{a}k^{m}\nabla _{m}\ell _{b}-\ell _{b}k_{n}\nabla _{a}\ell ^{n}
-\ell _{a}\ell _{b}k_{m}k_{n}\nabla ^{m}\ell ^{n}
\end{align}
From the above equation we can derive two very important identity:
\begin{align}\label{Paper06_App03_Eq08}
\left(\nabla _{a}\ell _{b}\right)\left(\nabla ^{a}\ell ^{b}\right)&=
\Theta _{ab}\Theta ^{ab}+\ell ^{a}\ell _{b}k^{m}\nabla _{m}\ell ^{b}k^{n}\nabla _{a}\ell _{n}
+\ell _{a}\ell ^{b}k_{n}\nabla ^{a}\ell ^{n}k^{m}\nabla _{m}\ell _{b}
\nonumber
\\
&=\Theta _{ab}\Theta ^{ab}+2\left(\tilde{\kappa}k^{m}\ell _{m}\right)\left(\kappa \ell _{n}k^{n}\right)
\nonumber
\\
&=\Theta _{ab}\Theta ^{ab}+2\kappa \tilde{\kappa}
\end{align}
In the same spirit we will arrive at
\begin{align}\label{Paper06_App03_Eq09}
\left(\nabla _{a}\ell _{b}\right)\left(\nabla ^{b}\ell ^{a}\right)&=
\Theta _{ab}\Theta ^{ab}+\ell ^{a}\ell _{b}k_{n}\nabla _{a}\ell ^{n}k^{m}\nabla ^{b}\ell _{m}
+\ell _{a}\ell ^{b}k^{m}\nabla _{m}\ell ^{a}k^{n}\nabla _{n}\ell _{b}
\nonumber
\\
&=\Theta _{ab}\Theta ^{ab}+\left(\tilde{\kappa}k^{m}\ell _{m}\right)\left(\tilde{\kappa}k^{n}\ell _{n}\right)+\left(\kappa \ell _{n}k^{n}\right)\left(\kappa \ell _{m}k^{m}\right)
\nonumber
\\
&=\Theta _{ab}\Theta ^{ab}+\kappa ^{2}+\tilde{\kappa}^{2}
\end{align}
The extrinsic curvature for null surfaces, i.e., $\Theta _{ab}$ can be given a very natural interpretation. This essentially follows from \cite{Parattu:2015gga}. There the expression for $\Theta _{ab}$ in terms of Lie variation of $q_{ab}$ along the null generator $\ell _{a}$ was obtained as,
\begin{align}\label{Paper06_AppNew01}
\Theta _{ab}=\frac{1}{2}q^{m}_{a}q^{n}_{b}\pounds _{\ell}q_{mn}
\end{align}
Now expanding out the Lie derivative term we obtain,
\begin{align}
\pounds _{\ell}q_{mn}=\ell ^{i}\partial _{i}q_{mn}+q_{ma}\partial _{n}\ell ^{a}+q_{an}\partial _{m}\ell ^{a}
\end{align}
Which on being substituted in \ref{Paper06_AppNew01} immediately leads to,
\begin{align}
\Theta _{ab}=\frac{1}{2}q^{m}_{a}q^{n}_{b}\ell ^{i}\partial _{i}q_{mn}+\frac{1}{2}q_{ai}q^{n}_{b}\partial _{n}\ell ^{i}+\frac{1}{2}q_{bi}q^{m}_{a}\partial _{m}\ell ^{i}
\end{align}
Now on the null surface $q_{ab}=q_{AB}$ as the only non-zero component. Hence the above equation can be written as,
\begin{align}
\Theta _{ab}=\Theta _{AB}=\frac{1}{2}\dfrac{d}{d\lambda}q_{AB}+\frac{1}{2}q_{AC}\partial _{B}\ell ^{C}+\frac{1}{2}q_{BC}\partial _{A}\ell ^{C}
\end{align}
On the null surface $q^{a}_{b}\ell ^{b}=0$, which in this coordinate system leads to $\ell ^{A}=0$ on the null surface. Since $\partial _{A}\ell ^{2}$ represent derivatives on the null surface it also vanishes. If $\ell ^{2}=0$ everywhere, then also $\ell ^{A}$ is identically zero everywhere. Hence we have 
\begin{align}\label{Paper06_AppNew02}
\Theta _{ab}=\frac{1}{2}\dfrac{d}{d\lambda}q_{AB}
\end{align}
There is another way to get this result. If $e^{a}_{A}$ are the basis vectors on the null surface and if $\ell _{a},e^{a}_{A}$ forms coordinate basis vectors, then $q_{AB}=q_{ab}e^{a}_{A}e^{b}_{B}$ is a scalar under 4-dimensional coordinate transformation. This immediately leads to the previous expression. For more discussions along identical lines see \cite{Parattu:2015gga}.

Now the expression for the quantity $\mathcal{S}$ can be obtained as
\begin{align}\label{Paper06_App03_Eq10}
\mathcal{S}&=\nabla _{i}\ell ^{j}\nabla _{j}\ell ^{i}-\left(\nabla _{i}\ell ^{i}\right)^{2}
\nonumber
\\
&=\Theta _{ab}\Theta ^{ab}+\kappa ^{2}+\tilde{\kappa}^{2}-\left(\Theta +\kappa +\tilde{\kappa}\right)^{2}
\nonumber
\\
&=\left(\Theta _{ab}\Theta ^{ab}-\Theta ^{2}\right)-2\Theta \left(\kappa +\tilde{\kappa}\right)
-2\kappa \tilde{\kappa}
\end{align}
Using the general expression for $R_{ab}\ell ^{a}\ell ^{b}$ we obtain the following form:
\begin{align}\label{Paper06_App03_Eq11}
R_{ab}\ell ^{a}\ell ^{b}&=\nabla _{i}\left(\ell ^{j}\nabla _{j} \ell ^{i}-\ell ^{i}\nabla _{j}\ell ^{j}\right)-\mathcal{S}
\nonumber
\\
&=\nabla _{i}\left(\ell ^{j}\nabla _{j} \ell ^{i}-\ell ^{i}\nabla _{j}\ell ^{j}\right)
-\left(\Theta _{ab}\Theta ^{ab}-\Theta ^{2}\right)+2\Theta \left(\kappa +\tilde{\kappa}\right)
+2\kappa \tilde{\kappa}
\end{align}
For the situation where, $\ell ^{2}=0$ everywhere we finally arrive at the following simplified expression
\begin{align}\label{Paper06_App03_Eq12}
R_{ab}\ell ^{a}\ell ^{b}&=-\left(\Theta _{ab}\Theta ^{ab}-\Theta ^{2}\right)+2\Theta \kappa
+\nabla _{i}\left(\kappa \ell ^{i}-\left[\Theta +\kappa \right]\ell ^{i}\right)
\nonumber
\\
&=-\left(\Theta _{ab}\Theta ^{ab}-\Theta ^{2}\right)+\Theta \kappa -\frac{1}{\sqrt{q}}\dfrac{d}{d\lambda}\left(\sqrt{q}\Theta \right)
\end{align}
Let us now try to derive the Raychaudhuri equation starting from the basic properties of null surfaces. We start with the following result
\begin{align}\label{Paper06_App03_Eq13}
\ell ^{a}\nabla _{a}\left(\nabla _{c}\ell _{d}\right)&=\ell ^{a}\nabla _{a}\nabla _{c}\ell _{d}
\nonumber
\\
&=R_{dbac}\ell ^{a}\ell ^{b}+\ell ^{a}\nabla _{c}\nabla _{a}\ell _{d}
\nonumber
\\
&=\nabla _{c}\left(\ell ^{a}\nabla _{a}\ell _{d}\right)-\nabla _{a}\ell _{d}\nabla _{c}\ell ^{a}-R_{bdac}\ell ^{b}\ell ^{a}
\end{align}
Then contraction of the indices $c,d$ leads to the following result
\begin{align}\label{Paper06_App03_Eq14}
\ell ^{a}\nabla _{a}\left(\nabla _{c}\ell ^{c}\right)&=\nabla _{c}\left(\ell ^{a}\nabla _{a}\ell ^{c}\right)-\nabla _{a}\ell _{b}\nabla ^{b}\ell ^{a}-R_{ab}\ell ^{a}\ell ^{b}
\end{align}
Otherwise we can rewrite it in a different manner which exactly coincide with \ref{Paper06_App03_Eq04}.
On using \ref{Paper06_App03_Eq09} and the decomposition: $\Theta _{ab}=(1/2)\Theta q_{ab}+\sigma _{ab}+\omega _{ab}$ we arrive at
\begin{align}\label{Paper06_App03_Eq14a}
\ell ^{a}\nabla _{a}\left(\nabla _{c}\ell ^{c}\right)&-\nabla _{c}\left(\ell ^{a}\nabla _{a}\ell ^{c}\right)
=-\Theta _{ab}\Theta ^{ab}-\kappa ^{2}-\tilde{\kappa}^{2}-R_{ab}\ell ^{a}\ell ^{b}
\nonumber
\\
&=-\frac{1}{2}\Theta ^{2}-\sigma ^{ab}\sigma _{ab}+\omega _{ab}\omega ^{ab}
-\kappa ^{2}-\tilde{\kappa}^{2}-R_{ab}\ell ^{a}\ell ^{b}
\end{align}
For the situation where, $\ell ^{2}=0$ the left hand side is just: $d\left(\Theta +\kappa\right)/d\lambda$ and the first term on the right hand side is $d\kappa/d\lambda+\kappa (\Theta +\kappa)$ the above equation leads to
\begin{align}\label{Paper06_App03_Eq15}
\dfrac{d\Theta}{d\lambda}&=\kappa \Theta +\kappa ^{2}-\nabla _{a}\ell _{b}\nabla ^{b}\ell ^{a}-R_{ab}\ell ^{a}\ell ^{b}
\nonumber
\\
&=\kappa \Theta -\Theta _{ab}\Theta ^{ab}-R_{ab}\ell ^{a}\ell ^{b}
\nonumber
\\
&=\kappa \Theta -\frac{1}{2}\Theta ^{2}-\sigma ^{ab}\sigma _{ab}+\omega _{ab}\omega ^{ab}
-R_{ab}\ell ^{a}\ell ^{b}
\end{align}
where to arrive at the last line we have used the following decomposition: $\Theta _{ab}=(1/2)\Theta q_{ab}+\sigma _{ab}+\omega _{ab}$. This is precisely the one obtained in \cite{Gravitation} though in a completely different manner.

The next object to consider is the quantity $\ell _{a}J^{a}(\ell)$. This can be obtained by using the identity for Noether current leading to,
\begin{align}\label{Paper06_App03_Eq16}
\frac{1}{A}\ell _{a}J^{a}(\ell)&=\nabla _{b}\left(\left\lbrace \ell ^{a}\ell ^{b}
-\ell ^{2}g^{ab}\right\rbrace \frac{\nabla _{a}A}{A^{2}}\right)
\nonumber
\\
&=\nabla _{b}\left[\frac{1}{A}\ell ^{b}\left(\kappa -\tilde{\kappa}\right)\right]
-\nabla _{b}\left(\ell ^{2}\frac{\nabla ^{b}A}{A^{2}}\right)
\nonumber
\\
&=\frac{1}{A}\nabla _{i}\left[\left(\kappa -\tilde{\kappa}\right)\ell ^{i}\right]
-\frac{1}{A}\left(\kappa -\tilde{\kappa}\right)^{2}-\frac{\nabla ^{b}A}{A^{2}}\nabla _{b}\ell ^{2}
\nonumber
\\
&=\frac{1}{A}\nabla _{i}\left[\left(\kappa -\tilde{\kappa}\right)\ell ^{i}\right]
-\frac{1}{A}\left(\kappa -\tilde{\kappa}\right)^{2}
-\frac{2}{A}\tilde{\kappa}\left(\kappa -\tilde{\kappa}\right)
\end{align}
This can be written in a slightly modified manner as,
\begin{align}\label{Paper06_App03_Eq17}
\ell _{a}J^{a}(\ell)&=\nabla _{i}\left[\left(\kappa -\tilde{\kappa}\right)\ell ^{i}\right]
-\left(\kappa ^{2} -\tilde{\kappa}^{2}\right)
\nonumber
\\
&=\ell ^{i}\nabla _{i}\left(\kappa -\tilde{\kappa}\right)
+\left(\kappa -\tilde{\kappa}\right)\left(\Theta +\kappa +\tilde{\kappa}\right)
-\left(\kappa ^{2} -\tilde{\kappa}^{2}\right)
\nonumber
\\
&=\dfrac{d}{d\lambda}\left(\kappa -\tilde{\kappa}\right)+\Theta \left(\kappa -\tilde{\kappa}\right)
\end{align}
The above expression can be simplified significantly by noting that $\Theta =d(\ln \sqrt{q})/d\lambda$, which leads to
\begin{align}\label{Paper06_App03_Eq17a}
\ell _{a}J^{a}(\ell)=\frac{1}{\sqrt{q}}\dfrac{d}{d\lambda}\left[\left(\kappa -\tilde{\kappa}\right)\sqrt{q}\right]
\end{align}
Again, we have
\begin{align}\label{Paper06_App03_Eq18}
D_{a}\left[\left(\kappa -\tilde{\kappa}\right)\ell ^{a}\right]
&=\left(g^{ab}+\ell ^{a}k^{b}+\ell ^{b}k^{a}\right)
\nabla _{a}\left[\left(\kappa -\tilde{\kappa}\right)\ell_{b}\right]
\nonumber
\\
&=\nabla _{a}\left[\left(\kappa -\tilde{\kappa}\right)\ell ^{a}\right]
+\left(\ell ^{a}k^{b}+\ell ^{b}k^{a}\right)
\left[\left(\kappa -\tilde{\kappa}\right)\nabla _{a}\ell _{b}
+\ell _{b}\nabla _{a} \left(\kappa -\tilde{\kappa}\right)\right]
\nonumber
\\
&=\nabla _{a}\left[\left(\kappa -\tilde{\kappa}\right)\ell ^{a}\right]
-\kappa \left(\kappa -\tilde{\kappa}\right)-\tilde{\kappa}\left(\kappa -\tilde{\kappa}\right)
-\dfrac{d}{d\lambda}\left(\kappa -\tilde{\kappa}\right)
\\
&=\left(\kappa -\tilde{\kappa}\right)\left(\Theta +\kappa +\tilde{\kappa}\right)-\left(\kappa ^{2}-\tilde{\kappa}^{2}\right)
\nonumber
\\
&=\left(\kappa -\tilde{\kappa}\right)\dfrac{d \ln \sqrt{q}}{d\lambda}
\end{align}
Thus we arrive at
\begin{align}\label{Paper06_App03_Eq19}
\ell _{a}J^{a}(\ell)&=D_{a}\left[\left(\kappa -\tilde{\kappa}\right)\ell ^{a}\right]
+\dfrac{d}{d\lambda}\left(\kappa -\tilde{\kappa}\right)
+\left(\kappa -\tilde{\kappa}\right)\left(\kappa +\tilde{\kappa}\right)
-\left(\kappa ^{2} -\tilde{\kappa}^{2}\right)
\nonumber
\\
&=D_{a}\left[\left(\kappa -\tilde{\kappa}\right)\ell ^{a}\right]
+\dfrac{d}{d\lambda}\left(\kappa -\tilde{\kappa}\right)
\end{align}
From the expression of the Noether current we get,
\begin{equation}\label{Paper06_App03_Eq20}
\ell _{a}J^{a}(\ell)=2R_{ab}\ell ^{a}\ell ^{b}+\ell _{a}g^{ij}\pounds _{\ell}N^{a}_{ij}
\end{equation}
The above equation can be used to write, $g^{ab}\pounds _{\ell}N^{c}_{ab}$ in terms of $\kappa$ and $R_{ab}$. For that purpose we use \ref{Paper06_Sec03_Eq02} and insert \ref{Paper06_App03_Eq17} leading to
\begin{align}\label{Paper06_App03_Eq21}
\ell _{a}g^{ij}\pounds _{\ell}N^{a}_{ij}&=\ell _{a}J^{a}(\ell)-2R_{ab}\ell ^{a}\ell ^{b}
\nonumber
\\
&=\dfrac{d}{d\lambda}\left(\kappa -\tilde{\kappa}\right)+\Theta \left(\kappa -\tilde{\kappa}\right)
-2R_{ab}\ell ^{a}\ell ^{b}
\nonumber
\\
&=\frac{1}{\sqrt{q}}\dfrac{d}{d\lambda}\left[\sqrt{q}\left(\kappa -\tilde{\kappa}\right)\right]
-2R_{ab}\ell ^{a}\ell ^{b}
\nonumber
\\
&=\frac{2}{\sqrt{q}}\dfrac{d}{d\lambda}\left(\sqrt{q}\kappa\right)-2R_{ab}\ell ^{a}\ell ^{b}
-\frac{1}{\sqrt{q}}\dfrac{d}{d\lambda}\left[\sqrt{q}\left(\kappa +\tilde{\kappa}\right)\right]
\end{align}
The above expression when integrated over the null surface with integration measure, $\sqrt{q}d^{2}xd\lambda$ and then being divided by $16\pi$ leads to
\begin{align}\label{Paper06_App03_Eq22}
\frac{1}{16\pi}\int d^{2}xd\lambda \sqrt{q}\left\lbrace \ell _{a}g^{ij}\pounds _{\ell}N^{a}_{ij}
+\frac{1}{\sqrt{q}}\dfrac{d}{d\lambda}\left[\sqrt{q}\left(\kappa +\tilde{\kappa}\right)\right]\right\rbrace
=\frac{1}{8\pi}\int d^{2}x\sqrt{q}\kappa \vert _{1}^{2}
-\frac{1}{8\pi}\int d^{2}xd\lambda \sqrt{q}R_{ab}\ell ^{a}\ell ^{b}
\end{align}
Then on using the field equation: $R_{ab}=8\pi\left[T_{ab}-(1/2)g_{ab}T\right]=8\pi \bar{T}_{ab}$ and then being substituted in \ref{Paper06_App03_Eq22} we arrive at the following expression
\begin{align}\label{Paper06_App03_Eq23}
\frac{1}{4}\int d^{2}x \sqrt{q}\left(\frac{\kappa}{2\pi}\right)\vert _{1}^{2}
-\int d^{2}xd\lambda \sqrt{q}~T_{ab}\ell ^{a}\ell ^{b}=
\frac{1}{16\pi}\int d^{2}xd\lambda \sqrt{q}\left\lbrace \ell _{a}g^{ij}\pounds _{\ell}N^{a}_{ij}
+\frac{1}{\sqrt{q}}\dfrac{d}{d\lambda}\left[\sqrt{q}\left(\kappa +\tilde{\kappa}\right)\right]\right\rbrace
\end{align}
where the last equality follows from the fact that $\ell ^{2}=0$ on the null surface. 
The above equation can be written in a more abstract form as
\begin{align}\label{Paper06_App03_Eq24}
\frac{1}{16\pi}\int d^{2}xd\lambda \sqrt{q} \ell _{a}g^{ij}\pounds _{\ell}N^{a}_{ij}&=
\left[\frac{1}{4}\int _{\lambda _{2}} d^{2}x \sqrt{q}\left(\frac{\kappa}{2\pi}\right)
-\frac{1}{4}\int _{\lambda _{1}} d^{2}x \sqrt{q}\left(\frac{\kappa}{2\pi}\right)\right]
\nonumber
\\
&-\int d^{2}xd\lambda \sqrt{q}~T_{ab}\ell ^{a}\ell ^{b}
-\frac{1}{16\pi}\int d^{2}x d\lambda \dfrac{d}{d\lambda}\left[\sqrt{q}\left(\kappa +\tilde{\kappa}\right)\right]
\end{align}
As an illustration when $\ell ^{2}=0$ everywhere, we have $\tilde{\kappa}=0$, then \ref{Paper06_App03_Eq24} leads to,
\begin{align}\label{Paper06_App03_Eq29}
\frac{1}{16\pi}\int d^{2}xd\lambda \sqrt{q} \ell _{a}g^{ij}\pounds _{\ell}N^{a}_{ij}
&=\frac{1}{2}\left[\int _{\lambda _{2}} d^{2}x \frac{\sqrt{q}}{4}\left(\frac{\kappa}{2\pi}\right)
-\int _{\lambda _{1}} d^{2}x \frac{\sqrt{q}}{4}\left(\frac{\kappa}{2\pi}\right)\right]
\nonumber
\\
&-\int d^{2}xd\lambda \sqrt{q}~T_{ab}\ell ^{a}\ell ^{b}
\end{align}
We will now try to obtain an expression for the quantity $\ell _{a}g^{ij}\pounds _{\ell}N^{a}_{ij}$ independently. For that we start with the symmetric and anti-symmetric part of the derivative $\nabla _{a}\ell _{b}$ such that:
\begin{align}\label{Paper06_App03_Eq30}
S^{ab}=\nabla ^{a}\ell ^{b}+\nabla ^{b}\ell ^{a},\qquad J^{ab}=\nabla ^{a}\ell ^{b}-\nabla ^{b}\ell ^{a}
\end{align}
Then we have the following result: $\nabla ^{a}\ell ^{b}=(1/2)(S^{ab}+J^{ab})$, which on being substituted in the identity:
\begin{align}\label{Paper06_App03_Eq31}
\nabla _{b}\left(\nabla ^{a}\ell ^{b}\right)-\nabla ^{a}\left(\nabla _{b}\ell ^{b}\right)=R^{a}_{b}\ell ^{b}
\end{align}
leads to the following identification:
\begin{align}\label{Paper06_App03_Eq32}
g^{ab}\pounds _{\ell}N^{c}_{ab}=-\nabla _{b}\left(S^{bc}-g^{bc}S\right)
\end{align}
Hence we arrive at the following relation:
\begin{align}\label{Paper06_App03_Eq33}
\ell _{a}g^{bc}\pounds _{\ell}N^{a}_{bc}&=-\ell _{a}\nabla _{b}\left[\nabla ^{a}\ell ^{b}+\nabla ^{b}\ell ^{a}-2g^{ab}\nabla _{c}\ell ^{c}\right]
\nonumber
\\
&=-\nabla _{b}\left[\ell ^{a}\nabla _{a}\ell ^{b}+\ell _{a}\nabla ^{b}\ell ^{a}-2\ell ^{b}\left(\nabla _{c}\ell ^{c}\right) \right]
+\nabla _{a}\ell _{b}\nabla ^{b}\ell ^{a}+\nabla _{a}\ell _{b}\nabla ^{a}\ell ^{b}
-2\left(\nabla _{c}\ell ^{c}\right)^{2}
\nonumber
\\
&=-\nabla _{b}\left[\ell ^{a}\nabla _{a}\ell ^{b}+\ell _{a}\nabla ^{b}\ell ^{a}-2\ell ^{b}\left(\nabla _{c}\ell ^{c}\right) \right]
+2\Theta _{ab}\Theta ^{ab}+\left(\kappa +\tilde{\kappa}\right)^{2}-2\left(\Theta +\kappa +\tilde{\kappa}\right)^{2}
\nonumber
\\
&=-\nabla _{b}\left[\ell ^{a}\nabla _{a}\ell ^{b}+\ell _{a}\nabla ^{b}\ell ^{a}-2\ell ^{b}\left(\nabla _{c}\ell ^{c}\right) \right]
+2\left(\Theta _{ab}\Theta ^{ab}-\Theta ^{2}\right)-4\Theta \left(\kappa +\tilde{\kappa}\right)
-\left(\kappa +\tilde{\kappa}\right)^{2}
\end{align}
For the case $\ell ^{2}=0$ the term within bracket can be written in a simplified manner such that Lie derivative term gets simplified leading to:
\begin{align}\label{Paper06_App03_Eq34}
\ell _{a}g^{bc}\pounds _{\ell}N^{a}_{bc}&=2\left(\Theta _{ab}\Theta ^{ab}-\Theta ^{2}\right)-4\Theta \kappa 
-\kappa ^{2}+\nabla _{b}\left[\left(2\Theta +\kappa \right)\ell ^{b}\right]
\nonumber
\\
&=2\left(\Theta _{ab}\Theta ^{ab}-\Theta ^{2}\right)-4\Theta \kappa -\kappa ^{2}
+\left(2\Theta +\kappa \right)\left(\Theta +\kappa \right)+\dfrac{d}{d\lambda}\left(2\Theta +\kappa \right)
\nonumber
\\
&=2\left(\Theta _{ab}\Theta ^{ab}-\Theta ^{2}\right)-\Theta \kappa +\dfrac{d}{d\lambda}\kappa
+\frac{2}{\sqrt{q}}\dfrac{d}{d\lambda}\left(\sqrt{q}\Theta \right)
\end{align}
If the null generator is affinely parametrized then $\kappa =0$ and \ref{Paper06_App03_Eq34} reduces to:
\begin{align}\label{Paper06_App03_Eq34a}
\ell _{a}g^{bc}\pounds _{\ell}N^{a}_{bc}=2\left(\Theta _{ab}\Theta ^{ab}-\Theta ^{2}\right)
+\frac{2}{\sqrt{q}}\dfrac{d}{d\lambda}\left(\sqrt{q}\Theta \right)
\end{align}
While for the null generator $\ell ^{a}$ in GNC we have (see \ref{Paper06_App05_Derv} \ref{Paper06_App05_Eq47}):
\begin{align}\label{Paper06_App03_Eq34b}
\ell _{a}g^{bc}\pounds _{\ell}N^{a}_{bc}=2\left(\Theta _{ab}\Theta ^{ab}-\Theta ^{2}\right)
+\frac{2}{\sqrt{q}}\dfrac{d ^{2}\sqrt{q}}{d\lambda ^{2}}+2\dfrac{d\kappa}{d\lambda}-\frac{2}{\sqrt{q}}\dfrac{d}{d\lambda}\left(\sqrt{q}\kappa \right)
\end{align}
Through the above analysis we have obtained expressions for $\ell _{a}J^{a}(\ell)$, $R_{ab}\ell ^{a}\ell ^{b}$ and $\ell _{a}g^{bc}\pounds _{\ell}N^{a}_{bc}$. 

It turns out from the above analysis that $\Theta _{ab}\Theta ^{ab}-\Theta ^{2}$ can be given a more physical meaning by considering Lie variation of gravitational momentum. This can be obtained by considering variation of the gravitational momentum first:
\begin{align}\label{Paper06_App03_Eq35}
q_{ab}\delta \Pi ^{ab}&=q_{ab}\delta \left[\sqrt{q}\left(\Theta ^{ab}-\Theta q^{ab}\right)\right]
\nonumber
\\
&=q_{ab}\sqrt{q}\delta \Theta ^{ab}-2\sqrt{q}\delta \Theta -\sqrt{q}\Theta q_{ab}\delta q^{ab}-\Theta \delta \sqrt{q}
\nonumber
\\
&=\sqrt{q}q_{ab}\delta \Theta ^{ab}-2\sqrt{q}\delta \Theta +\Theta \delta \sqrt{q}
\end{align}
Now specializing to Lie variation we arrive at:
\begin{align}\label{Paper06_App03_Eq36}
-q_{ab}\pounds _{\ell}\Pi ^{ab}&=-\Theta \pounds _{\ell}\sqrt{q}-\sqrt{q}q_{ab}\pounds _{\ell}\Theta ^{ab}+2\sqrt{q}\pounds _{\ell}\Theta 
\nonumber
\\
&=-\sqrt{q}\pounds _{\ell}\Theta +\sqrt{q}\Theta ^{ab}\pounds _{\ell}q_{ab}-\Theta \pounds _{\ell}\sqrt{q}+2\sqrt{q}\pounds _{\ell}\Theta 
\nonumber
\\
&=2\sqrt{q}\left(\Theta ^{ab}\Theta _{ab}-\Theta ^{2}\right)+\pounds _{\ell}\left(\sqrt{q}\Theta \right)
\nonumber
\\
&=\sqrt{q}\ell _{a}g^{bc}\pounds _{\ell}N^{a}_{bc}-\dfrac{d ^{2}\sqrt{q}}{d\lambda ^{2}}
\end{align}
where in the last line we have used \ref{Paper06_App03_Eq34a}. Here the quantities $\Theta _{ab}$ and $\Theta$ can be defined as: $\Theta _{ab}=(1/2)\pounds _{\ell}q_{ab}$ and $\Theta =\pounds _{\ell}\ln \sqrt{q}$. In GNC parametrization: $(1/2)\pounds _{\ell}q_{ab}=(1/2)\partial _{u}q_{ab}$ and $\pounds _{\ell}\ln \sqrt{q}=\partial _{u}\ln \sqrt{q}$. Thus the Lie variation of gravitational momentum for affine parametrization is directly related to $\mathcal{D}$, i.e. to $(\Theta _{ab}\Theta ^{ab}-\Theta ^{2})$. 

For non-affine parametrization the gravitational momentum associated with null surfaces can be taken as $\Pi ^{ab}=\sqrt{q}(\Theta ^{ab}-(\Theta +\kappa)q^{ab})$. Then we readily arrive at the Lie variation expression:
\begin{align}\label{Paper06_App03_Eq37}
-q_{ab}\pounds _{\ell}\Pi ^{ab}&=2\sqrt{q}\left(\Theta ^{ab}\Theta _{ab}-\Theta ^{2}\right)+\pounds _{\ell}\left(\sqrt{q}\Theta \right)+2\sqrt{q}\pounds _{\ell}\kappa
\end{align}
Since $\kappa$ is a scalar Lie variation term can be written as: $\pounds _{\ell}\kappa =d\kappa/d\lambda$, where $\lambda$ is the parameter along $\ell ^{a}$. If we consider the null generator $\ell ^{a}$ from GNC then we arrive at (see \ref{Paper06_App03_Eq34b}):
\begin{align}\label{Paper06_App03_Eq38}
-q_{ab}\pounds _{\ell}\Pi ^{ab}=\sqrt{q}\ell _{a}g^{bc}\pounds _{\ell}N^{a}_{bc}-\dfrac{d ^{2}\sqrt{q}}{d\lambda ^{2}}+\frac{2}{\sqrt{q}}\dfrac{d}{d\lambda}\left(\sqrt{q}\kappa \right)
\end{align}
while if we have defined the conjugate momenta $\Pi ^{ab}$ as $\Pi ^{ab}=\sqrt{q}(\Theta ^{ab}-\Theta q^{ab})$, then the above relation could have been written as,
\begin{align}\label{Paper06_NewIntr}
-q_{ab}\pounds _{\ell}\Pi ^{ab}=\sqrt{q}\ell _{a}g^{bc}\pounds _{\ell}N^{a}_{bc}-\dfrac{d ^{2}\sqrt{q}}{d\lambda ^{2}}+2\kappa \dfrac{d}{d\lambda}\left(\sqrt{q}\right)
\end{align}

\section{Derivation of Various Expressions Used in Text}\label{Paper06_App05_Derv}

This appendix will contain derivations of most of the results that we have used in the main text. The derivations will be arranged in the same order as that in the main text. First we will present derivations related to the Navier-Stokes equation and then we will present the requisite derivations of subsequent sections.
\subsection{Derivation Regarding Navier-Stokes Equation}

The first thing to compute is the Lie derivative of the object $N^{c}_{ab}$. This can be obtained starting from the first principle, i.e., using expression for $N^{c}_{ab}$ in terms of $\Gamma ^{a}_{bc}$ and then using Lie variation of the connection. This immediately leads to:
\begin{align}\label{Paper06_App05_Eq01}
\pounds _{v}N^{a}_{bc}&=Q^{ad}_{be}\pounds _{v}\Gamma ^{e}_{cd}+Q^{ad}_{ce}\pounds _{v}\Gamma ^{e}_{bd}
\nonumber
\\
&=Q^{ad}_{be}\left(\nabla _{c}\nabla _{d}v^{e}+R^{e}_{~dmc}v^{m}\right)+Q^{ad}_{ce}\left(\nabla _{b}\nabla _{d}v^{e}+R^{e}_{~dmb}v^{m}\right)
\nonumber
\\
&=\frac{1}{2}\left(\delta ^{a}_{b}\nabla _{c}\nabla _{d}v^{d}+\delta ^{a}_{c}\nabla _{b}\nabla _{d}v^{d}\right)-\frac{1}{2}\left(\nabla _{b}\nabla _{c}v^{a}+\nabla _{c}\nabla _{b}v^{a}\right)
\nonumber
\\
&-\frac{1}{2}\left(R^{a}_{~bmc}+R^{a}_{~cmb}\right)v^{m}
\end{align}
In the above expression the second term in the last line can be written as:
\begin{align}\label{Paper06_App05_Eq02}
\left(\nabla _{b}\nabla _{c}v^{a}+\nabla _{c}\nabla _{b}v^{a}\right)&=2\partial _{b}\partial _{c}v^{a}+2\Gamma ^{a}_{bd}\partial _{c}v^{d}+2\Gamma ^{a}_{cd}\partial _{b}v^{d}-2\Gamma ^{d}_{bc}\partial _{d}v^{a}
\nonumber
\\
&+v^{d}\left(\partial _{b}\Gamma ^{a}_{cd}+\partial _{c}\Gamma ^{a}_{bd}\right)-2\Gamma ^{d}_{bc}\Gamma ^{a}_{de}v^{e}
+\Gamma ^{a}_{bd}\Gamma ^{d}_{ce}v^{e}+\Gamma ^{a}_{cd}\Gamma ^{d}_{be}v^{e}
\end{align}
In order to compute the Lie variation of $N^{c}_{ab}$ along the transverse direction we need the two objects $\pounds _{\ell}N^{A}_{ur}$ and $\pounds _{\ell}N^{A}_{BC}$. For the evaluation of $\pounds _{\ell}N^{A}_{ur}$ the following identities will be useful:
\begin{align}\label{Paper06_App05_Eq03}
\left(\nabla _{b}\nabla _{c}\ell ^{a}+\nabla _{c}\nabla _{b}\ell ^{a}\right)^{A}_{ur}&=2\partial _{u}\partial _{r}\ell ^{A}+2\Gamma ^{A}_{ud}\partial _{r}\ell ^{d}+2\Gamma ^{A}_{rd}\partial _{u}\ell ^{d}-2\Gamma ^{d}_{ur}\partial _{d}\ell ^{A}
\nonumber
\\
&+v^{d}\left(\partial _{u}\Gamma ^{A}_{rd}+\partial _{r}\Gamma ^{A}_{ud}\right)-2\Gamma ^{d}_{ur}\Gamma ^{A}_{de}\ell ^{e}
+\Gamma ^{A}_{ud}\Gamma ^{d}_{re}v^{e}+\Gamma ^{A}_{rd}\Gamma ^{d}_{ue}\ell ^{e}
\nonumber
\\
&=2\partial _{u}\beta ^{A}+4\alpha \Gamma ^{A}_{ur}+2\beta ^{B}\Gamma ^{A}_{uB}-2\beta ^{A}\Gamma ^{r}_{ur}+\partial _{u}\Gamma ^{A}_{ru}+\partial _{r}\Gamma ^{A}_{uu}
\nonumber
\\
&-2\Gamma ^{d}_{ur}\Gamma ^{A}_{du}+\Gamma ^{A}_{ud}\Gamma ^{d}_{ur}+\Gamma ^{A}_{rd}\Gamma ^{d}_{uu}
\end{align}
and
\begin{align}\label{Paper06_App05_Eq04}
\left(\nabla _{b}\nabla _{c}\ell ^{a}+\nabla _{c}\nabla _{b}\ell ^{a}\right)^{A}_{ur}&+\left[\left(R^{a}_{~bmc}+R^{a}_{~cmb}\right)\ell ^{m}\right]^{A}_{ur}=2\partial _{u}\beta _{A}+4\alpha \Gamma ^{A}_{ur}+2\beta ^{B}\Gamma ^{A}_{uB}-2\beta ^{A}\Gamma ^{r}_{ur}
\nonumber
\\
&+\partial _{u}\Gamma ^{A}_{ru}+\partial _{r}\Gamma ^{A}_{uu}-2\Gamma ^{d}_{ur}\Gamma ^{A}_{du}+\Gamma ^{A}_{ud}\Gamma ^{d}_{ur}+\Gamma ^{A}_{rd}\Gamma ^{d}_{uu}+R^{A}_{~uur}
\nonumber
\\
&=2\partial _{u}\beta ^{A}+2\beta ^{B}\Gamma ^{A}_{uB}+2\partial _{u}\Gamma ^{A}_{ur}
\nonumber
\\
&=\partial _{u}\beta ^{A}+\beta ^{B}q^{AC}\partial _{u}q_{BC}
\end{align}
While for $\pounds _{\ell}N^{A}_{BC}$ we have:
\begin{align}\label{Paper06_App05_Eq05}
\left(\nabla _{b}\nabla _{c}\ell ^{a}+\nabla _{c}\nabla _{b}\ell ^{a}\right)^{A}_{BC}&+\left[\left(R^{a}_{~bmc}+R^{a}_{~cmb}\right)\ell ^{m}\right]^{A}_{BC}=-2\beta ^{A}\Gamma ^{r}_{BC}+\partial _{B}\Gamma ^{A}_{uC}+\partial _{C}\Gamma ^{A}_{uB}-2\Gamma ^{d}_{BC}\Gamma ^{A}_{ud}
\nonumber
\\
&+\Gamma ^{A}_{Bd}\Gamma ^{d}_{uC}+\Gamma ^{A}_{Cd}\Gamma ^{d}_{uB}+\partial _{u}\hat{\Gamma}^{A}_{BC}-\partial _{C}\Gamma ^{A}_{Bu}+\Gamma ^{A}_{ud}\Gamma ^{d}_{BC}
\nonumber
\\
&-\Gamma ^{A}_{Cd}\Gamma ^{d}_{uB}+\partial _{u}\hat{\Gamma }^{A}_{BC}-\partial _{B}\Gamma ^{A}_{Cu}+\Gamma ^{A}_{ud}\Gamma ^{d}_{BC}-\Gamma ^{A}_{Bd}\Gamma ^{d}_{uC}
\nonumber
\\
&=-2\beta ^{A}\Gamma ^{r}_{BC}+2\partial _{u}\hat{\Gamma}^{A}_{BC}
\end{align}
These are the expressions used to get expressions in \ref{Paper06_NS}. From the vector $\Omega _{a}$ given in \ref{Paper06_Sec04_Eq02} we can calculate the Lie variation along $\ell ^{a}$ leading to,
\begin{align}\label{Paper06_App05_Eq06}
\pounds _{\ell}\Omega _{n}&=\ell ^{m}\partial _{m}\Omega _{n}+\Omega _{m}\partial _{n}\ell ^{m}
\nonumber
\\
&=\left(0,\frac{1}{2}\beta _{A}\beta ^{A},\frac{1}{2}\partial _{u}\beta _{A}\right)
\end{align}
and equivalently,
\begin{align}\label{Paper06_App05_Eq07}
\pounds _{\ell}\Omega ^{n}&=\ell ^{m}\partial _{m}\Omega ^{n}-\Omega ^{m}\partial _{m}\ell ^{n}
\nonumber
\\
&=\left(0,0,\frac{1}{2}\partial _{u}\beta ^{A}\right)
\end{align}
Also,
\begin{align}\label{Paper06_App05_Eq08}
D_{m}\Theta ^{m}_{a}&=\partial _{B}\Theta ^{B}_{A}+\partial _{C}\ln \sqrt{q}\Theta ^{C}_{A}-\hat{\Gamma} ^{C}_{AB}\Theta ^{B}_{C}
\nonumber
\\
&=\frac{1}{2}\partial _{B}\left(q^{BC}\partial _{u}q_{AC}\right)+\frac{1}{2}q^{CD}\partial _{C}\ln \sqrt{q}\partial _{u}q_{AD}-\frac{1}{2}q^{BD}\partial _{u}q_{CD}\hat{\Gamma}^{C}_{AB}
\end{align}
Using these results we finally obtain,
\begin{align}\label{Paper06_App05_Eq09}
q^{n}_{a}\pounds _{\ell}\Omega _{n}&+D_{m}\Theta ^{m}_{a}+\Theta \Omega _{a}-D_{a}\left(\Theta +\alpha \right)=\frac{1}{2}\partial _{u}\beta _{A}+\partial _{B}\left(\frac{1}{2}q^{BC}\partial _{u}q_{AC}\right)
\nonumber
\\
&+\frac{1}{2}q^{CD}\partial _{u}q_{AD}\partial _{C}\ln \sqrt{q}-\frac{1}{2}q^{BD}\partial _{u}q_{CD}\hat{\Gamma}^{C}_{AB} 
+\partial _{u}\ln \sqrt{q}\frac{1}{2}\beta _{A}-\partial _{A}\partial _{u}\ln \sqrt{q}-\partial _{A}\alpha 
\end{align}
In raising the free index of the above equation the following identities can be useful
\begin{subequations}
\begin{align}
&-q^{CD}q^{AB}\partial _{u}q_{BC}\partial _{D}\ln \sqrt{q}=\partial _{u}\left(q^{AD}\partial _{D}\ln \sqrt{q}\right)
-q^{AD}\partial _{u}\partial _{D}\ln \sqrt{q}
\label{Paper06_App05_Eq10a}
\\
\partial _{u}\left(q^{AD}\partial _{D}\ln \sqrt{q}\right)&-q^{AB}\partial _{u}q^{CF}\partial _{C}q_{BF}-q^{AB}\partial _{D}\left(q^{CD}\partial _{u}q_{BC}\right)
\nonumber
\\
&=-\partial _{u}\left(q^{BC}\hat{\Gamma}^{A}_{BC}\right)+q^{CF}\partial _{u}q^{AB}\partial _{C}q_{BF}-q^{AB}\partial _{D}q^{CD}\partial _{u}q_{BC}
\nonumber
\\
&=-\partial _{u}\left(q^{BC}\hat{\Gamma}^{A}_{BC}\right)
\label{Paper06_App05_Eq10b}
\\
q^{AB}q^{CD}\partial _{u}q_{ED}\hat{\Gamma}^{E}_{BC}&=\hat{\Gamma}^{A}_{FC}\partial _{u}q^{CF}-q^{AB}\partial _{C}q_{BF}\partial _{u}q^{CF}
\label{Paper06_App05_Eq10c}
\end{align}
\end{subequations}
Moreover we also have:
\begin{align}\label{Paper06_App05_Eq11}
R_{ab}\ell ^{a}q^{b}_{c}&=R_{uA}=G_{uA}=\frac{1}{2}\partial _{u}\beta _{A}-\partial _{A}\alpha +\frac{1}{2}q^{BC}\partial _{u}\partial _{B}q_{CA}+\frac{1}{2}\partial _{B}q^{BC}\partial _{u}q_{AC}-\partial _{u}\partial _{A}\ln \sqrt{q}
\nonumber
\\
&+\frac{1}{2}\beta _{A}\partial _{u}\ln \sqrt{q}+\frac{1}{2}q^{BC}\partial _{u}q_{AC}\partial _{B}\ln \sqrt{q}-\frac{1}{2}q^{BD}\partial _{u}q_{CD}\hat{\Gamma}^{C}_{~AB} 
\end{align}
It can be checked that this expression coincides exactly with \ref{Paper06_Sec04_Eq08b} in \ref{Paper06_NS} as it should.

To bring out the physics associated with Noether current and its various projections we compute the Noether potential and hence the Noether current completely in GNC for the vector $\xi ^{a}$. To start with we provide all the components of the tensor $\nabla _{a}\xi _{b}$, which are:
\begin{align}
\left(\nabla _{a}\xi _{b}\right)_{uu}&=-r\partial _{u}\alpha;\qquad \left(\nabla _{a}\xi _{b}\right)_{ur}=\alpha +r\partial _{r}\alpha ;\qquad \left(\nabla _{a}\xi _{b}\right)_{ru}=-\alpha -r\partial _{r}\alpha
\label{Paper06_App05_Eq12a}
\\
\left(\nabla _{a}\xi _{b}\right)_{uA}&=r\partial _{A}\alpha -r\partial _{u}\beta _{A};\qquad \left(\nabla _{a}\xi _{b}\right)_{Au}=-r\partial _{A}\alpha ;\qquad \left(\nabla _{a}\xi _{b}\right)_{rA}=-\frac{1}{2}\beta _{A}-\frac{1}{2}r\partial _{r}\beta _{A}
\label{Paper06_App05_Eq12b}
\\
\left(\nabla _{a}\xi _{b}\right)_{rr}&=0 ;\qquad \left(\nabla _{a}\xi _{b}\right)_{Ar}=\frac{1}{2}\beta _{A}+\frac{1}{2}r\partial _{r}\beta _{A};
\nonumber
\\
\left(\nabla _{a}\xi _{b}\right)_{AB}&=\frac{1}{2}\partial _{u}q_{AB}+\frac{1}{2}r\left(\partial _{A}\beta _{B}-\partial _{B}\beta _{A}\right);\qquad \left(\nabla _{a}\xi _{b}\right)_{BA}=\frac{1}{2}\partial _{u}q_{AB}-\frac{1}{2}r\left(\partial _{A}\beta _{B}-\partial _{B}\beta _{A}\right)
\label{Paper06_App05_Eq12c}
\end{align} 
Then components of Noether potential $J_{ab}=\nabla _{a}\xi _{b}-\nabla _{b}\xi _{a}$ have the following expression:
\begin{align}
J_{uu}&=0;\qquad J_{ur}=2\alpha +2r\partial _{r}\alpha ;\qquad J_{uA}=2r\partial _{A}\alpha -r\partial _{u}\beta _{A}
\label{Paper06_App05_Eq13a}
\\
J_{rA}&=-\beta _{A}-r\partial _{r}\beta _{A};\qquad J_{AB}=r\left(\partial _{A}\beta _{B}-\partial _{B}\beta _{A}\right)
\label{Paper06_App05_Eq13b}
\end{align}
The upper components of Noether potential can be obtained as
\begin{align}
J^{uu}&=0;\qquad J^{ur}=-2\alpha -2r\partial _{r}\alpha -r\beta _{A}\beta ^{A}-r^{2}\beta ^{A}\partial _{r}\beta _{A}
\label{Paper06_App05_Eq14a}
\\
J^{uA}&=-\beta ^{A}-rq^{AB}\partial _{r}\beta _{B};\qquad J^{rr}=r^{3}\beta ^{A}\beta ^{B}\left(\partial _{A}\beta _{B}-\partial _{B}\beta _{A}\right)
\label{Paper06_App05_Eq14b}
\\
J^{rA}&=2rq^{AB}\partial _{B}\alpha -rq^{AB}\partial _{u}\beta _{B}-2r^{2}\alpha q^{AB}\partial _{r}\beta _{B}-r^{3}\beta ^{2}q^{AB}\partial _{r}\beta _{B}
\nonumber
\\
&-r^{2}q^{AB}\beta ^{C}\left(\partial _{B}\beta _{C}-\partial _{C}\beta _{B}\right)+2r^{2}\beta ^{A}\partial _{r}\alpha -r^{3}\beta ^{A}\beta ^{B}\partial _{r}\beta _{B}
\label{Paper06_App05_Eq14c}
\\
J^{AB}&=-r\beta ^{A}q^{BC}\left(\beta _{C}+r\partial _{r}\beta _{C}\right)+r\beta ^{B}q^{AC}\left(\beta _{C}+r\partial _{r}\beta _{C}\right)+rq^{AC}q^{BD}\left(\partial _{C}\beta _{D}-\partial _{D}\beta _{C}\right)
\label{Paper06_App05_Eq14d}
\end{align}
Using the above components of Noether potential the components of Noether current can be obtained as
\begin{align}
J^{u}(\xi)&=-4\partial _{r}\alpha -\beta ^{2}-2\alpha \partial _{r}\ln \sqrt{q}-\frac{1}{\sqrt{q}}\partial _{A}\left(\sqrt{q}\beta ^{A}\right)
\label{Paper06_App05_Eq15a}
\\
J^{r}(\xi)&=2\alpha \partial _{u}\ln \sqrt{q}+2\partial _{u}\alpha 
\label{Paper06_App05_Eq15b}
\\
J^{A}(\xi)&=\frac{1}{\sqrt{q}}\partial _{u}\left(\sqrt{q}\beta ^{A}\right)+q^{AB}\partial _{u}\beta _{B}-2q^{AB}\partial _{B}\alpha 
\label{Paper06_App05_Eq15c}
\end{align}
Note that $k_{a}J^{a}(\xi)=-J^{u}(\xi)$, $q^{a}_{b}J^{b}(\xi)=J^{A}(\xi)$ and finally $\ell _{a}J^{a}(\xi)=J^{r}(\xi)$. As we will see all of them matches with our desired expressions. Also in the stationary limit we have $\partial _{u}\alpha=\partial _{u}\beta _{A}=\partial _{u}q_{AB}=0$, which in particular tells us that $J^{r}=0$. Hence in the static limit Noether current is on the null surface since its component along $k^{a}$ (which is $-\ell _{a}J^{a}(\xi)$) vanishes. Also in this case we have:
\begin{align}\label{Paper06_App05_Eq16}
\left(\nabla _{b}\nabla _{c}\xi ^{a}+\nabla _{c}\nabla _{b}\xi ^{a}\right)^{A}_{ur}&+\left[\left(R^{a}_{~bmc}+R^{a}_{~cmb}\right)\xi ^{m}\right]^{A}_{ur}=
\partial _{u}\Gamma ^{A}_{ru}+\partial _{r}\Gamma ^{A}_{uu}-\Gamma ^{d}_{ur}\Gamma ^{A}_{du}+\Gamma ^{A}_{rd}\Gamma ^{d}_{uu}+R^{A}_{~uur}
\nonumber
\\
&=2\partial _{u}\Gamma ^{A}_{ur}=-\partial _{u}\beta ^{A}
\end{align}
as well as,
\begin{align}\label{Paper06_App05_Eq17}
\left(\nabla _{b}\nabla _{c}\xi ^{a}+\nabla _{c}\nabla _{b}\xi ^{a}\right)^{A}_{BC}&+\left[\left(R^{a}_{~bmc}+R^{a}_{~cmb}\right)\xi ^{m}\right]^{A}_{BC}=\partial _{B}\Gamma ^{A}_{uC}+\partial _{C}\Gamma ^{A}_{uB}-2\Gamma ^{d}_{BC}\Gamma ^{A}_{ud}
\nonumber
\\
&+\Gamma ^{A}_{Bd}\Gamma ^{d}_{uC}+\Gamma ^{A}_{Cd}\Gamma ^{d}_{uB}+\partial _{u}\hat{\Gamma}^{A}_{BC}-\partial _{C}\Gamma ^{A}_{Bu}+\Gamma ^{A}_{ud}\Gamma ^{d}_{BC}
\nonumber
\\
&-\Gamma ^{A}_{Cd}\Gamma ^{d}_{uB}+\partial _{u}\hat{\Gamma }^{A}_{BC}-\partial _{B}\Gamma ^{A}_{Cu}+\Gamma ^{A}_{ud}\Gamma ^{d}_{BC}-\Gamma ^{A}_{Bd}\Gamma ^{d}_{uC}
\nonumber
\\
&=2\partial _{u}\hat{\Gamma}^{A}_{BC}
\end{align}
Using these two results we arrive at:
\begin{align}
\pounds _{\xi}N^{A}_{ur}&=\frac{1}{2}\partial _{u}\beta ^{A}
\label{Paper06_App05_Eq18a}
\\
\pounds _{\xi}N^{A}_{BC}&=\frac{1}{2}\delta ^{A}_{B}\partial _{C}\Theta +\frac{1}{2}\delta ^{A}_{C}\partial _{B}\Theta -\partial _{u}\hat{\Gamma}^{A}_{BC}
\label{Paper06_App05_Eq18b}
\end{align}
These are the expressions used in \ref{Paper06_NS}.
\subsection{Derivation Regarding Spacetime Evolution}

We need to consider the object $\ell _{a}g^{ij}\pounds _{\ell}N^{a}_{ij}$ in GNC. This in turn requires us to obtain expressions for $\pounds _{\ell}N^{r}_{ur}$ and $\pounds _{\ell}N^{r}_{AB}$. Then using the identity for Lie variation of $N^{c}_{ab}$ we can obtain both the Lie variations. For that purpose we have:
\begin{align}
\frac{1}{2}\Big(\delta ^{a}_{b}\nabla _{c}\nabla _{d}\ell ^{d}&+\delta ^{a}_{c}\nabla _{b}\nabla _{d}\ell ^{d}\Big)^{r}_{ur}=\frac{1}{2}\partial _{u}\Theta +\partial _{u}\alpha
\label{Paper06_App05_Eq45a}
\\
\Big(\nabla _{b}\nabla _{c}\ell ^{a}&+\nabla _{c}\nabla _{b}\ell ^{a}\Big)^{r}_{ur}=2\partial _{u}\alpha 
\label{Paper06_App05_Eq45b}
\\
\Big[-\frac{1}{2}\Big(R^{a}_{~bmc}&+R^{a}_{~cmb}\Big)\ell ^{m}\Big]^{r}_{ur}=0
\label{Paper06_App05_Eq45c}
\\
\Big(\nabla _{b}\nabla _{c}\ell ^{a}&+\nabla _{c}\nabla _{b}\ell ^{a}\Big)^{r}_{AB}=\alpha \partial _{u}q_{AB}-\frac{1}{2}q^{CD}\partial _{u}q_{AC}\partial _{u}q_{BD}
\label{Paper06_App05_Eq45d}
\\
\Big[-\frac{1}{2}\Big(R^{a}_{~bmc}&+R^{a}_{~cmb}\Big)\ell ^{m}\Big]^{r}_{AB}=-\frac{1}{2}\alpha \partial _{u}q_{AB}+\frac{1}{2}\partial _{u}^{2}q_{AB}-\frac{1}{4}q^{CD}\partial _{u}q_{AC}\partial _{u}q_{BD}
\label{Paper06_App05_Eq45e}
\end{align}
This immediately leads to
\begin{align}
\pounds _{\ell}N^{r}_{ur}&=\frac{1}{2}\partial _{u}^{2}\ln \sqrt{q}
\label{Paper06_App05_Eq46a}
\\
\pounds _{\ell}N^{r}_{AB}&=-\alpha \partial _{u}q_{AB}+\frac{1}{2}\partial _{u}^{2}q_{AB}
\label{Paper06_App05_Eq46b}
\end{align}
Combining all the pieces and using the results $\Theta =\partial _{u}\ln \sqrt{q}$ and $\Theta _{AB}=(1/2)\partial _{u}q_{AB}$, which is the only non-zero component of $\Theta _{ab}$, \cite{Parattu:2015gga} we finally obtain
\begin{align}\label{Paper06_App05_Eq47}
\ell _{a}g^{ij}\pounds _{\ell}N^{a}_{ij}&=2\pounds _{\ell}N^{r}_{ur}+q^{AB}\pounds _{\ell}N^{r}_{AB}
\nonumber
\\
&=-2\alpha \partial _{u}\ln \sqrt{q}+2\partial _{u}^{2}\ln \sqrt{q}-\frac{1}{2}\partial _{u}q_{AB}\partial _{u}q^{AB}
\nonumber
\\
&=2\partial _{u}\alpha +2\left(\Theta _{ab}\Theta ^{ab}-\Theta ^{2}\right)+\frac{2}{\sqrt{q}}\dfrac{d ^{2}\sqrt{q}}{du^{2}} -\frac{2}{\sqrt{q}}\dfrac{d}{du}\left(\sqrt{q}\alpha \right)
\end{align}
which can also be obtained from a completely different viewpoint. For sake of completeness we will illustrate the alternative methods as well. For the null vector $\ell ^{a}$ in the adapted GNC system we have:
\begin{align}\label{Paper06_App05_Eq48}
\left(\ell ^{c}\nabla _{c}\ell ^{a}\right)^{u}&=\alpha +r\beta ^{2}+\mathcal{O}(r^{2});\qquad 
\left(\ell ^{c}\nabla _{c}\ell ^{a}\right)^{r}=r\partial _{u}\alpha +2r\alpha ^{2}+\mathcal{O}(r^{2})
\nonumber
\\
\left(\ell ^{c}\nabla _{c}\ell ^{a}\right)^{A}&=r\alpha \beta ^{A}+rq^{CA}\partial _{C}\alpha +\mathcal{O}(r^{2})
\end{align}
Hence on $r=0$ surface, we have $\kappa =\alpha$, as well as, $\tilde{\kappa}=-(1/2)k^{a}\nabla _{a}\ell ^{2}=\alpha$. Now we will use the Raychaudhuri equation to get $R_{ab}\ell ^{a}\ell ^{b}$ and hence the Lie variation term. In this case we have, $du=d\lambda$, thus Raychaudhuri equation reduces to the following form (see \ref{Paper06_App03_Eq14})
\begin{align}\label{Paper06_App05_Eq49}
\ell ^{a}\nabla _{a}\left(\Theta +2\alpha\right)&=\nabla _{c}\left(\ell ^{a}\nabla _{a}\ell ^{c}\right)-\nabla _{a}\ell _{b}\nabla ^{b}\ell ^{a}-R_{ab}\ell ^{a}\ell ^{b}
\end{align}
Where, the $\Theta +2\alpha$ term comes from $\nabla _{i}\ell ^{i}$. Then we have,
\begin{align}\label{Paper06_App05_Eq50}
\nabla _{c}\left(\ell ^{a}\nabla _{a}\ell ^{c}\right)&=\partial _{c}\left(\ell ^{a}\nabla _{a}\ell ^{c}\right)+\ell ^{a}\nabla _{a}\ell ^{c}\partial _{c}\ln \sqrt{q}
\nonumber
\\
&=2\alpha ^{2}+\alpha \partial _{u}\ln \sqrt{q}+2\partial _{u}\alpha
\end{align}
Thus non zero components of $B_{ab}=\nabla _{a}\ell _{b}$ are as follows:
\begin{align}\label{Paper06_App05_Eq51}
B_{ur}=\alpha ;\qquad B_{rA}=\frac{1}{2}\beta _{A};\qquad B_{AC}=\frac{1}{2}\partial _{u}q_{AC}
\end{align}
From which it can be easily derived that, $B_{ab}B^{ba}=2\alpha ^{2}-(1/4)\partial _{u}q_{AB}\partial _{u}q^{AB}$. Thus we obtain
\begin{align}\label{Paper06_App05_Eq52}
R_{ab}\ell ^{a}\ell ^{b}&=-\partial _{u}\Theta +2\alpha ^{2}+\Theta \alpha 
-B_{ab}B^{ba}
\nonumber
\\
&=\alpha \Theta -\frac{1}{2}q^{AB}\partial _{u}^{2}q_{AB}
-\frac{1}{4}\partial _{u}q_{AB}\partial _{u}q^{AB}
\nonumber
\\
&=\alpha \Theta -\frac{1}{\sqrt{q}}\partial _{u}^{2}\sqrt{q}+\left(\partial _{u}\ln \sqrt{q}\right)^{2}-\Theta _{ab}\Theta ^{ab}
\end{align}
where $\Theta _{ab}$ has the only non-zero component to be, $\Theta _{AB}=(1/2)\partial _{u}q_{AB}$. For the GNC null normal $\ell _{a}$, the Noether current vanishes, such that Lie variation of $N^{a}_{bc}$ turns out to have the following expression
\begin{align}\label{Paper06_App05_Eq53}
\ell _{a}g^{ij}\pounds _{\ell}N^{a}_{ij}&=-2R_{ab}\ell ^{a}\ell ^{b}
\nonumber
\\
&=2\partial _{u}\alpha +2\left(\Theta _{ab}\Theta ^{ab}-\Theta ^{2}\right)+\frac{2}{\sqrt{q}}\dfrac{d ^{2}\sqrt{q}}{du^{2}} -\frac{2}{\sqrt{q}}\dfrac{d}{du}\left(\sqrt{q}\alpha \right)
\end{align} 
The components of $S_{ab}=\nabla _{a}\ell _{b}+\nabla _{b}\ell _{a}$ In GNC are as follows:
\begin{align}\label{Paper06_App05_Eq54}
S_{uu}&=2r\partial _{u}\alpha -4r\alpha ^{2}+\mathcal{O}(r^{2});\qquad
S_{ur}=2\alpha +2r\partial _{r}\alpha +r\beta ^{2}+\mathcal{O}(r^{2})
\nonumber
\\
S_{uA}&=-r\beta ^{B}\partial _{u}q_{AB}+2r\partial _{A}\alpha -2r\alpha \beta _{A}+\mathcal{O}(r^{2});\qquad
S_{rr}=0
\nonumber
\\
S_{rA}&=\beta _{A}+r\partial _{r}\beta _{A}-r\beta ^{C}\partial _{r}q_{CA}+\mathcal{O}(r^{2})
\nonumber
\\
S_{AB}&=\partial _{u}q_{AB}+2r\alpha \partial _{r}q_{AB}+r\left(D_{A}\beta _{B}+D_{B}\beta _{A}\right)
+\mathcal{O}(r^{2})
\end{align}
Thus the trace at $r=0$ leads to: $S=4\alpha +2\partial _{u}\ln \sqrt{q}$. Thus we arrive at the following expression (see \ref{Paper06_App03_Eq32} of \ref{Paper06_App03_General})
\begin{align}\label{Paper06_App05_Eq55}
\ell _{a}g^{ij}\pounds _{\ell}N^{a}_{ij}&=2\partial _{u}\left(\Theta +2\alpha \right)
-\partial _{b}S^{rb}-\Gamma ^{r}_{bc}S^{bc}-S^{rc}\partial _{c}\ln \sqrt{q}
\end{align}
Then the upper components of $S_{ab}$ necessary for the above computation are the followings:
\begin{align}\label{Paper06_App05_Eq56}
S^{ur}&=S_{ur}+r\beta ^{A}S_{rA}=2\alpha +2r\partial _{r}\alpha +2r\beta ^{2}+\mathcal{O}(r^{2})
\nonumber
\\
S^{rr}&=2r\partial _{u}\alpha +4r\alpha ^{2}+\mathcal{O}(r^{2})
\nonumber
\\
S^{rA}&=4\alpha r\beta ^{A}+2rq^{AB}\partial _{A}\alpha 
-2r\alpha \beta ^{A}+\mathcal{O}(r^{2})
\end{align}
The mixed components leads to nothing new so we have not presented them. From \ref{Paper06_App05_Eq55} the expression for Lie derivative can be obtained as:
\begin{align}\label{Paper06_App05_Eq57}
\ell _{a}g^{ij}\pounds _{\ell}N^{a}_{ij}&=2\partial _{u}\left(\Theta +2\alpha \right)
-\partial _{u}S^{ru}-\partial _{r}S^{rr}-\partial _{A}S^{rA}+4\alpha ^{2}+2\Theta _{ab}\Theta ^{ab}
-2\alpha \Theta 
\nonumber
\\
&=2\left(\Theta _{ab}\Theta ^{ab}-\Theta ^{2}\right)+\frac{2}{\sqrt{q}}\dfrac{d}{d\lambda}\left(\sqrt{q}\Theta \right)-2\alpha \Theta +4\partial _{u}\alpha +4\alpha ^{2}
-4\partial _{u}\alpha -4\alpha ^{2}
\nonumber
\\
&=2\left(\Theta _{ab}\Theta ^{ab}-\Theta ^{2}\right)+\frac{2}{\sqrt{q}}\dfrac{d}{d\lambda}\left(\sqrt{q}\Theta \right)-2\alpha \Theta
\end{align}
which exactly matches with \ref{Paper06_App05_Eq53}. The same can be ascertained for \ref{Paper06_App05_Eq52} by computing $R_{ab}\ell ^{a}\ell ^{b}$ on the null surface i.e. in the $r\rightarrow 0$ limit, directly leading to:
\begin{align}\label{Paper06_App05_Eq58}
R_{ab}\ell ^{a}\ell ^{b}&=R_{uu}=\partial _{a}\Gamma ^{a}_{uu}-\partial _{u}\Gamma ^{a}_{ua}
+\Gamma ^{a}_{uu}\Gamma ^{b}_{ab}-\Gamma ^{a}_{ub}\Gamma ^{b}_{ua}
\nonumber
\\
&=\partial _{u}\Gamma ^{u}_{uu}+\partial _{r}\Gamma ^{r}_{uu}+\partial _{A}\Gamma ^{A}_{uu}
-\partial _{u}^{2}\ln \sqrt{q}+\Gamma ^{u}_{uu}\partial _{u}\ln \sqrt{q}-\Gamma ^{a}_{ub}\Gamma ^{b}_{ua}
\nonumber
\\
&=2\alpha ^{2}-\partial _{u}^{2}\ln \sqrt{q}+\alpha \partial _{u}\ln \sqrt{q}
-\Gamma ^{u}_{ub}\Gamma ^{b}_{uu}-\Gamma ^{r}_{ub}\Gamma ^{b}_{ur}-\Gamma ^{A}_{ub}\Gamma ^{b}_{uA}
\nonumber
\\
&=-\partial _{u}^{2}\ln \sqrt{q}+\alpha \partial _{u}\ln \sqrt{q}-\Theta _{ab}\Theta ^{ab}
\end{align}
which under some manipulations will match exactly with \ref{Paper06_App05_Eq52}. Then in GNC we obtain in identical fashion, the following expression for heat density,
\begin{align}\label{Paper06_App05_Eq59}
\mathcal{S}&=\nabla _{i}\ell _{j}\nabla ^{j}\ell ^{i}-\left(\nabla _{i}\ell ^{i}\right)^{2}
\nonumber
\\
&=\left(2\alpha ^{2}-(1/4)\partial _{u}q _{AB}\partial _{u}q ^{AB}\right)
-\left(\Theta +2\alpha \right)^{2}
\nonumber
\\
&=-2\alpha ^{2}-4\alpha \Theta -\Theta ^{2}+\Theta _{ab}\Theta ^{ab}
\end{align}
This on integration over the null surface leads to,
\begin{align}\label{Paper06_App05_Eq60}
\frac{1}{8\pi}\int du d^{2}x\sqrt{q}\mathcal{S}
=\frac{1}{8\pi}\int dud^{2}x\sqrt{q}\left(\Theta _{ab}\Theta ^{ab}-\Theta ^{2}\right)
-\frac{1}{4\pi}\int du d^{2}x\sqrt{q}\alpha ^{2}-4\int d^{2}xTds
\end{align}
Let us now write the integral form of $R_{ab}\ell ^{a}\ell ^{b}$, for that we note the integration measure to be $dud^{2}x\sqrt{q}$. Thus on integration with proper measure and $(1/8\pi)$ factor leads to
\begin{align}\label{Paper06_App05_Eq61}
\frac{1}{8\pi}\int dud^{2}x\sqrt{q}R_{ab}\ell ^{a}\ell ^{b}
&=-\frac{1}{8\pi}\int du d^{2}x\sqrt{q}\mathcal{D}
-\frac{1}{8\pi}\dfrac{d\mathcal{A}_{\perp}}{d\lambda}\Big \vert _{1}^{2}
+\int d^{2}x Ts\vert _{1}^{2}-\int d^{2}x sdT
\end{align}
which can be written in a slightly modified manner as:
\begin{align}\label{Paper06_App05_Eq62}
\frac{1}{8\pi}\int dud^{2}x\sqrt{q}R_{ab}\ell ^{a}\ell ^{b}
&=-\frac{1}{8\pi}\int du d^{2}x\sqrt{q}\mathcal{D}
-\frac{1}{8\pi}\dfrac{d\mathcal{A}_{\perp}}{d\lambda}\Big \vert _{1}^{2}
+\int d^{2}x Tds
\end{align}
Also the Lie variation term (with all the surface contributions kept) on being integrated over the null surface we obtain
\begin{align}\label{Paper06_App05_Eq63}
\frac{1}{16\pi}\int dud^{2}x\sqrt{q}&\times \ell _{a}g^{ij}\pounds _{\ell}N^{a}_{ij}
=\frac{1}{8\pi}\int du d^{2}x\sqrt{q}\mathcal{D}
+\frac{1}{8\pi}\dfrac{d\mathcal{A}_{\perp}}{d\lambda}\Big \vert _{1}^{2}
-\int d^{2}x~\left(\frac{\alpha}{2\pi}\right) d\left(\frac{\sqrt{q}}{4}\right)
\nonumber
\\
&=-\int d^{2}x~Tds+\frac{1}{8\pi}\int du d^{2}x\sqrt{q}\mathcal{D}
+\frac{1}{8\pi}\dfrac{d\mathcal{A}_{\perp}}{d\lambda}\Big \vert _{1}^{2}
\end{align}
To calculate Lie variation for $\xi ^{a}$ we need to calculate $\nabla _{a}\xi _{b}+\nabla _{b}\xi _{a}=S_{ab}$. This tensor has the following components:
\begin{align}\label{Paper06_App05_Eq64}
S_{uu}&=-2r\partial _{u}\alpha,\qquad S_{ur}=0
\nonumber
\\
S_{uA}&=-r\partial _{u}\beta _{A},\qquad S_{rr}=0
\nonumber
\\
S_{rA}&=0.\qquad S_{AB}=\partial _{u}q_{AB}
\end{align}
Thus in the null limit obtained from the relation: $r\rightarrow 0$ we arrive at the result that all the components of $S_{ab}$ vanishes except for the $S_{AB}$ components. If we want to satisfy the Killing condition for $\xi ^{a}$ on the null surface we would require $\partial _{u}q_{AB}=0$. From the above relations it is clear that $\nabla _{a}\xi ^{a}=\Theta$. Moreover we also have,
\begin{subequations}
\begin{align}
\kappa &=-k_{b}\xi ^{a}\nabla _{a}\xi ^{b}=-\Gamma ^{b}_{ac}k_{b}\xi ^{a}\xi ^{c}
=\Gamma ^{u}_{uu}=\alpha
\label{Paper06_App05_Eq65a}
\\
\tilde{\kappa}&=-\frac{1}{2}k_{b}\nabla ^{b}\xi ^{2}=\frac{1}{2}\partial _{r}\left(-2r\alpha\right)
=-\alpha
\label{Paper06_App05_Eq65b}
\end{align}
\end{subequations}
which shows that for $\xi ^{a}$, $\kappa =\tilde{\kappa}$. Thus even without the condition $\partial _{u}q_{AB}=0$, we arrive at the relation $\kappa =-\tilde{\kappa}=\alpha$. Moreover Lie variation of $N^{a}_{bc}$ along $\xi ^{a}$ can be obtained by computing the following objects:
\begin{align}
\frac{1}{2}\Big(\delta ^{a}_{b}\nabla _{c}\nabla _{d}\xi ^{d}&+\delta ^{a}_{c}\nabla _{b}\nabla _{d}\xi ^{d}\Big)^{r}_{ur}=\frac{1}{2}\partial _{u}\Theta 
\label{Paper06_App05_New01}
\\
\Big(\nabla _{b}\nabla _{c}\xi ^{a}&+\nabla _{c}\nabla _{b}\xi ^{a}\Big)^{r}_{ur}=-2\partial _{u}\alpha 
\label{Paper06_App05_New02}
\\
\Big[-\frac{1}{2}\Big(R^{a}_{~bmc}&+R^{a}_{~cmb}\Big)\xi ^{m}\Big]^{r}_{ur}=0
\label{Paper06_App05_New03}
\\
\Big(\nabla _{b}\nabla _{c}\xi ^{a}&+\nabla _{c}\nabla _{b}\xi ^{a}\Big)^{r}_{AB}=-\alpha \partial _{u}q_{AB}-\frac{1}{2}q^{CD}\partial _{u}q_{AC}\partial _{u}q_{BD}
\label{Paper06_App05_New04}
\\
\Big[-\frac{1}{2}\Big(R^{a}_{~bmc}&+R^{a}_{~cmb}\Big)\xi ^{m}\Big]^{r}_{AB}=-\frac{1}{2}\alpha \partial _{u}q_{AB}+\frac{1}{2}\partial _{u}^{2}q_{AB}-\frac{1}{4}q^{CD}\partial _{u}q_{AC}\partial _{u}q_{BD}
\label{Paper06_App05_New05}
\end{align}
which can be used to obtain the Lie variation term associated with $\xi ^{a}$ as,
\begin{align}\label{Paper06_App05_NewLie}
\ell _{a}g^{ij}\pounds _{\xi}N^{a}_{ij}&=2\partial _{u}\alpha+2\left(\Theta _{ab}\Theta ^{ab}-\Theta ^{2}\right)+\frac{2}{\sqrt{q}}\partial _{u}^{2}\sqrt{q}
\nonumber
\\
&=\frac{2}{\sqrt{q}}\partial _{u}\left(\alpha \sqrt{q}\right)+\ell _{a}g^{ij}\pounds _{\ell}N^{a}_{ij}
\end{align}
Then using the momentum $\Pi ^{ab}=\sqrt{q}[\Theta ^{ab}-q^{ab}(\Theta +\kappa)]$ conjugate to the induced metric $q_{ab}$ from \ref{Paper06_App03_Eq37} we immediately arrive at,
\begin{align}\label{Paper06_New_10}
-q_{ab}\pounds _{\xi}\Pi ^{ab}=\sqrt{q}\ell _{a}g^{ij}\pounds _{\xi}N^{a}_{ij}-\dfrac{d^{2}\sqrt{q}}{d\lambda ^{2}}
\end{align}
These expressions are used to obtain \ref{Paper06_Sec06_Eq16}. Also the variational principles in this context are:
\begin{subequations}
\begin{align}
Q_{1}&=\int d\lambda d^{2}x\sqrt{q}\left(-\frac{1}{8\pi}R_{ab}\ell ^{a}\ell ^{b}
+T_{ab}\ell ^{a}\ell ^{b}\right)
\nonumber
\\
&=\int d\lambda d^{2}x\sqrt{q}\Big[\frac{1}{8\pi}\mathcal{D}
+T_{ab}\ell ^{a}\ell ^{b}\Big]
-\int d^{2}x~Tds+\frac{1}{8\pi}\dfrac{d\mathcal{A}_{\perp}}{d\lambda}\Big \vert _{1}^{2}
\label{Paper06_App05_Eq69a}
\\
Q_{2}&=\int d\lambda d^{2}x \sqrt{q}\left[\frac{1}{16\pi}\ell _{a}g^{ij}\pounds _{\xi}N^{a}_{ij}
+T_{ab}\ell ^{a}\ell ^{b}\right]
\nonumber
\\
&=\int d\lambda d^{2}x \sqrt{q}\Big[\frac{1}{8\pi}\mathcal{D}
+T_{ab}\ell ^{a}\ell ^{b}\Big]+\int d^{2}x~sdT+\frac{1}{8\pi}\dfrac{d\mathcal{A}_{\perp}}{d\lambda}\Big \vert _{1}^{2}
\label{Paper06_App05_Eq69b}
\end{align}
\end{subequations}
These are the expressions used in \ref{Paper06_Evo}.


\bibliography{Gravity_1_full,Gravity_2_partial}

\bibliographystyle{./utphys1}
\end{document}